\documentclass[]{interact}

\usepackage{epstopdf}
\usepackage[caption=false]{subfig}
\usepackage{svg}
\usepackage{mathtools}
\usepackage[english]{babel}
\usepackage[autostyle]{csquotes}
\usepackage[export]{adjustbox}
\usepackage{array}
\usepackage{placeins}
\usepackage{amsbsy}
\usepackage{graphicx}
\usepackage{epstopdf}
\AppendGraphicsExtensions{.tif}

\usepackage[numbers,sort&compress]{natbib}
\bibpunct[, ]{[}{]}{,}{n}{,}{,}
\renewcommand\bibfont{\fontsize{10}{12}\selectfont}
\makeatletter
\def\NAT@def@citea{\def\@citea{\NAT@separator}}
\makeatother

\theoremstyle{plain}

\theoremstyle{definition}

\theoremstyle{remark}

\begin{document}


\title{Circulation in turbulent flow through a contraction}

\author{
\name{Vivek Mugundhan\textsuperscript{a,b} and Sigurdur T. Thoroddsen\textsuperscript{a}\thanks{CONTACT S. T. Thoroddsen. Email: Sigurdur.Thoroddsen@KAUST.edu.sa}}
\textsuperscript{a}\affil{Division of Physical Sciences and Engineering, King Abdullah University of Science and Technology (KAUST), Thuwal 23955-6900, Saudi Arabia; \textcolor{black}{\textsuperscript{b}Department of Mechanical Engineering, Amrita School of Engineering Coimbatore, Amrita Vishwa Vidyapeetham, India.}}
}
\maketitle

\begin{abstract}
\textcolor{black}{We study experimentally the statistical properties and evolution of circulation in a turbulent flow passing through a smooth 2-D contraction.  The turbulence is generated with an active grids to reach $Re_{\lambda} \simeq 220$ at the inlet to the 2.5:1 contraction.  We employ time-resolved 3-D Lagrangian Particle Tracking technique with the Shake-The-Box algorithm to obtain volumetric velocity fields which we use to calculate the simultaneous circulation in three perpendicular planes.  Forming a circulation vector and studying the PDFs of the relative strength of its components, we can quantify how the mean strain enhances and orients coherent vortical structures with the streamwise direction.  This is further studied with streamwise space and time correlations of the circulations over a range of loop sizes.  The streamwise component of the circulation, over same-size square loops, shows increased integral length, while the other two components are less affected.  The circulation around the compressive direction weakens and reaches prominent negative correlation values, suggesting buckling or sharp reorientation of transverse vortices.  
The PDFs of circulation transit from non-Gaussian to Gaussian behavior as the loop size is increased from dissipative to large scales.}
\end{abstract}


\section{Introduction}

The quest for universality in the statistical description of turbulence has lead to studies on the characteristics of velocity structure functions \textcolor{black}{based on the cascade ideas of Richardson and Kolmogorov \cite{Davidson_Book}.  The statistics of circulation have been added to the arsenal used for these studies \cite{Migdal1994,Sreenivasan1995,Sreenivasan1996}. }  
By virtue of the Stokes theorem, circulation statistics help in obtaining the characteristics of vorticity, without directly computing or measuring the vorticity field. 
Measurements of vorticity involve gradients of velocities, its deductions is therefore less accurate compared to the velocity measurements themselves.
Hence there are advantages in using circulation to characterize coherent vortical structures from experiments, which often do not resolve the finest scales.
\textcolor{black}{With the advent of volumetric velocity measurements new avenues have opened up for the study of circulation which were inaccessible with the traditional hot-wire single-point studies of turbulence.  Herein we exploit volumentric Lagrangian Particle Tracking methods to go beyond planar PIV to measure time-resolved circulations simultaneously in three perpendicular planes.  We do this for turbulent flow subjected to mean strain as it is passed through a 2-D contraction, while previous studies of circulation statistics have been focused on homogeneous and isotropic configurations.  We can thereby study how the applied mean strain enhances and orients the direction components of the circulation vector, which we interpret as being characteristic of coherent vortical structures.  Below we list previous studies of turbulence in contractions and then describe earlier work on circulation in turbulence.}

\textcolor{black}{The contraction is an important geometry of study in turbulence which has wide range of engineering applications, e.g. in ducting and pipe connections.
Early experiments in axisymmetric contractions are those of Uberoi \cite{Uberoi1956}, who investigated effect of three different axisymmetric contraction-ratios 
of \textcolor{black}{$c=$} ${\it {4, 9, 16}}$, for a square cross-section duct. 
Following this were the experiments by Hussain and Ramjee \cite{Hussain1976}, wherein they studied the effect of contraction wall shapes with the same \textcolor{black}{$c=$} 11.
Tan-Atichat {\it et al.} \cite{Tanatichat1980} performed experiments with effect of axisymmetric contractions on grid turbulence for \textcolor{black}{$c$} ranging from 2$-$36, and Ayyalasomayajula and Warhaft \cite{Ayyalasomayajula2006} investigated grid-generated turbulence subjected to mean strain using an axisymmetric contraction (4:1) in a wind tunnel.}
Recently, Ertun\c{c} and Durst \cite{Ertunc2008} have questioned the earlier experimental work which is based on using two-component $\times$-configuration hot-wires.
They performed experiments in contraction with \textcolor{black}{$c=$} 14.75 and showed the continuous decrease of streamwise fluctuations, by carefully accounting for possible errors involved in hot-wire measurements.
Brown {\it et al.} \cite{Brown2006} used two-component Laser-Doppler Velocimetry (LDV) to study turbulence in a planar contraction.
\textcolor{black}{Thoroddsen and Van Atta \cite{Thoroddsen1995a} investigated the effect of a 2-D contraction ($c=$ 2.5) in a thermally-stratified wind tunnel, with a strong linear temperature gradient. In their subsequent work \cite{Thoroddsen1995b} they studied turbulence in a constant-area vertical expansion.}

Circulation over a closed contour $C$ cutting through a flow domain, is obtained by integrating the velocity vector field $\bf{u}$ projected onto the tangent of the curve along the closed contour, given by

\begin{equation}
\Gamma_A=\oint_{C}{\bf{u}}\cdot dl=\oint_{A}{\boldsymbol{\omega}\cdot dA}.
\label{eqn:gamma_defn}
\end{equation}
Here, $A$ is the area enclosed by the closed contour $C$ and $\boldsymbol{\omega}=\nabla\times\bf{u}$ is the vorticity field. 
As the contour can be characterized by a linear dimension $r$, we also denote $\Gamma_A$ calculated by \textcolor{black}{Eq. (\ref{eqn:gamma_defn})} as $\Gamma_r$. 
In this equation, $dl$ and $dA$ represent the contour elemental tangent length and the elemental area enclosed by the contour respectively.

The earliest studies of circulation in turbulent flows has been by Migdal \cite{Migdal1994} and Sreenivsan et al. \cite{Sreenivasan1995,Sreenivasan1996}. 
Some of the main characteristics of the probability density function (PDF) of circulation obtained in homogeneous-isotropic turbulence are: (i) the PDF of circulation satisfies the area rule \cite{Migdal1994} i.e, its form is only a function of the size of area enclosed by the contour but not the contour's shape; (ii) by the law of large numbers, the PDF should be Gaussian for large values of $A$, given that the dimensions of the contour are in the inertial range (IR).
Another aspect of interest is the scaling of circulation with $r$ and their comparison with Kolmogorv's scaling of the velocity increments.
According to Kolmogorov theory \cite{Migdal1994,K41}, the $n$-th moment of circulation scales with $r$ as, 

\begin{equation}
\left \langle {\left (\Gamma_A  \right )}^n \right \rangle\sim r^{\lambda_n} \sim r^{4n/3} .
\label{eqn:gamma_scaling}
\end{equation}

As $A=r^2$, we can write this scaling as, $\left \langle {\left (\Gamma_A  \right )}^n \right \rangle\sim A^{\lambda_n/2}$.
This scaling is strictly applicable when the circulation PDFs are Gaussian distributed.

The circulation properties and their scaling have been investigated in experimental \cite{Sreenivasan1995,Zhou2008,Thoroddsen1996}, numerical simulations \cite{Umeki1993,Benzi1997,Yoshida2000,Iyer2019,Iyer2021} and theoretical studies \cite{Migdal1994,Moriconi1998}.
Moriconio and Takakura \cite{Moriconi1998} obtained statistics using saddle-point based Martin-Siggia-Rose (MSR) technique.  
Migdal \cite{Migdal1994} first started studies on the statistical properties of circulation.
He proposed that the circulation PDF depends on the area of the loop but not the shape.
Umeki \cite{Umeki1993} showed that the circulation second moment for a figure-eight loop (loop consisting of two square loops with one common vertex) scales with the scalar area of the constituent loops, for 2-D decaying isotropic turbulence. 
He suggested that the area rule be called scalar-area rule and modifications to the functional forms used in the circulation statistics to match the observed behavior.
Benzi et al. \cite{Benzi1997} also showed that the circulation PDF depends on the area and not the shape for 3-D shear flows using his simulations at Taylor-Reynolds number Re$_\lambda \sim$ 40.
Recently, Iyer et al. \cite{Iyer2019} showed that the PDF of circulation is independent of dimensions of rectangular loops over the entire PDF even at very high Re$_\lambda$ using DNS, which goes against Migdal's arguments that area rule is applicable only for the tails.
They performed high resolution simulations for Re$_\lambda$ in the range 140 to 1300 using meshes up to $16384^3$.
At higher Re$_\lambda$, area rule becomes applicable over wider range of scales in the IR.
They also showed that the circulation computed over a figure-eight loop scaled with the scalar sum of area of the loops.
\textcolor{black}{In their subsequent work \cite{Iyer2021}, they test the validity of this area rule for non-planar loops.
For non-planar loops, the area taken is the minimal surface enclosed by the loop.}
\textcolor{black}{The PDFs of circulation 
differed for planar and non-planar loops having the same area and hence the area rule is not true.
However, the PDF of the circulation normalized by the 8-th moment, shows a good collapse for both planar and non-planar loops.}  

The circulation PDF is Gaussian for large loops with its dimension in the IR, however deviates from the Gaussian behavior as the loop size tends to the dissipation scales.
Sreenivasan {\it et al.} \cite{Sreenivasan1995} reports this in their seminal early study wherein they explored the circulation properties in the turbulent wake of a cylinder at Re$_\lambda \leq$ 40, using two-dimensional particle image velocimetry (PIV).
They observed that the PDF deviated from Gaussian and approached exponential form for smaller loop areas.
Later, Zhou et al. \cite{Zhou2008} experimentally measured the PDFs and the moments of velocity and circulation structure functions in turbulent Rayleigh-B\'{e}nard convection in a cylindrical convection cell using \textcolor{black}{PIV} technique.
The measurements were carried out in central region of the cells, in a vertical plane, far from the boundaries and hence they approximated the flow to be homogeneous and isotropic.
They showed the flatness of the PDF tended to the Gaussian value 3 with increase in the characteristic dimension.
Thoroddsen \cite{Thoroddsen1996} presented an approach based on conditional averaging, to study the spatial structure of dissipation in the context of validation of Kolmogorov’s refined similarity hypothesis. 
He presented the PDF of a large-scale vorticity about the transverse direction for grid turbulence at Re$_\lambda=$230. 
The vorticity is calculated using velocities measured across finite distances using multiple probes consisting of 4 single-wire and one X-wire probe. 
This quantity being computed over finite distance can be viewed as circulation of loop with a characteristic dimension to be of the order of the probe spacing. 
In line with this analogy, he reported the large-scale vorticity is nearly-Gaussian distributed with a flatness of 3.2.

\textcolor{black}{The transition of circulation PDF from non-Gaussian to Gaussian has also been observed in direct numerical simulations (DNS) \cite{Sreenivasan1996,Umeki1993,Benzi1997,Yoshida2000,Iyer2019,Iyer2021}.}
Umeki \cite{Umeki1993} reports this behavior at Re$_\lambda=$ 100 in 2-D decaying isotropic turbulence.
Cao et al. \cite{Sreenivasan1996} showed this for a more isotropy-controlled homogeneous-isotropic turbulence at $Re_\lambda=$ 101, 181 and 216.
They showed that the flatness of the PDF tends to Gaussian value of 3, as the contour approaches the integral scale, but remains higher than that of the velocity increments, indicating circulation to be more intermittent. 
This higher intermittency of circulation PDF was also reported by Yoshida and Hatakeyama \cite{Yoshida2000} in their DNS data at $Re_\lambda=$ 90.
Using a threshold value for vorticity magnitude, they decomposed vorticity into strong and weak components.
Circulation computed based on the weak vorticity showed a self-similar behavior with loop size and was nearly Gaussian.
The flatness of the distribution was close to 3 when threshold vorticity magnitude is chosen to twice its rms value.
On the other hand, circulation associated with the strong component showed stretched tails and departure from the self-similarity. 
They concluded that the circulation associated with strong-vortical activity contributed to the non-Gaussian behavior.
They proposed a model which matches the asymptotic behavior of the strong component circulation.

\textcolor{black}{Recent studies by Iyer {\it et al.} \cite{Iyer2019,Iyer2021} showed that for rectangular loops of different aspect ratios, but the same area, the circulation PDFs collapse into a single curve, for loops contained inside the IR.}
However, this fails when one of the dimension lies outside the IR.
It is to be noted that unlike the previous studies, they present PDF for circulation $\Gamma$ normalized by the product of root-mean-square velocity fluctuation and the integral length scale.
They showed that the flatness of the distribution is 3 at large scales, increases to a higher value and remains fairly constant in the IR.
At small scales, the flatness value approaches its dissipation-limit which is given by the much larger flatness values of the vorticity component normal to the circulation plane.
They observed that for the highest Re$_\lambda $, the circulation flatness remained high, but constant with loop size in the IR, which indicates lower intermittency of circulation.
By extrapolation, they predicted the intermittency-free limit to be at Re$_\lambda=$ 1900.
One of the important finding in this work is the scaling properties of circulation's moments. 
The moments have a simpler bifractal behavior - (a) lower moments $n<3$, $\lambda_n=$ 1.367$n$ scaling which is very close to the Kolmogorov scaling, 
and (b) for $n>3$ (up to 10) has sub-Kolomogorov scaling with a fractal dimension of 2.2.

\textcolor{black}{Circulation statistics have attracted attention in turbulence after the theoretical work of Migdal \cite{Migdal1994}, following which Sreenivasan and his co-workers \cite{Sreenivasan1995,Sreenivasan1996,Iyer2019,Iyer2021} have explored the characteristics of circulation with experiments and DNS.
The major issue that has been in focus is the validity of the area rule.
However, we see from this literature survey that most of the studies have focused on the idealized isotropic turbulence and the experimental studies are based on two-dimensional PIV techniques, providing $\Gamma$ only in one plane.}
The flow measured in these studies, which include wakes and thermal-convection cells which are far from isotropic on the large scales.
In the current study, we look at circulation statistics in a non-canonical turbulent flow - strained flow through a smooth contraction. 
We employ the state-of-the-art \textcolor{black}{ time-resolved 3-D Lagrangian Particle Tracking (LPT) technique using high-speed cameras to track particle trajectories. 
The particle velocities are then mapped on a grid in volumetric slice to obtain the Eulerian velocity field.}
This enables us to present circulation over loops contained in mutually perpendicular planes, and such results have not been reported.
By virtue of the measurement technique, we present the relative strengths of the circulations computed on the three mutually-perpendicular planes with center aligned with the center of the contraction.
\textcolor{black}{Such 3-D measurement techniques make it possible to compute circulation in non-planar and complex loops and hence verify the limits of the area rule.}
The effect of straining on the evolution of circulation statistics like its rms and PDF are presented.
We obtain the space and space-time correlations of the circulations and deduce from there an integral-scale based on the spatial correlation.
The moments of the PDFs and the scaling exponents for varying size of loops are also  presented.

\begin{figure}
\begin{center}
\includegraphics[width= 0.97\textwidth]{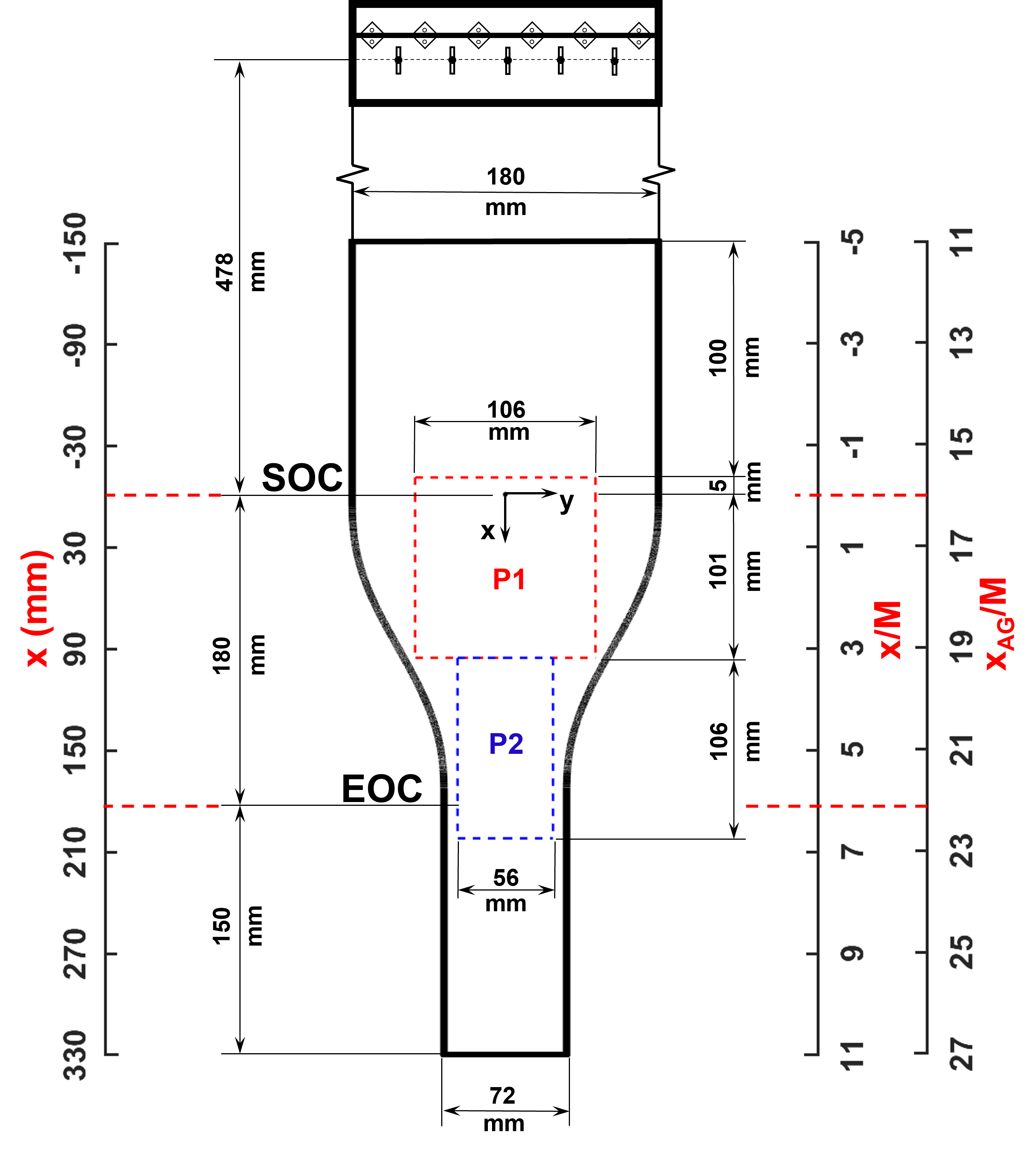}
\caption{
\textcolor{black}{Measurement regions are shown with respect to the coordinate axis positioned at the {\it Start of Contraction} ($SOC$). 
$EOC$ denotes the {\it End of Contraction}.
Regions P1 and P2 approximately measure 106 mm ${\times}$ 106 mm and 56 mm ${\times}$ 106 mm respectively.  
Here $x_{AG}$ represents the distance from the bottom shaft of the active grid.
The flow is along the positive $x$ coordinate.}}
\label{fig:cont}
\end{center}
\end{figure}

\section{Experimental methodology}
\subsection{Experimental setup}
We study the evolution of circulation through a smooth 2-D contraction (contraction ratio, \textcolor{blue}{$c=$} 2.5) in a constant-head vertical water tunnel. 
The 2D schematic of the contraction with the measurement regions is shown in Figure \ref{fig:cont}. 
\textcolor{black}{Water from the overhead tank flows through a straight vertical inlet duct of cross section 260 mm$ {\times}$ 260 mm, where the background turbulence is suppressed with flow conditioners.
Following this, the flow is guided smoothly into the active grid module through a 203-mm long converging section which reduces the cross section to 180 mm$ {\times}$ 180 mm.
The active grid with spacing of $M=$ 30 mm greatly enhances the turbulence level.
Downstream of the grid, water flows through a 240-mm straight section which homogenizes the flow before it enters the contraction.
The $180$-mm long contraction smoothly reduces the cross section from $180 \times 180$ mm to $72 \times 180$ mm.
Constant head is maintained in the overhead tank using a centrifugal pump. 
The level in the overhead tank and thus the flow rate through the tunnel  is controlled using ball valves, after the pump. 
All experiments were performed with the valve set to $\left \langle U_{in} \right \rangle \approx$ 0.3 ms$^{-1}$ at the inlet to the contraction.
A detailed description, illustrations of the experimental setup and the active grid used in this study can be found in our previous work, Mugundhan {\it et al.} \cite{Mugundhan2020}.
Note that, the modified setup with a 240-mm extension inserted between the grid and the contraction, as discussed in \cite{Mugundhan2020}, is used in this study.
} 
All measurements are performed 5 minutes after a steady flow is achieved in the loop.  The experiments are performed at 21$^{\circ}$C, giving water density of $\rho=$998 kg/m$^3$ and dynamics viscosity of $\mu=$9.79$\times$10$^{-4}$ Pa-s.

For this study, we choose two grid rotation protocols $-$ the synchronous mode 1 (S1) and the random mode (R). 
Of the modes employed in \cite{Mugundhan2020}, it is seen that modes S1 and R injected the highest level of turbulence.
In mode S1, all shafts rotate synchronously at a constant speed of $N=$4 rps (Rossby number, $Ro=2\left<U_{in}\right>/M\Omega \approx$0.80, where $\Omega=2\pi N$) with alternate shafts rotating in the opposite direction.
In the random mode, the shafts are rotated in the single-random protocol at a constant speed of $N=$3.5 rps ($Ro\approx$0.91).
The cruise time varies randomly between the period of 180$^{\circ}$ to 540$^{\circ}$ of a rotation. 
This ensures that the average cruise time equals the period of one complete revolution.
\textcolor{black}{The random mode results in the highest Taylor Reynolds number, of $Re_{\lambda} \sim 220$ at the start of contraction.}

\subsection{Imaging, calibration and seeding particles}
\textcolor{black}{We use the LaVision Tomo-PIV imaging system to acquire the particle images and obtain volumetric flow field.}
Particles are illuminated using a dual-cavity Nd-YLF, 527 nm green laser and LaVision volume optics (see \cite{Mugundhan2020} for details on the illumination setup).
\textcolor{black}{We use the 3-D LPT technique with the shake-the-box (STB) algorithm \cite{Schanz2016} to track the particles. 
The particle velocities are then mapped on to a volumetric grid to obtain the three-dimensional Eulerian velocity field.}
Region P1 is inside the contraction, with its top portion 5 mm above the start of the contraction ({\it SOC}), and region P2 is at the bottom of the contraction, which includes the exit and the following straight section.
\textcolor{black}{The volumetric regions P1 and P2 measure $106 \times 106 \times 24$ mm$^3$ and  $56 \times 106 \times 24$ mm$^3$ respectively.
Even though the illumination volume was 24 mm thick, only the velocity calculations on a grid of thickness $\approx$ 18$-$19 mm was used in the statistics.
The spurious vectors seen occasionally close to the edges were therefore not included in the analysis.}

Imaging is done using four high-speed video cameras (Phantom v2640) and 85 mm Nikkor tilt lenses at an aperture of f/22. 
The tilt lenses are employed to ensure that a greater depth of the measurement volume remains in focus. 
The cameras can go up to 6600 fps at full 2048$\times$1952 resolution.
\textcolor{black}{The arrangement of the cameras is similar to that used in the measurements in \cite{Mugundhan2020}.
Cameras 1 \& 2 are placed on one side of the contraction viewing its front side, while 
Cameras 3 \& 4 are placed on the opposite side.
For good reconstruction quality, we maintain optimum angles between the cameras.
Cameras viewing from opposite sides are positioned so that they do not have co-linear lines of sight.
For position P1, the angles between C1/C2 and C3/C4 are 32${^{\circ}}$ and 30${^{\circ}}$ respectively, while for the narrower section in P2 they are 29${^{\circ}}$ and 27${^{\circ}}$.
In P1, cameras C1, C2, C3, C4 make angles 106${^{\circ}}$, 74${^{\circ}}$, 75${^{\circ}}$ and 105${^{\circ}}$ with the laser respectively.}
In region P1, images are captured with a resolution of 2048$\times$1952 px at 1300 fps. 
However, in position \textcolor{black}{P2} at the contraction exit, images are captured at a lower resolution of 1024$\times$1952 px at the same 1300 fps. 
\textcolor{black}{Imaging at P1 and P2 require major changes in the laser optics and the cameras.
Hence, experiments in the two measurement regions are run on separate days.}

Spatial calibration is done using two 11.8 mm thick 4-plane calibration plate from LaVision (Number: 106-10). 
The whole calibration plate of size 106 mm $\times$ 106 mm was used for spatial calibration in region P1. 
For the smaller region P2, a modified plate with a reduced size of 56 mm $\times$ 106 mm was used.
Calibration is performed with the test section filled with water and normal lighting.
Third-order polynomials that map the physical coordinates to the image coordinates are obtained using the LaVision DaVis software. 
The calibration is followed by volume self-calibration, which accounts for any misalignments such as slight changes in the position of cameras and other optical disturbances during the experiment. 
We use Tween-80 surfactant-treated fluorescent red or orange polyethylene microspheres (from Cospheric) as seeding particles. 
The particles have diameters in the range $63-106$ ${\mu}$m, which are close to neutrally buoyant with a density of 1.05 g/cm$^3$ which gives a Stokes number of 2${\times}$10$^{-4}$.

\subsection{Velocity computation using STB}
We use shake-the-box (STB) algorithm for the velocity computations.
\textcolor{black}{STB is a 4-D Lagrangian particle-tracking velocity measurement {algorithm} that can handle high seeding density in the flow \cite{Schanz2016,Schanz2013a}.}
The raw images acquired by the cameras are pre-processed by subtracting a sliding minimum over 5 pixels and normalizing intensities with a local average. 
In the next step, the volume self-calibration is done to improve the original calibration using the pre-processed images. 
This step accounts for any misalignments in the camera positions and other optical disturbances that could occur during the conduct of the experiment. 
The STB algorithm tracks the particles in time and hence makes the Lagrangian velocities available, which is then mapped on to a regular Eulerian grid. 
The three-dimensional vectors are smoothed over 3$\times$ voxels for spatial and time continuity of the vectors. 
We use the the STB algorithm implemented in the LaVision’s DaVis (version 10.1.1) software in this study.
The algorithm results in about 115,000 and 30,000 tracks in regions P1 and P2 respectively. 
We use velocities mapped on a grid of \textcolor{black}{size 48$\times$48$\times$48 voxels} with an overlap of 75\% and a time filter length of 5 time steps for the circulations statistics. 
\textcolor{black}{We use 3-D velocity grids of size 176$\times$165$\times$33 \& 57$\times$168$\times$35 in regions P1 and P2 respectively.
This gives us 0.96 M vectors \& 0.34 M vectors in regions P1 and P2 respectively.
On an average there are 10$-$15 particles identified per 48$\times$48$\times$48 voxels which is used for velocity calculations.
The particle concentration varied between 0.03$-$0.06 particles-per-pixel.}
The grid resolution with this setting is $\delta$= 0.64 mm, which $\sim$ 2.1 $-$ 2.4 times the smallest Kolmogorov scale. 
This choice of grid resolution was based on our systematic study on the effect of resolution and time filter length on the velocities in \cite{Mugundhan2020}.
\textcolor{black}{The uncertainty in the velocity measurements and the convergence of our statistics are included in Appendix E.}

\begin{table}
\centering
\begin{tabular}{l*{3}{c}r}
Parameters & Sync Mode 1  & Random \\ 
\vspace{0.1in}\\
\hline \\
${{\left \langle u^2 \right \rangle}^{1/2} / \left \langle U_{in} \right \rangle}$ ${{\%}}$ & 5.3 & 6.4 \\
${k}$ (m$^2$ s$^{-2}$) & ${4.83 \times10^{-4}}$ & ${6.12 \times10^{-4}}$\\
${\varepsilon}$ (m$^2$ s$^{-3}$) & ${1.76 \times10^{-4}}$ & ${1.20 \times10^{-4}}$ \\
${L}$ (mm) & 14.49 & 25.80 \\
${\lambda}$ (mm) & 5.98 & 8.65 \\
${\eta}$ (mm) & 0.271 & 0.298 \\
${Re_{L}}$ & 325 & 652 \\
${Re_{\lambda}}$ & 134 & 218 \\
$\delta/\eta$ & 2.4 & 2.1 \\
$S^{*}=Sk/{\varepsilon}$ & 2.5 & 4.4 \\
\vspace{0.1in}\\
\hline 
\end{tabular}\vspace{0.1in}\\
\caption{Turbulence flow parameters at the start of contraction  ($x_{AG}/M=16$). Here ${\left \langle U_{in} \right \rangle}$ is the mean inlet velocity, $u$ is the streamwise velocity fluctuation, ${k}$ is the turbulence kinetic energy, ${\varepsilon}$ is the dissipation rate, ${L}$ is the integral length scale, ${\lambda}$ is the Taylor microscale based on the streamwise correlation function $f$, ${\eta}$ is the Kolmogorov length scale, ${Re_{L}}$, ${Re_{\lambda}}$ are Reynolds numbers based on $L$ and ${\lambda}$, $\delta$ is the velocity-grid resolution and $S^{*}$ is turbulence-to-mean-strain time ratio.}
\label{tab:parameters}
\end{table}

\begin{figure}[htbp]
\begin{center}
  \centering
  \includegraphics[width= 1.0\textwidth]{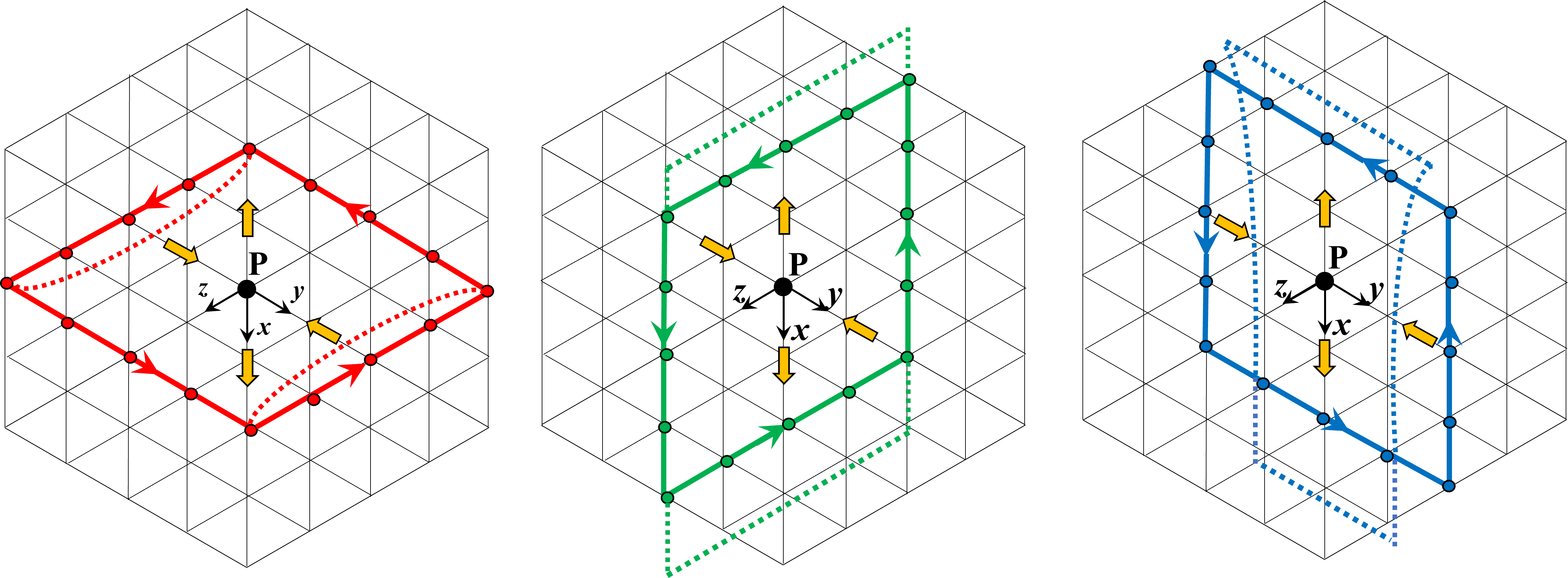}
  \caption{The mutually perpendicular square loops used to compute circulations about point $P$. 
  The loops shown in the left, middle and the right are used to calculate $\Gamma_{x}$, $\Gamma_{y}$ and $\Gamma_{z}$ respectively.
  A cubical 5$\times$5$\times$5 Eulerian grid is shown in the figure as an example and the circles represent the velocity locations on the grid. 
  The arrows along the side of the loop indicate the direction of integration.
  The stretching and compression directions are indicated by the bold arrows.
  The dotted lines indicate a deformed fluid element which initially matches with the square loop.}
\label{fig:loops}
\end{center}
\end{figure}

\subsection{Turbulence parameters}

The important turbulence parameters computed for a point on the centerline and at $SOC$ ($x=$ 0) is tabulated in Table \ref {tab:parameters}. 
The notation used in this paper is the following: the instantaneous velocity $U=\left \langle U \right \rangle+u$, $V=\left \langle V \right \rangle+v$, $W=\left \langle W \right \rangle+w$,
where $\left \langle U \right \rangle$, $\left \langle V \right \rangle$ and $\left \langle W \right \rangle$ are the time averaged velocities and, $u$, $v$ and $w$ denote the fluctuations. 
The values presented are averaged over independent realizations. 
In each mode, 4 cases are recorded in region P1 and 6 cases in region P2. 
Each realization occupied the full memory of the four cameras ($\approx$ 1 terabyte). 

The mean streamwise velocity at the inlet to the contraction varied between 0.29 and 0.31 ms$^{-1}$ in the experiments. 
As seen in our previous work, the random mode results in the highest $Re_{\lambda}$ which is $\sim$ 220. 
The lengths scales presented are based on the spatial velocity correlation function $f$.
The Taylor microscale estimated as $\sqrt{u^2/{\left \langle \partial u/\partial x \right \rangle}^2 }$ \cite{Pope2000} gives slightly smaller values for $Re_{\lambda}$. 
Using this estimate, the largest of our $Re_{\lambda}$ at the start of contraction, which occurs for the random mode, reduces from 218 to 170.

\begin{figure}
\begin{tabular}{ p{0.06cm} c c }
{} & Synchronous Mode & Random Mode\\
{} & {} & {}\\ 
\adjustbox{valign=t}{(a)} & \includegraphics[width=0.49\textwidth,valign=t]{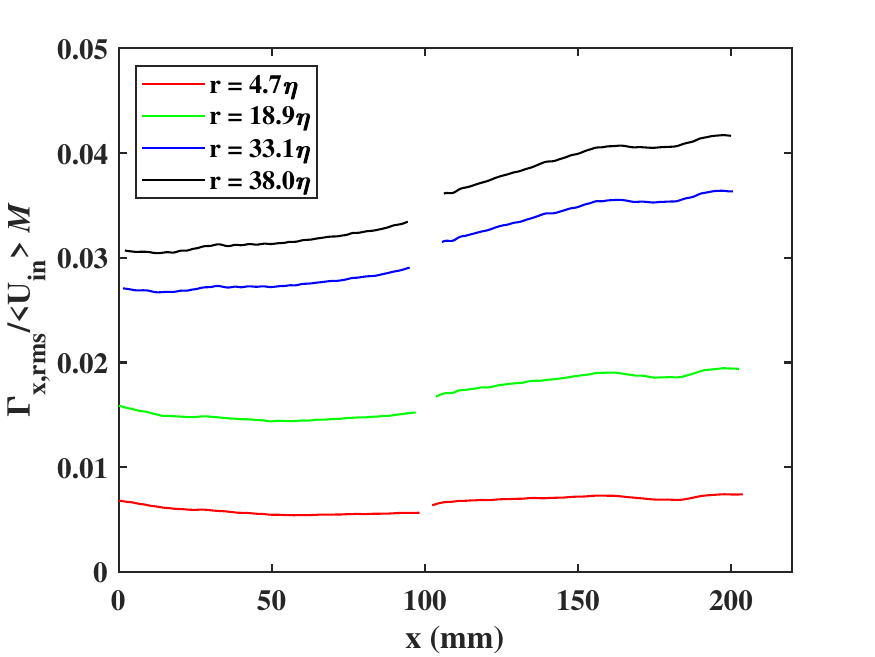} & \includegraphics[width=0.49\textwidth,valign=t]{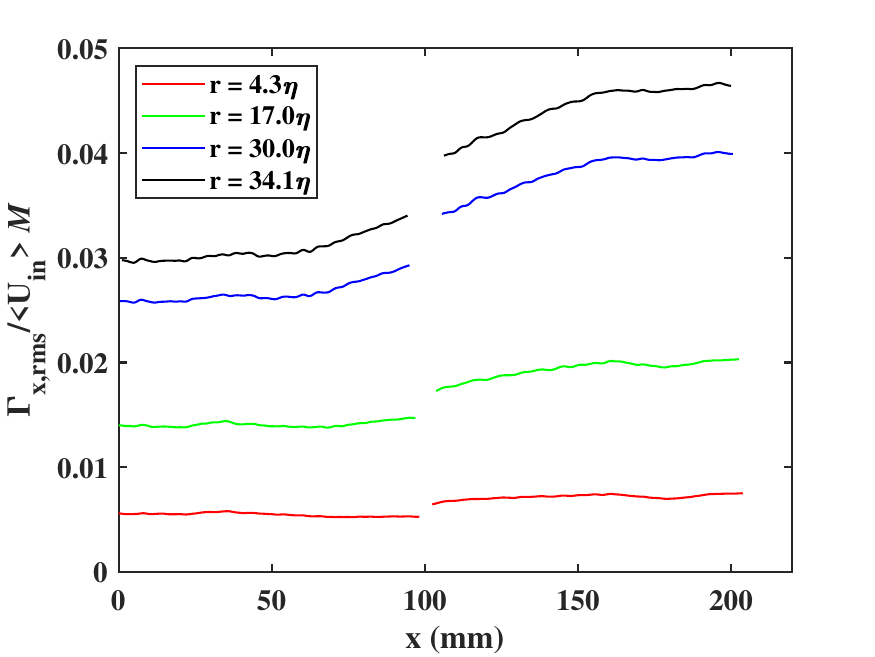} \\
\adjustbox{valign=t}{(b)} & \includegraphics[width=0.49\textwidth,valign=t]{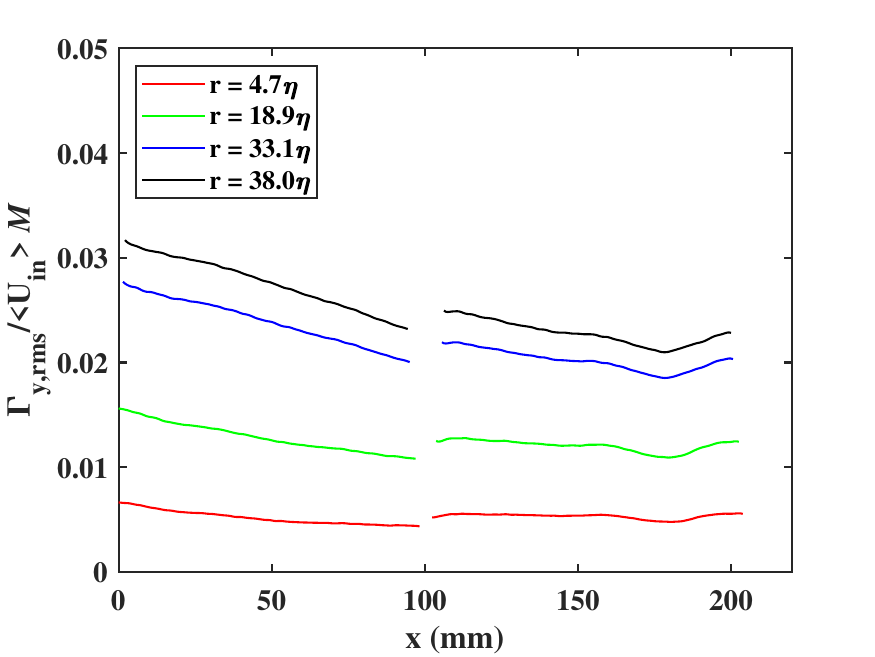} & \includegraphics[width=0.49\textwidth,valign=t]{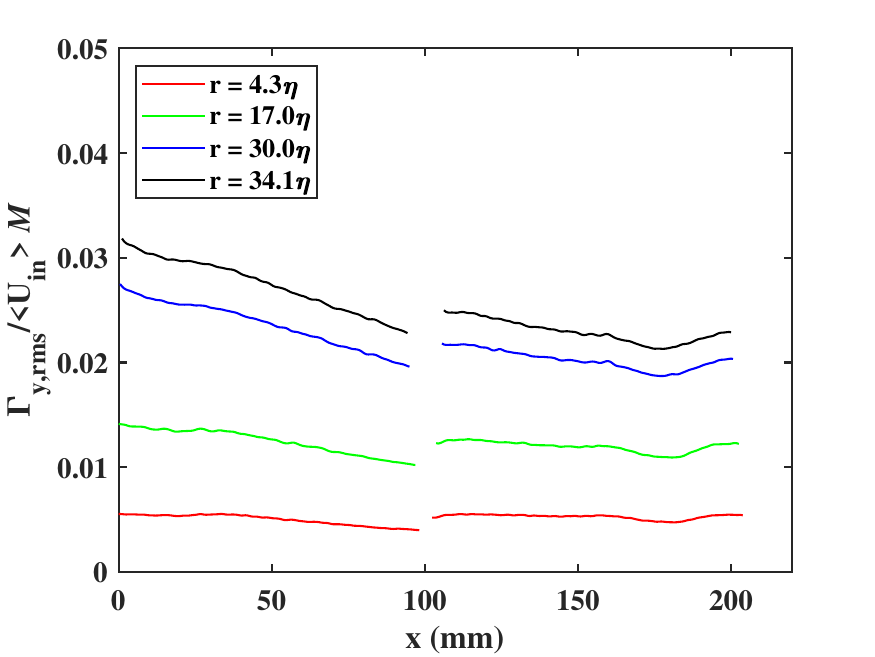} \\
\adjustbox{valign=t}{(c)} & \includegraphics[width=0.49\textwidth,valign=t]{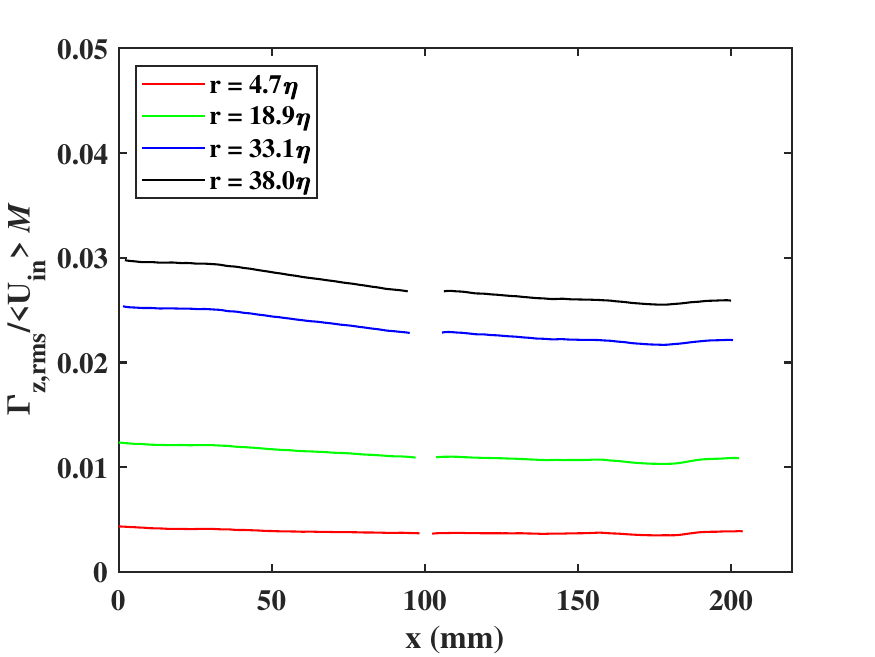} & \includegraphics[width=0.49\textwidth,valign=t]{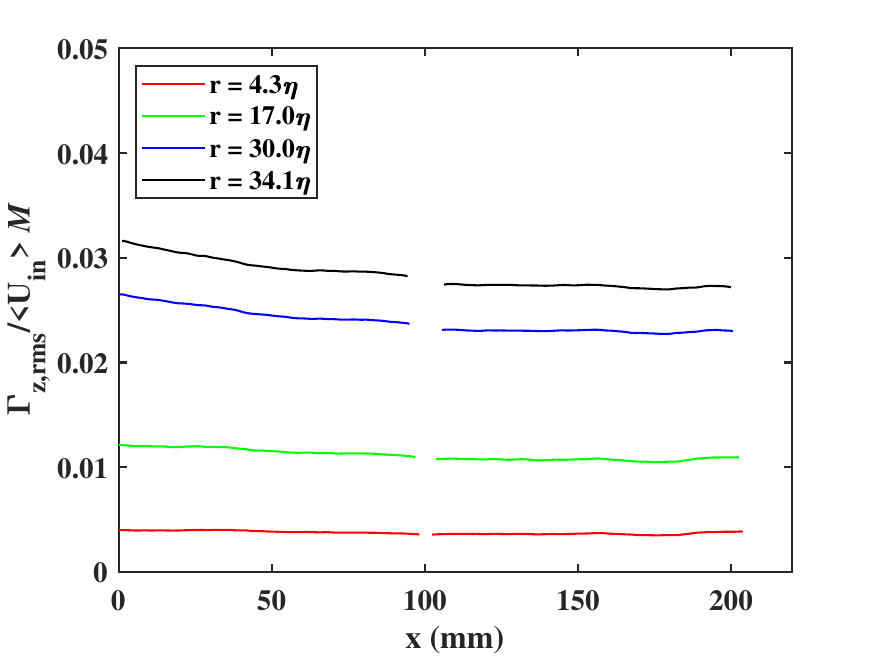} \\
\end{tabular}
\caption{Streamwise variation of circulation r.m.s along the centerline for synchronous mode S1 (left) and the random mode R (right). Circulations computed in the perpendicular planes: (a)  $\Gamma_{x,rms}$ (Straining direction); (b) $\Gamma_{y,rms}$ (Compression direction); and (c) $\Gamma_{z,rms}$.
$\left \langle U_{in} \right \rangle$ is the the area-averaged mean inlet velocity in the straight section and $M$ is the mesh size. 
}
\label{fig:RMSCirc}
\end{figure}

\section{Results and Discussion}
\subsection{Streamwise variation of circulation}

We present the effect of straining on the circulation r.m.s in this section. 
We use the velocity field obtained from the STB calculations to determine the circulation by integration along a square loop of dimension $r$ and centered about a point $P$ as shown in Figure \ref{fig:loops}.
The planes are shown in a subset of the regular Eulerian grid consisting of 5$\times$5$\times$5 grid points. 
The nomenclature used for circulation is based on the unit normal of the loop. 
For example, the circulation calculated by performing the line integration (Eq. \ref{eqn:gamma_defn}) of the velocity vector along the square loop on the extreme left in Figure \ref{fig:loops} is denoted as $\Gamma_{x}$, with $-x$ being the normal direction.
Note that the flow through the contraction is along the $+x$ direction. 
While calculating $\bf{u}\cdot dl$ over the line element $dl$ connecting two adjacent nodes, average of these velocities is used. 
The circulations $\Gamma_{x}$, $\Gamma_{y}$ and $\Gamma_{z}$ may be viewed as the rotation strength of large structures of dimensions comparable to the size of the loop . In the limiting case of the loop size shrinking to point, the circulation in these planes represents the corresponding components of the local vorticity.

Figure \ref{fig:RMSCirc} shows the streamwise variation of the r.m.s circulation $\Gamma_{x,rms}$,  $\Gamma_{y,rms}$ and $\Gamma_{z,rms}$ for the synchronous and the random modes. 
The r.m.s values are normalized using the mean velocity at the inlet to the contraction $\left \langle U_{in} \right \rangle$ and the mesh size $M$. The results shown for a particular measurement region is the average of five realizations. 
Circulation is computed for four square loops centered about the centerline of the contraction and having sizes $r=$ 2.6 mm, 5.1 mm, 8.9 mm and 10.2 mm, which cover the range of scales from near dissipative to large scales. 
This corresponds to a $r/\eta=$ 5 $-$ 40.
Note that the Taylor microscale is in the range 5 $-$ 9 mm. 

\begin{figure}
\begin{center}
\begin{tabular}{ c c }
(a) & (b)\\
\includegraphics[width=0.5\textwidth,]{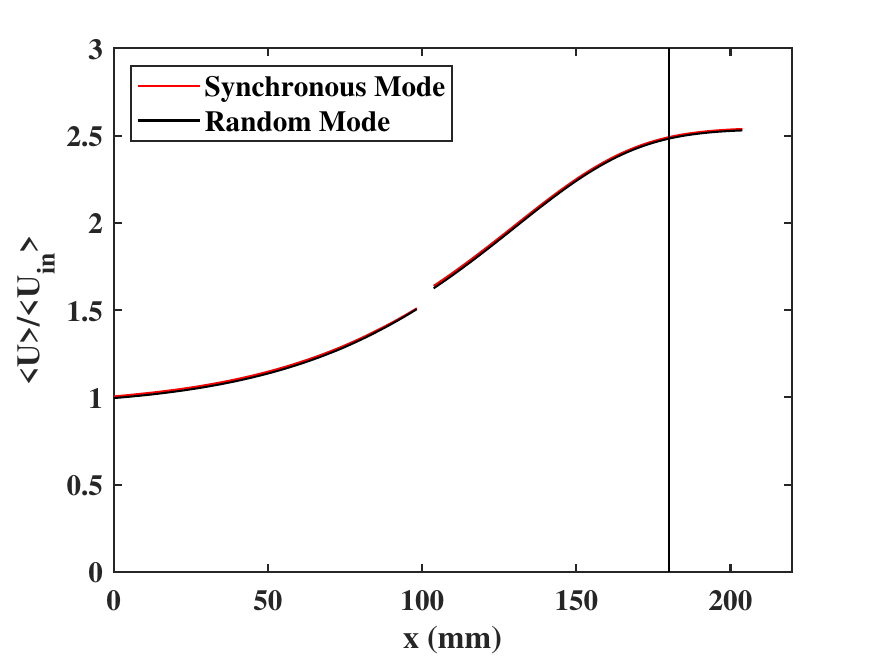} & 
\includegraphics[width=0.5\textwidth,]{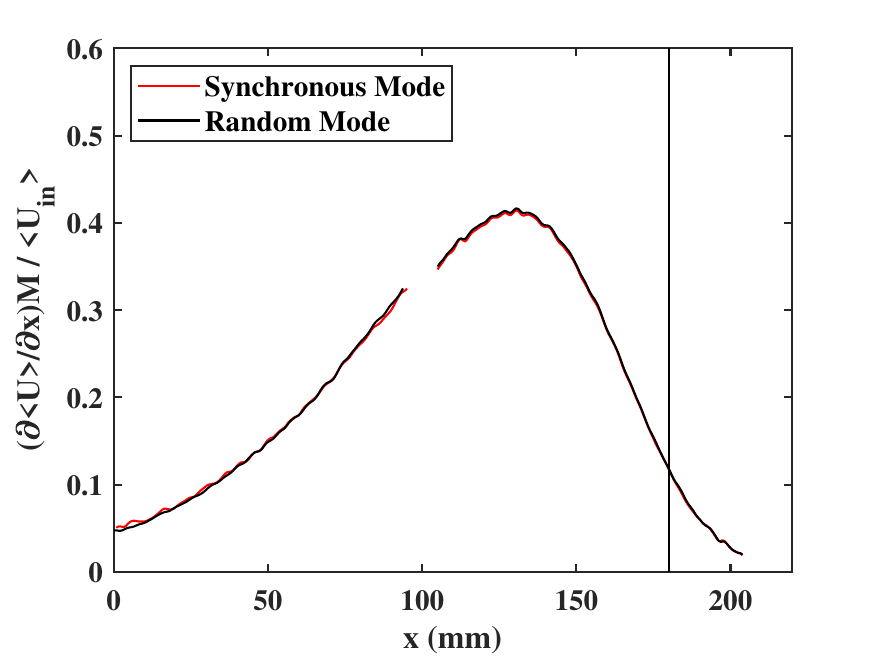} \\   
\end{tabular}
\end{center}
\caption{\textcolor{black}{Streamwise variation of (a) the streamwise velocity $\left \langle U \right \rangle$, and (b) the normalized mean strain $\partial \left \langle U \right \rangle/\partial x$ along the centerline, for the synchronous and random grid-rotation modes. Here $\left \langle U_{in} \right \rangle$ is the mean velocity at the inlet to the contraction and $M$ is the grid mesh size.
The vertical line marks the end of contraction.}}
\label{fig:strainrate}
\end{figure}

We see a slight discontinuity in the profiles between the measurement regions. 
The measurements in regions \textcolor{black}{P1 and P2} were made with different optical/camera set-up and alignment on different days with major changes in the optical system, camera orientation and different calibrations. 
As pointed out in \cite{Mugundhan2020}, the transverse inhomogeneity from the active grid and the mismatch of illuminated volumes depth could be another reason for the discontinuity in profiles. 
However, the variation is similar to that seen with the r.m.s vorticity. 
We presented the streamwise variation of all three components of velocity and vorticity for the same contraction in our earlier study \cite{Mugundhan2020}. 
In Figure \ref{fig:RMSCirc} we see that the circulation $\Gamma_{x,rms}$ increases with $x$ initially up to $x\approx$ 150 mm, following which it remains approximately constant to $x\approx$ 175 mm.
On the other hand, $\Gamma_{y,rms}$ decays till its minimum at $x\approx$ 175 mm, but then shows an increase until the exit.
This is because of the mean strain experienced by the vortical structures. 
\textcolor{black}{The streamwise variation of the mean velocity and the dominant mean strain $\partial \left \langle U \right \rangle/\partial x$, along the centerline of the volume, is shown in Figure \ref{fig:strainrate}.
The flow accelerates through the contraction reaching 2.5 times the inlet velocity at $x=$ 180 mm as seen in Figure \ref{fig:strainrate}(a).
The mean strain increases, reaching a maximum at $x = 130$ mm, following which it decreases till the end of the measurement region as seen in \ref{fig:strainrate}(b).}
\textcolor{black}{The contour and vector plots of the time-averaged velocity fields in measurement regions P1 and P2 are shown in Figure C1.}
The vortex tubes aligned along the flow direction gets stretched by the mean strain, whereas those along the transverse direction get compressed by the conservation of angular momentum. 
This explains the increase in $\Gamma_{x,rms}$ and decrease in $\Gamma_{y,rms}$. 
\textcolor{black}{The dip observed in $\Gamma_{x,rms}$ and $\Gamma_{y,rms}$ is at $x \approx$ 175 mm slightly before the exit of the contraction which is at $x =$ 180 mm.
Hence, we think that the maximum strain that occurs at $x \approx$ 130 mm, does not coincide with the dip, which may occur where the return to isotropy starts at the end of the contraction.
The subsequent increase for $x>$ 180 mm following the exit of the contraction, indicates return to isotropy.}
\textcolor{black}{This increase towards the exit is prominent in $\Gamma_{y,rms}$ when compared to $\Gamma_{x,rms}$.
Weak contractions are sometimes used after the turbulence-generating grids to improve isotropy \cite{ComteBellot1966,Uberoi1956}.
Uberoi \cite{Uberoi1956} in his experiments with 4:1 axisymmetric contraction reports the tendency of streamwise r.m.s velocity $u_{rms}$ to increase towards the transverse r.m.s velocity $v_{rms}$ after exiting the contraction.}
Figure \ref{fig:RMSCirc}(c) shows the circulation around the $z-$axis $\Gamma_{z,rms}$ is  relatively constant along the flow direction because of the complete lack of mean strain in the $z$ direction in our 2-D contraction. 

The profiles obtained with different loop sizes are qualitatively the same except that the larger loops result in higher r.m.s values. 
The circulation strength increases with loop area as circulation is the area integral of vorticity normal to the area of integration. 
With larger loops, circulation accounts for a large-scale rotation and hence results in higher values of fluctuations. 
It can also be seen that the initial grid protocol has minimum effect on the circulation r.m.s variation. 
However, random mode shows slightly higher r.m.s values as it is seen to inject higher levels of turbulence compared to the synchronous mode.

\begin{figure}
\begin{tabular}{ p{0.06cm} c c }
{} & $r\sim$ 25$\eta$ & $r\sim$ 50$\eta$ \\
{} & {} & {}\\ 
\adjustbox{valign=t}{(a)} & \includegraphics[width=0.49\textwidth,valign=t]{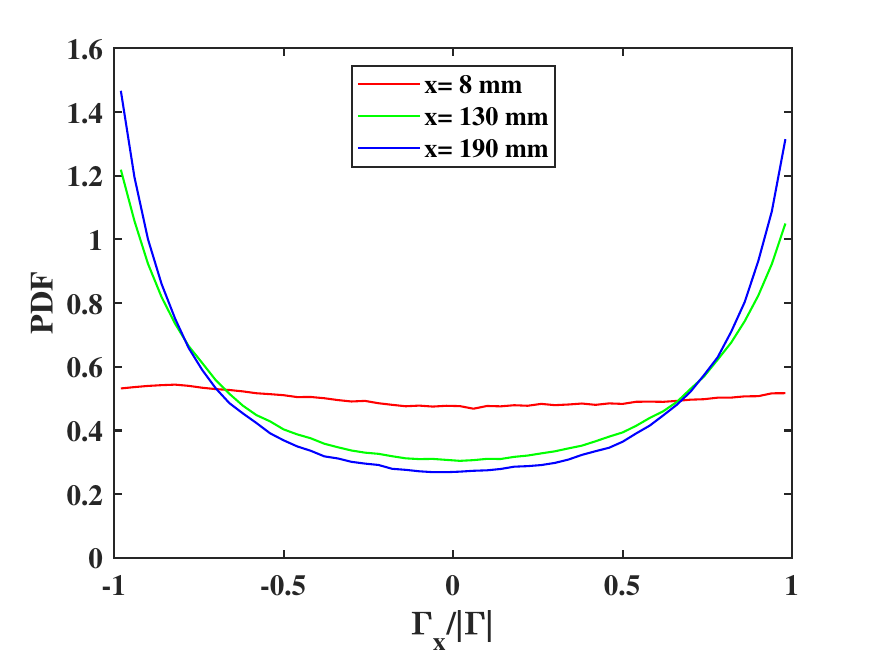} &  \includegraphics[width=0.49\textwidth,valign=t]{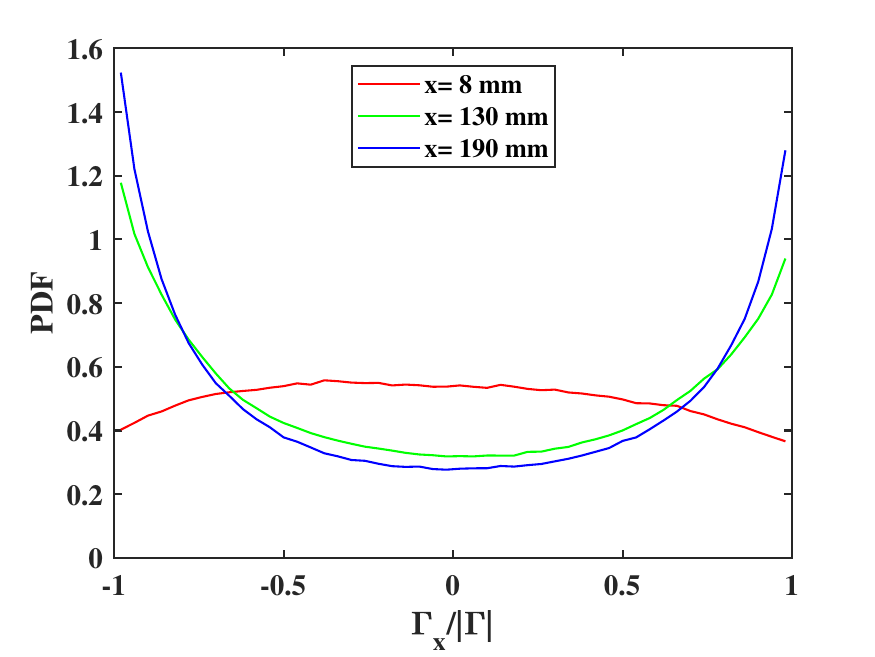}\\
\adjustbox{valign=t}{(b)} & \includegraphics[width=0.49\textwidth,valign=t]{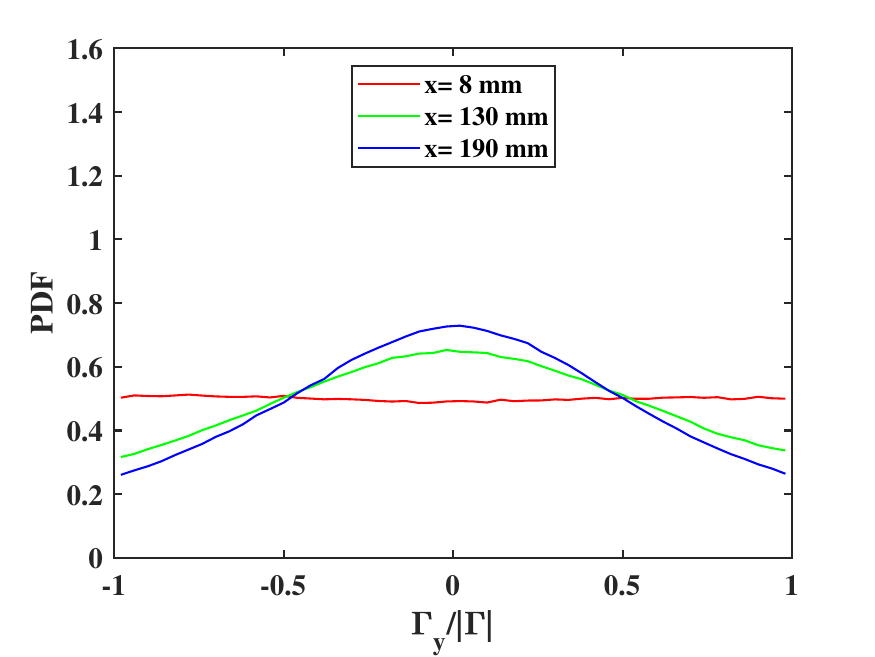} & \includegraphics[width=0.49\textwidth,valign=t]{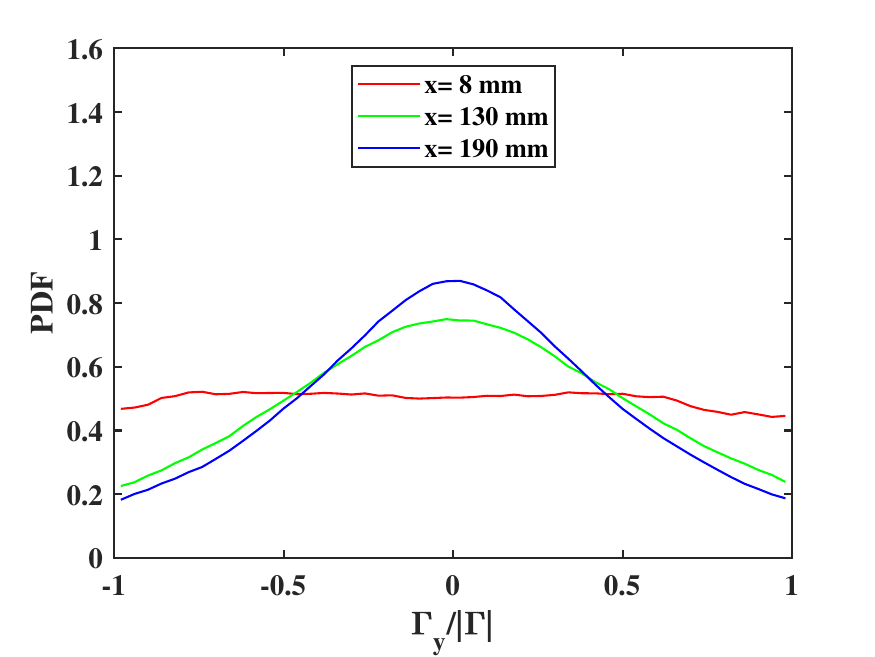}\\
\adjustbox{valign=t}{(c)} & \includegraphics[width=0.49\textwidth,valign=t]{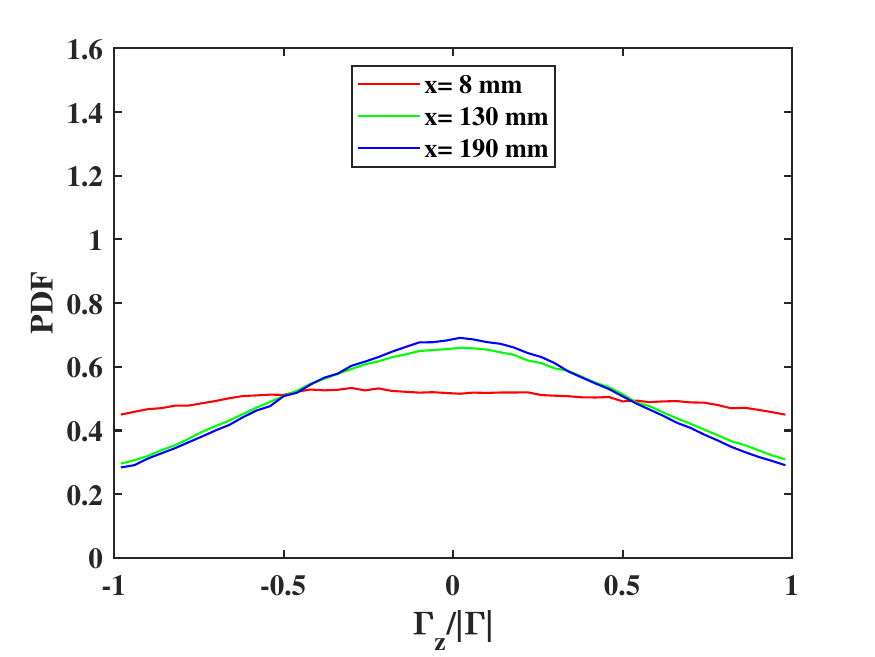} & \includegraphics[width=0.49\textwidth,valign=t]{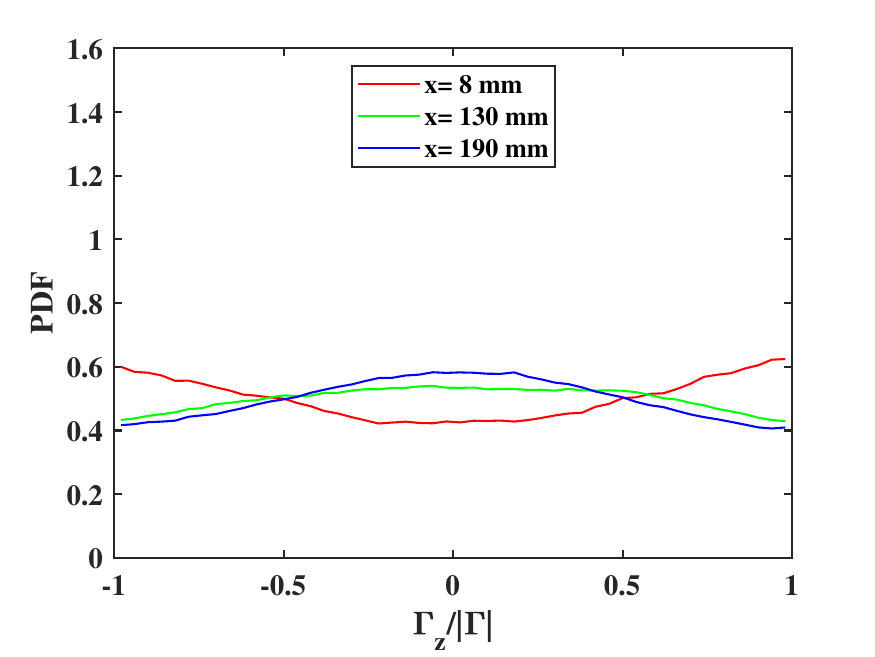}\\
\end{tabular}
\caption{The streamwise variation of probability density function (PDF) of the relative circulation strengths of $\Gamma_{x}/\left | \Gamma \right |$ (a), $\Gamma_{y}/\left | \Gamma \right |$ (b) and $\Gamma_{z}/\left | \Gamma \right |$ (c) measured with the random mode, loop sizes $r\sim$ 25$\eta$ (left) and $r\sim$ 50$\eta$ (right). The circulation strength, $\left | \Gamma \right |$ is calculated as $\left (\Gamma_{x}^2+\Gamma_{y}^2+\Gamma_{z}^2\right )^{0.5}$. The streamwise variation is shown near the start of contraction $(SOC)$ at $x=$ 8 mm, near the maximum straining location at $x=$ 130 mm and after the end of contraction $(EOC)$ at $x=$ 190 mm. The PDF is computed using 50 bins. At the inlet (red curves) all the three circulation components are evenly distributed while inside the contraction the circulation aligns in the streamwise stretching direction and it's strength in the compression direction becomes small.}
\label{fig:PDFCircSt_R}
\end{figure}

\subsection{Relative strength of circulation components}
We can consider the circulations in the three perpendicular planes, as the components of a `vector'.  The relative strength of these circulations then gives indication of the local orientation of coherent vortices.  
In Figure \ref{fig:PDFCircSt_R} we present this in terms of PDF of relative strengths of components and how they change in the streamwise direction. 
The strengths of these circulation components are normalized by its instantaneous absolute value given by 
\begin{equation}
\left | \Gamma \right |=\sqrt{\Gamma_{x}^2+\Gamma_{y}^2+\Gamma_{z}^2}
\label{eqn:abs_gamma}
\end{equation}
This is computed at three streamwise locations, first near the inlet ($SOC$), at the maximum straining location and after its end ($EOC$).

Circulation computed over a closed contour is the area-integral of the vorticity enclosed by the contour. 
Hence the relative strength of the circulation components can be viewed as the instantaneous orientation of the vortical tube/filament enclosed by the contour.
A single vortex tube aligned exactly with the streamwise direction would give $\Gamma_{x}/\left | \Gamma \right |= \pm$ 1 and zero for the other two components.
Sreenivasan {\it et al.} \cite{Sreenivasan1995} have argued that circulation can take adequate account of the contribution of filament-like structures unlike  structure functions, which are just one-dimensional cuts of the three dimensional turbulence.

The left and right columns of Figure \ref{fig:PDFCircSt_R} show the PDF of the relative strength computed with the random mode, for two loop sizes, in the IR, of $r\sim$ 25$\eta$ and $\sim$ 50$\eta$ respectively. 
We see at the inlet (red curves) that the orientation is random with uniform PDF for all three components of $\Gamma$.
In contrast, as we move through the contraction the streamwise circulation PDF, in Figure \ref{fig:PDFCircSt_R}(a), tends to peak at $\pm$1 and this peak value increases slightly downstream from $x=130$ to 190 mm.
This indicates greater relative strength of the streamwise circulation ($\Gamma_{x}$) compared to the circulation in the other two planes.  
This also implies the preferential alignment of coherent vortical structures in the $x-$direction. 
The transverse compression of the contraction reduces the strength of $\Gamma_{y}$, eventually making $\Gamma_{y}\simeq 0$  more probable at the end of the contraction (blue curve), as seen in Figure \ref{fig:PDFCircSt_R}(b).
On the other hand, the distribution of $\Gamma_{z}/\left | \Gamma \right |$ undergoes little change as we move along the contraction as seen in Figure \ref{fig:PDFCircSt_R}(c).
\textcolor{black}{However, for the smaller loop $r\sim$ 25$\eta$, $\Gamma_z$ is only slightly less uniform compared to that for $r\sim$ 50$\eta$.
The circulation tends to be more dominant in the $x$ direction, rather than the transverse $y-$ or $z-$directions.}
\textcolor{black}{At the inlet (red curves) all the three circulation components are approximately evenly distributed while inside the contraction the strongest circulation aligns in the streamwise stretching direction and it's strength in the compression direction becomes small.}
For similar conditions, we showed in \cite{Mugundhan2020} that this preferential alignment exists with the vorticity vector and the coherent vortical structures identified based on the vorticity magnitude. 

The PDFs in Figure \ref{fig:PDFCircSt_R} are close to symmetric from the beginning of the contraction, as is expected with the random mode changing the grid-rotation direction, yielding a homogeneous distribution.
On the other hand, the synchronous mode's PDFs exhibit a significant asymmetry. 
The results for the synchronous mode are therefore relegated to the Appendix section.

To further characterize this non-homogeneity we present the 2-D contour plots of the average and r.m.s vorticity for both the synchronous and random modes in Appendix Figures B1 and B2.
The contours are shown in the mid $x-y$ plane of the measurement region P1.
We see much `better' transverse homogeneity with the random mode when compared to the synchronous mode.
However, keep in mind that the color is selected to highlight rather small non-homogeneities in the mean fields.

\begin{figure}
\begin{tabular}{ p{0.06cm} c c }
{} & $x=$10 mm & $x=$130 mm\\
{} & {} & {}\\ 
\adjustbox{valign=t}{(a)} & \includegraphics[width=0.48\textwidth,valign=t]{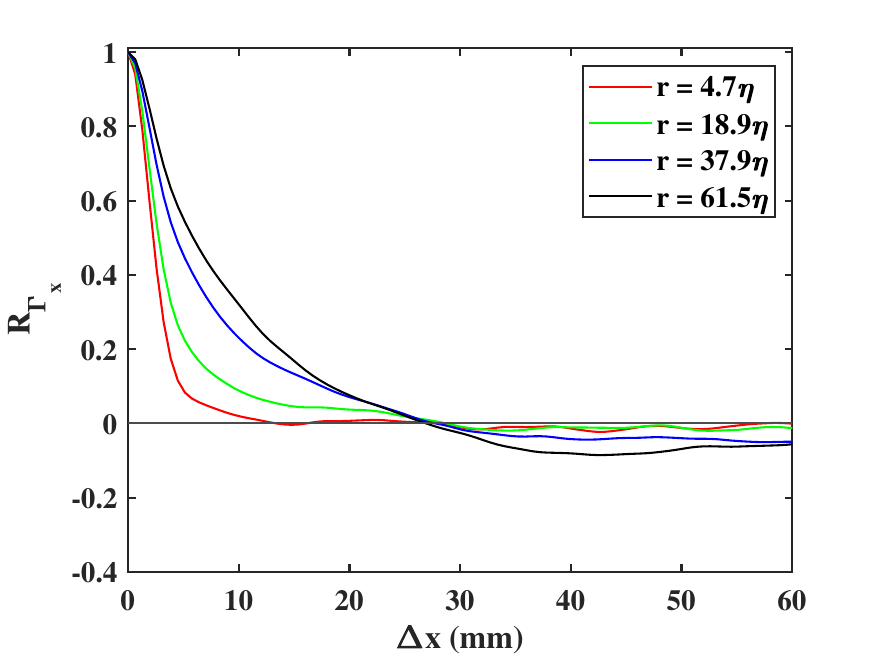}&
\includegraphics[width=0.48\textwidth,valign=t]{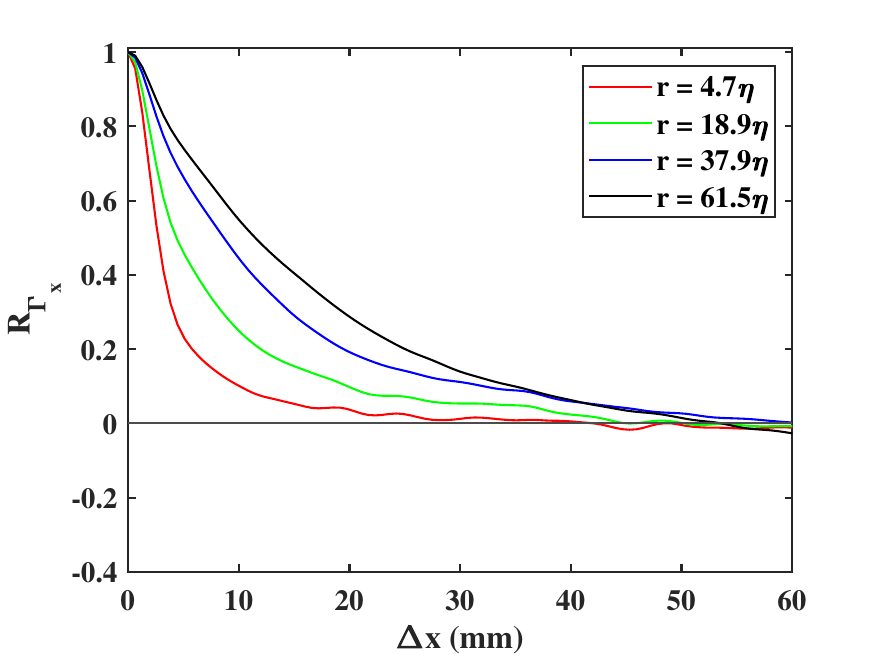}\\
\adjustbox{valign=t}{(b)} & \includegraphics[width=0.48\textwidth,valign=t]{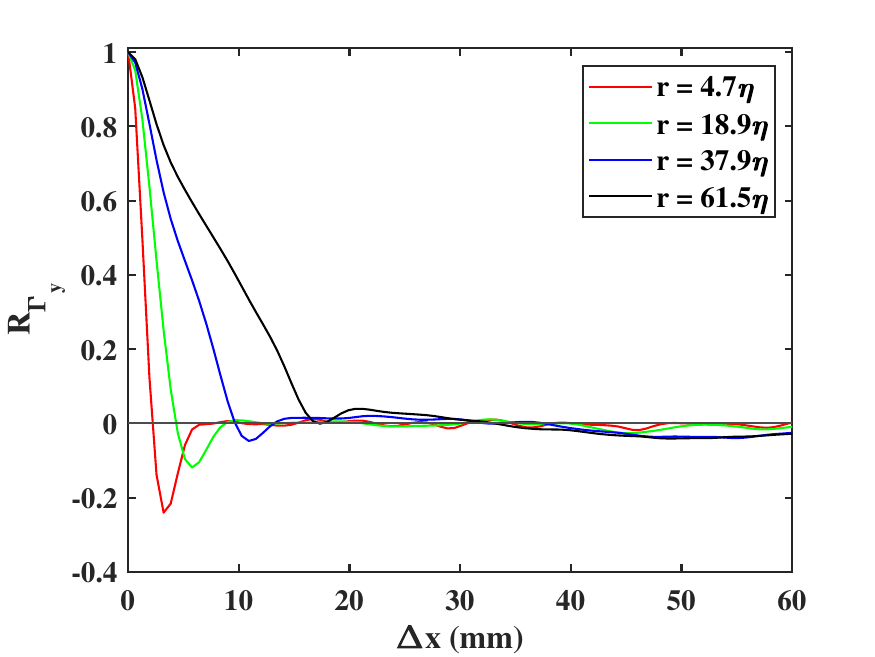}&
\includegraphics[width=0.48\textwidth,valign=t]{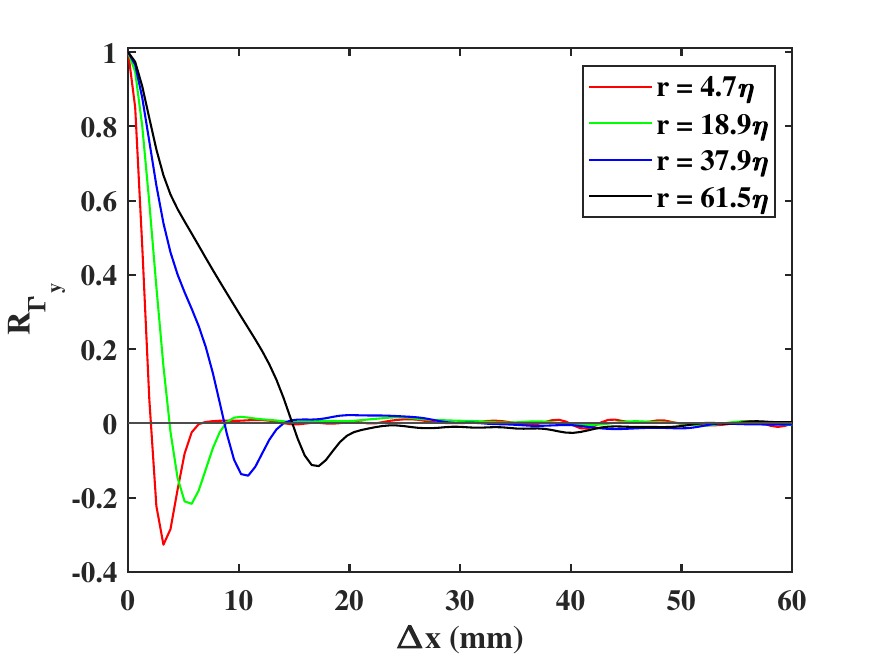}\\
\adjustbox{valign=t}{(c)} & \includegraphics[width=0.48\textwidth,valign=t]{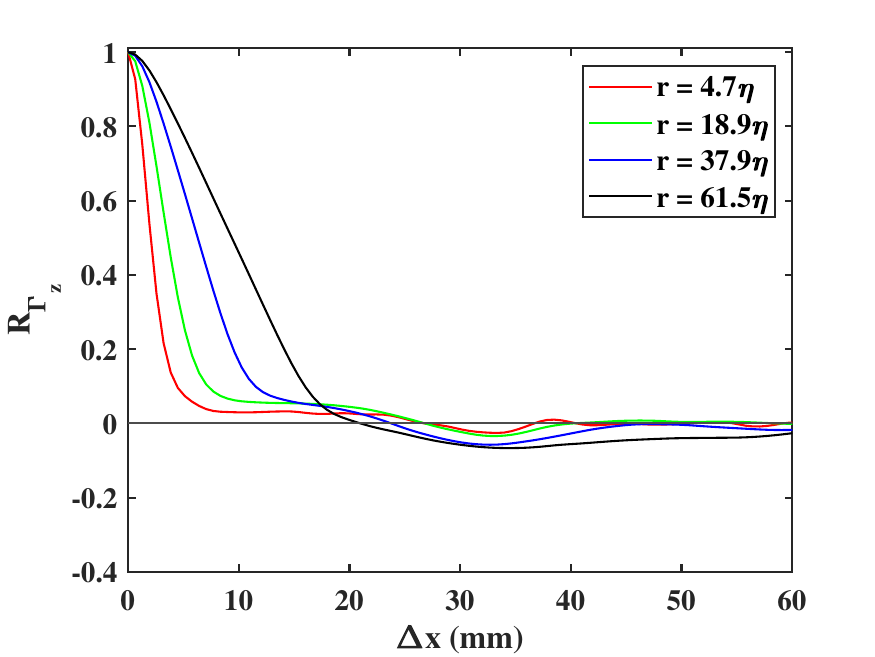}&
\includegraphics[width=0.48\textwidth,valign=t]{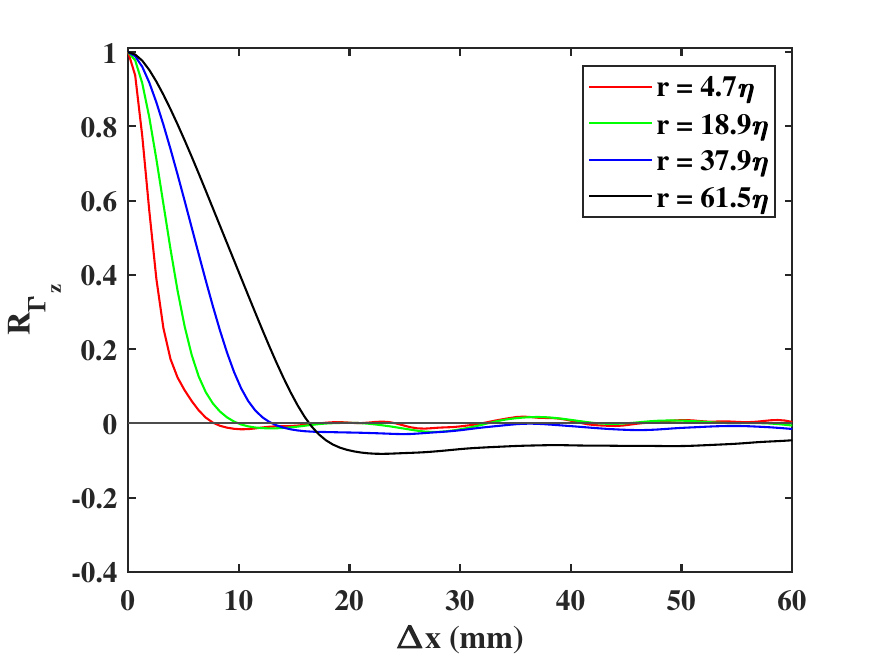}\\
\end{tabular}

\caption{The spatial correlation of circulations in the three perpendicular planes: (a) $R_{\Gamma_{x}}$, (b) $R_{\Gamma_y}$, and (c) $R_{\Gamma_z}$ for mode S1 and different loop sizes. 
The correlation is computed between points on the centerline of the measurement region by varying their streamwise separation $\Delta x$ with $x_1=$10 mm (left) which is near the $\it{SOC}$ and at $x_1=$ 130 mm (right) which is downstream of the maximum strain location.}
\label{fig:Corr}
\end{figure}

\begin{figure}
\begin{tabular}{ p{0.06cm} c c }
{} & Synchronous Mode & Random Mode\vspace{-0.05in}\\
{} & {} & {}\\ 
\adjustbox{valign=t}{(a)} & \includegraphics[width=0.48\textwidth,valign=t]{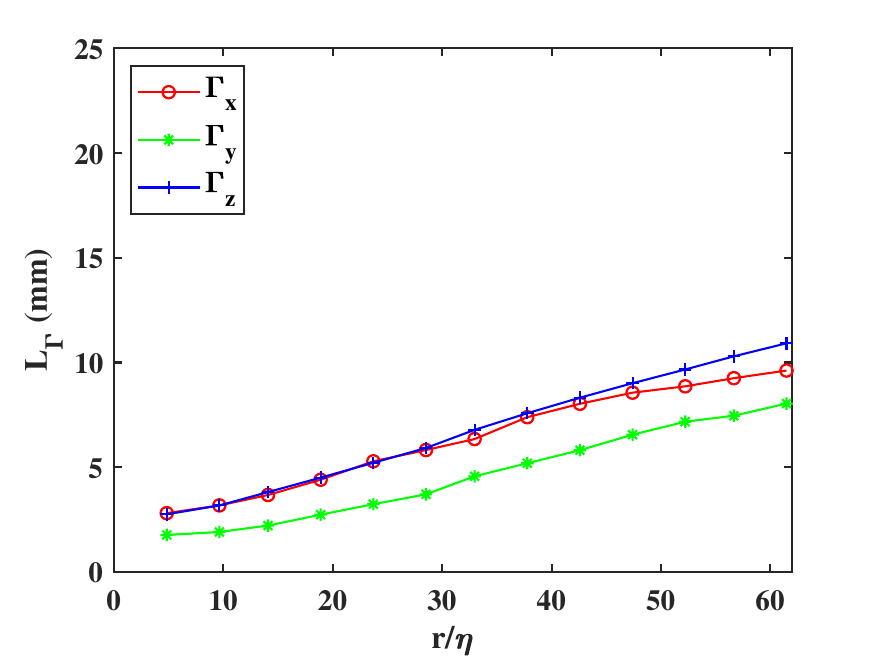}&
\includegraphics[width=0.48\textwidth,valign=t]{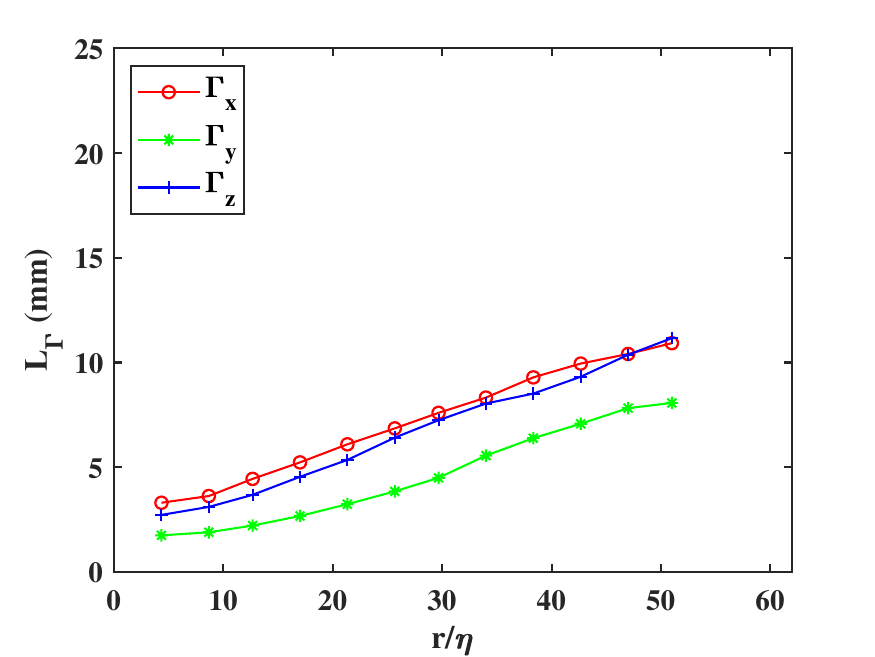}\\
\adjustbox{valign=t}{(b)} & \includegraphics[width=0.48\textwidth,valign=t]{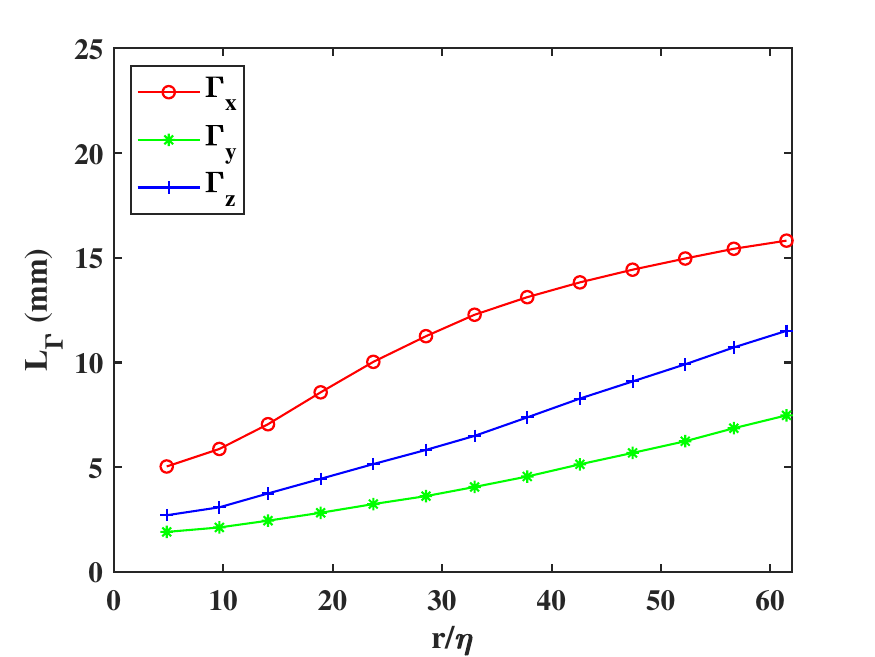}&
\includegraphics[width=0.48\textwidth,valign=t]{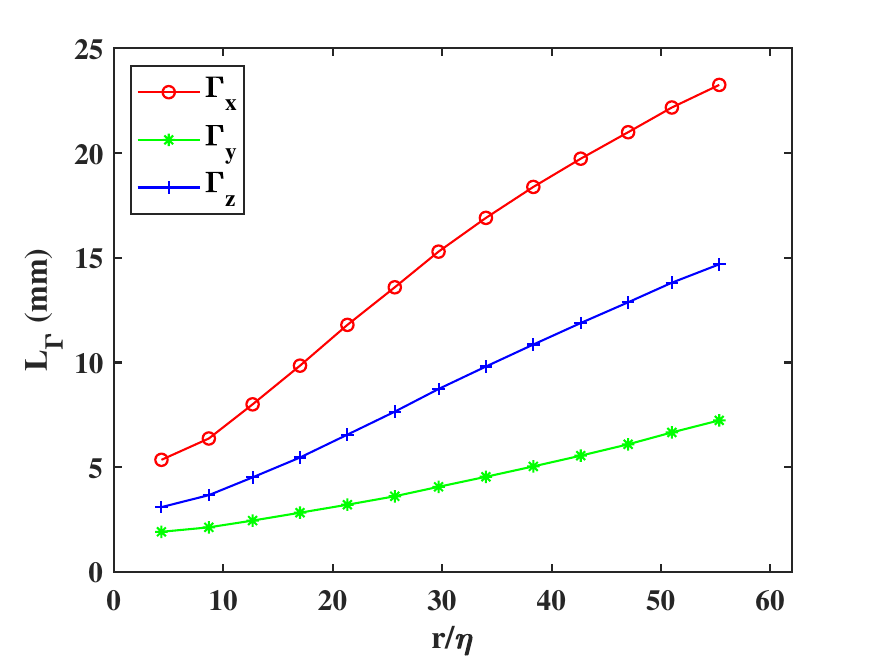}\vspace{-0.01in}\\
\end{tabular}
\caption{\textcolor{black}{The variation of the circulation integral scale $L_{\Gamma}$ with loop size. 
The integral scale corresponds to the correlations obtained with the first points at $x_1=$10 mm (a) and at $x_1=$130 mm (b).
The integral scale is computed by integrating the absolute values of the spatial correlation functions ($R_{\Gamma_{x}}$, $R_{\Gamma_y}$, $R_{\Gamma_z}$) from $\Delta x=$ 0 to the point which corresponds to the second zero crossover.}}
\label{fig:IntScale}
\end{figure}

\subsection{\textcolor{black}{Integral Length Scale of $\Gamma$}}
In turbulence, correlation functions of velocity components separated in space or in time are used to characterize the associated length and time scales. 
In line with this, we define a correlation function $R_{\Gamma}$ which correlates circulation around two central points separated by $\Delta x$ in the streamwise $x-$direction only, for example,
the correlation function for $\Gamma_{x}$ is defined as
\begin{equation}
R_{\Gamma_{x}}(x_1)=\frac{\left \langle \gamma_{x}(x_1,t)\gamma_{x}(x_1+\Delta x,t)\right \rangle}{\left \langle \gamma_{x}(x_1,t)^2\right \rangle},
\label{eqn:corr_defn}
\end{equation}
{\textcolor{black}{where the instantaneous $\Gamma_{x}(x,t)$ is decomposed into its mean and fluctuation components, $\left \langle\Gamma_{x}(x)\right \rangle+\gamma_{x}(x,t)$.
The correlation functions $R_{\Gamma_{x}}$, $R_{\Gamma_y}$ and $R_{\Gamma_z}$ are calculated having the first point $x_1$ fixed and by varying the parameter $\Delta x$.}
To understand the effect of straining, we compare these correlations at $x_1=$ 10 mm, which is just inside the contraction and at $x_1=$ 130 mm, which is slightly downstream of the maximum straining location.
We present the correlations over a wide range of square-loops sizes between $\approx $ 5$\eta$ and 60$\eta$ for the synchronous mode in Figure \ref{fig:Corr}.
The space shift $\Delta x$ is always in the $x$ direction and hence $R_{\Gamma_{x}}$ is a longitudinal correlation whereas $R_{\Gamma_y}$, $R_{\Gamma_z}$ are the transverse cross-spatial correlations.
We see that as $\Delta x$ increases, the initially positively correlated circulation crosses over to negative values. 
The first crossover happens over a shorter $\Delta x$ with $\Gamma$ over small loops when compared to $\Gamma$ over large loops.
For example, the crossover of $R_{\Gamma_{x}}$ occurs at $\Delta x\approx$ 15 \textcolor{black}{mm} with $r=$ 4.7$\eta$ and at $\Delta x\approx$ 28 \textcolor{black}{mm} with $r=$ 61.5$\eta$ (see Figure \ref{fig:Corr}(a) on the left side).
This is true with all the three correlation functions. 
As pointed out earlier, $\Gamma$ computed over small loops approaches the behavior of vorticity.
On the other hand, $\Gamma$ over large loops encompasses the vorticity over the loop area and hence is representative of large coherent structures. 
Hence, $R_{\Gamma}$ over larger loops exhibit positive correlation over a longer distance.
In certain cases, we do see \textcolor{black}{an} early second crossover of the function to positive values.
This second crossover is very evident with $R_{\Gamma_y}$ and it occurs over a small $\Delta x$ as seen in Figure \ref{fig:Corr} (b).
\textcolor{black}{It is interesting to note that the first crossover point of $R_{\Gamma_y}$ at $x=$ 130 mm approximately matches the loop size $r$.
The crossovers are observed at $\Delta x\approx$ 2, 3, 8.5 \& 15 mm for $r=$ 1.3, 5, 7.7 and 16.6 mm respectively.}
We have seen that the corresponding component of vorticity $\omega_y$ decreases due to the compression applied by the contraction in the $y-$direction.
\textcolor{black}{This compression manifests in sudden change in $R_{\Gamma_y}$ from positive to negative, which could heuristically be associated with buckling of vortices in the $y-$direction.}

\textcolor{black}{At the maximum straining location at $x=$ 130 mm, the $R_{\Gamma_x}$ correlation  exhibits a larger correlation length, with the first crossover at a larger $\Delta x$ than for the other two components.}
This shows that the straining causes an increase in the correlation length which represents existence of long coherent structures with majority of them aligned in the streamwise direction \cite{Mugundhan2020}. 
\textcolor{black}{However, strain appears to have little influence on the other correlations $R_{\Gamma_y}$ and $R_{\Gamma_z}$, which are similar for $x=8$ and $x=130$ mm.}

\textcolor{black}{We define a new integral length scale $L_{\Gamma}$ based on $R_{\Gamma}$ as,}
\begin{equation}
L_{\Gamma}=\int_{0}^{l_c}\left | R_{\Gamma} \right |dx,
\label{eqn:intL_defn}
\end{equation}
\textcolor{black}{where, $l_c$ is the length of the second crossover point from the origin.} 
\textcolor{black}{We consider up to the second crossover point as there exists a \enquote{good} correlation even after the first crossover though the correlation is negative.
To take this negative-correlation region also into account in characterizing the integral length scale, we use the absolute value of $R_\Gamma$ in Eq. \ref{eqn:intL_defn}.
We present the variation of $L_\Gamma$ with loop size, with $x_1=$ 10 and 130 mm for both the modes in Figure \ref{fig:IntScale}.
We see that $L_\Gamma$ based on all the three correlation functions increases with increase in the loop size for both the modes.
This increase with $r$ is fairly linear at the inlet to the contraction.
But there is greater increase seen with $L_{\Gamma_{x}}$ after the maximum straining location as seen in Figure \ref{fig:IntScale}(b). 
The random mode has slightly higher values of $L_\Gamma$ at the inlet to the contraction.
This can also be seen in the integral scale summarized in Table \ref{tab:parameters}, which is computed using the correlation of streamwise velocity. 
This difference could be due to the difference in the rotation speed of the motors.
Due to a longer duration of structure interaction with the flaps at lower rotation speeds, the random mode yields higher values of integral scales.
The vortical structures get stretched and tend to align with the $x$ direction. 
It is to be noted that the contraction stretches these large vortical structure as they move downstream.
It also tends their principal axis to align in the $x-z$ plane perpendicular to the $y$ direction, with a higher probability to align with the $x$ direction.
This is the reason for the following order of the integral scales: $L_{\Gamma_{x}} > L_{\Gamma_{z}} > L_{\Gamma_{y}}$.} 

\subsection{Space-time correlation of circulation}
\begin{figure}
\begin{tabular}{ p{0.06cm} c c }
{} & $x=$10 mm & $x=$130 mm\\
{} & {} & {}\\ 
\adjustbox{valign=t}{(a)} & \includegraphics[width=0.48\textwidth,valign=t]{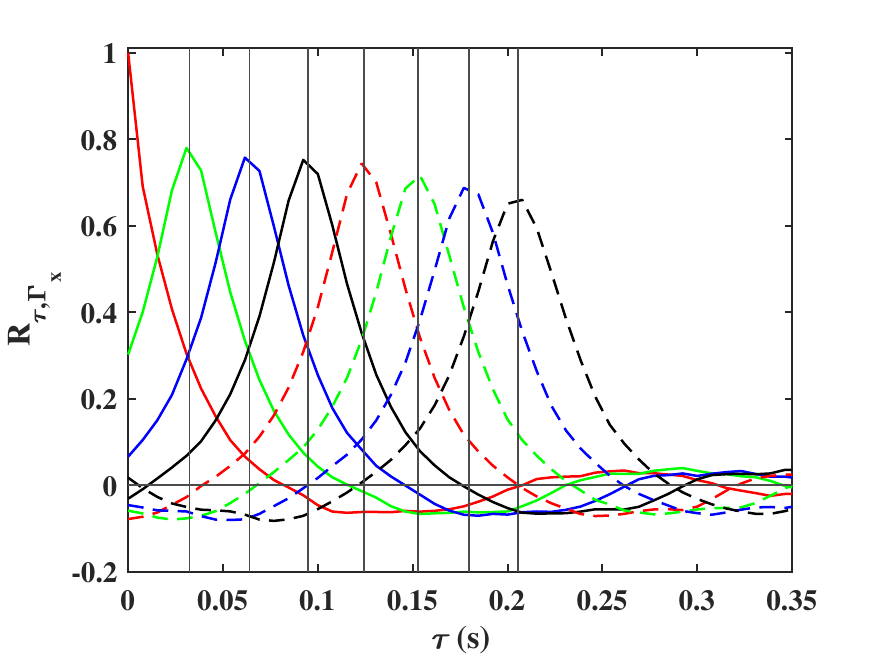}&
\includegraphics[width=0.48\textwidth,valign=t]{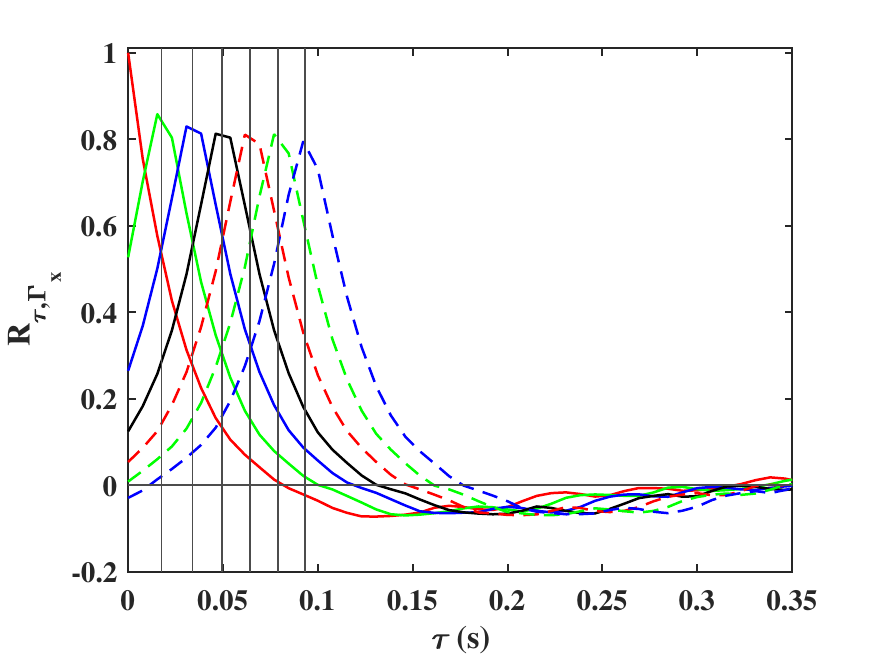}\\
\adjustbox{valign=t}{(b)} & \includegraphics[width=0.48\textwidth,valign=t]{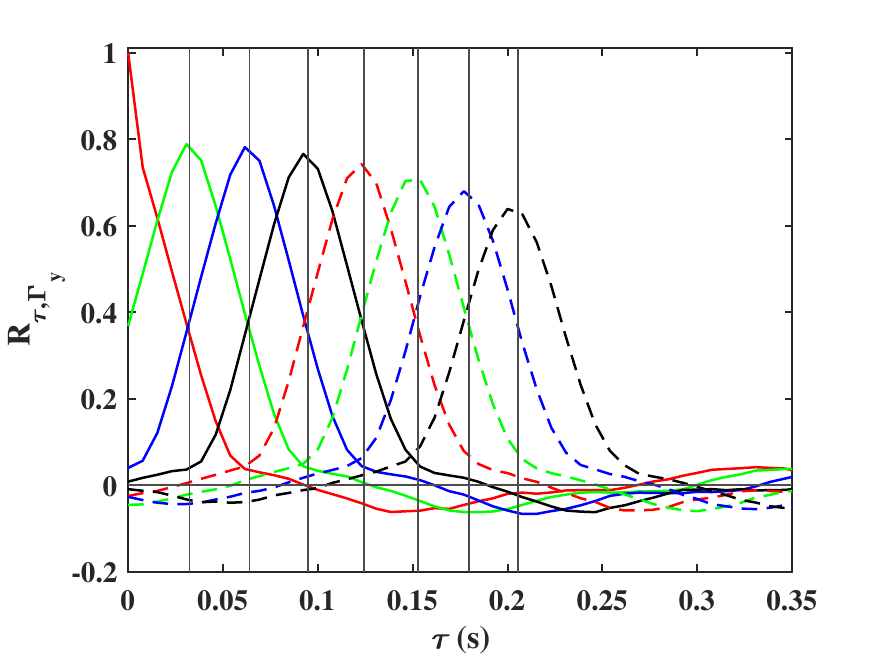}&
\includegraphics[width=0.48\textwidth,valign=t]{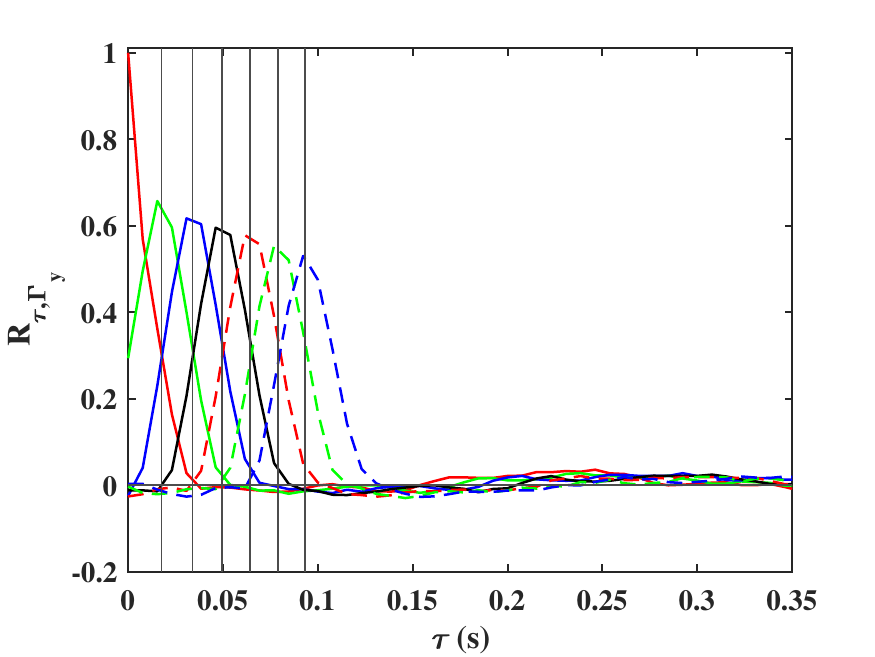}\\
\adjustbox{valign=t}{(c)} & \includegraphics[width=0.48\textwidth,valign=t]{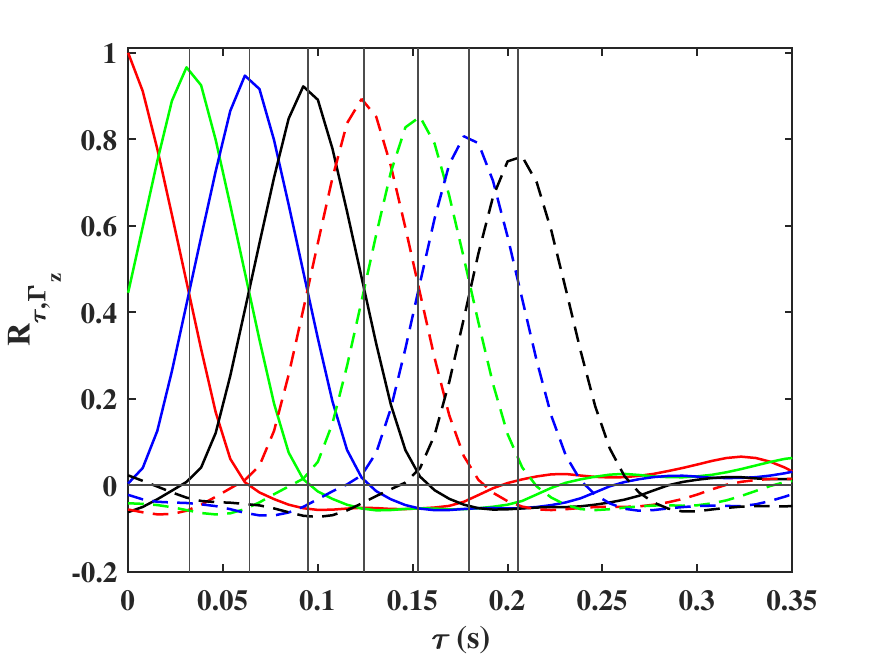}&
\includegraphics[width=0.48\textwidth,valign=t]{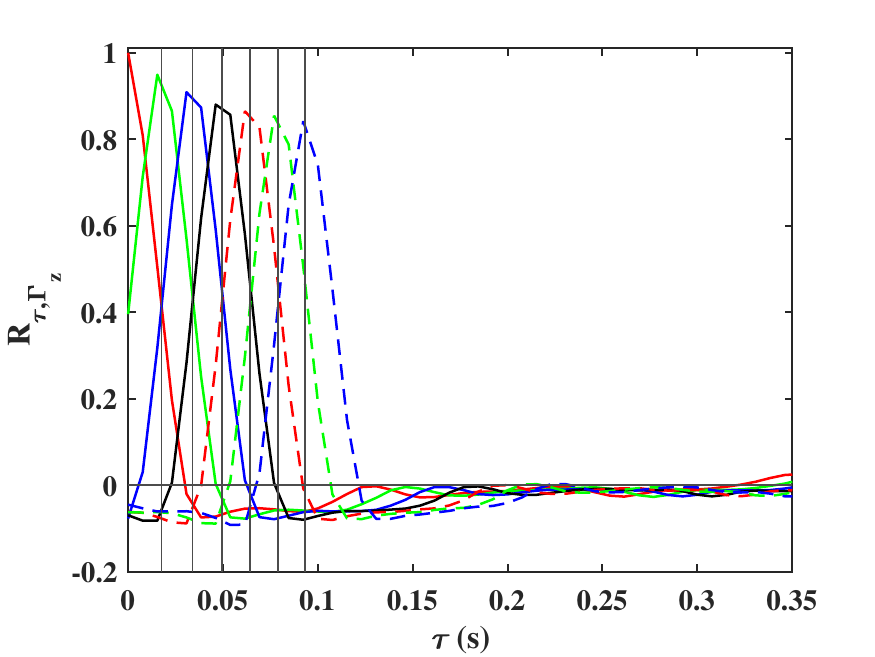}\\
\end{tabular}

\caption{The space-time correlation of circulations: (a) $R_{\tau,\Gamma_x}$, (b) $R_{\tau,\Gamma_y}$, and (c) $R_{\tau,\Gamma_z}$ for the Synchronous grid-rotation mode and loop size of $r/\eta=$ 61.5. 
The correlation is between the first points $x_1=$10 mm (left), $x_1=$130 mm (right) and points with $\Delta x=$ 0 ({\color{red}{---}}), 10.2 mm ({\color{green}{---}}), 20.5 mm ({\color{blue}{---}}), 30.8 mm ({\color{black}{---}}), 41.1 mm ({\color{red}${--}$}), 51.3 mm ({\color{green}${--}$}), 61.6 mm ({\color{blue}${--}$}) and 71.9 mm ({\color{black}${--}$}).
The vertical lines represent the predicted location of the maximum correlation based on the integrated advection time $\tau_{adv}$ from Eq. (\ref{eqn:inttau_adv}).}
\label{fig:TSCorrS1SQ27}
\end{figure}

\begin{figure}
\begin{tabular}{ p{0.06cm} c c }
{} & $x=$10 mm & $x=$130 mm\\
{} & {} & {}\\
\adjustbox{valign=t}{(a)} & \includegraphics[width=0.48\textwidth,valign=t]{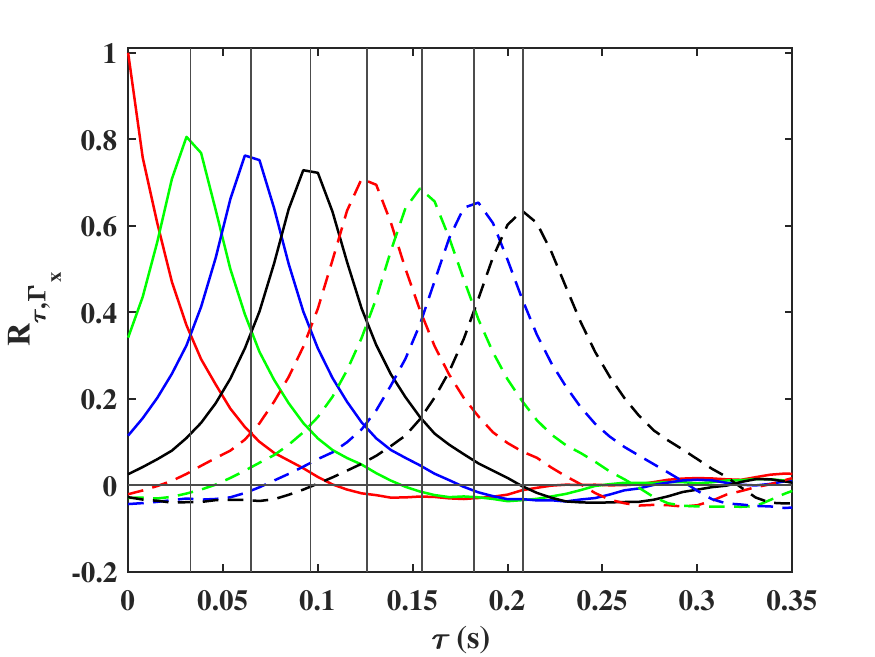}&
\includegraphics[width=0.48\textwidth,valign=t]{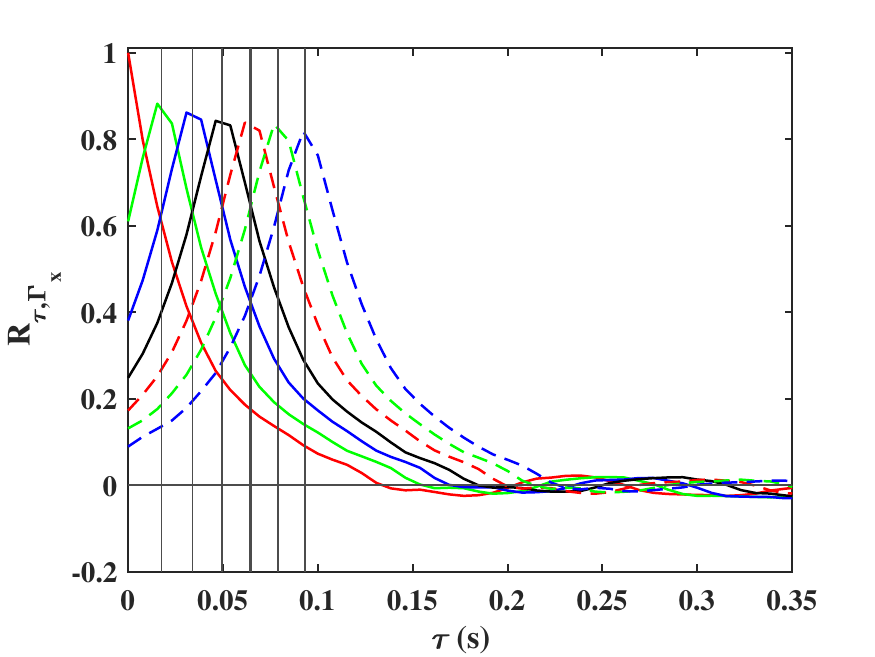}\\
\adjustbox{valign=t}{(b)} & \includegraphics[width=0.48\textwidth,valign=t]{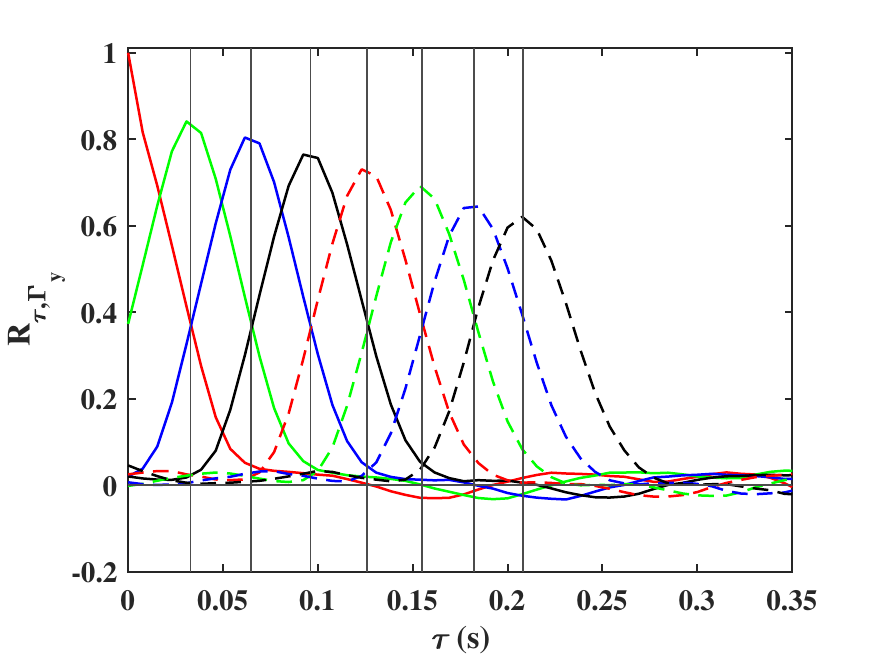}&
\includegraphics[width=0.48\textwidth,valign=t]{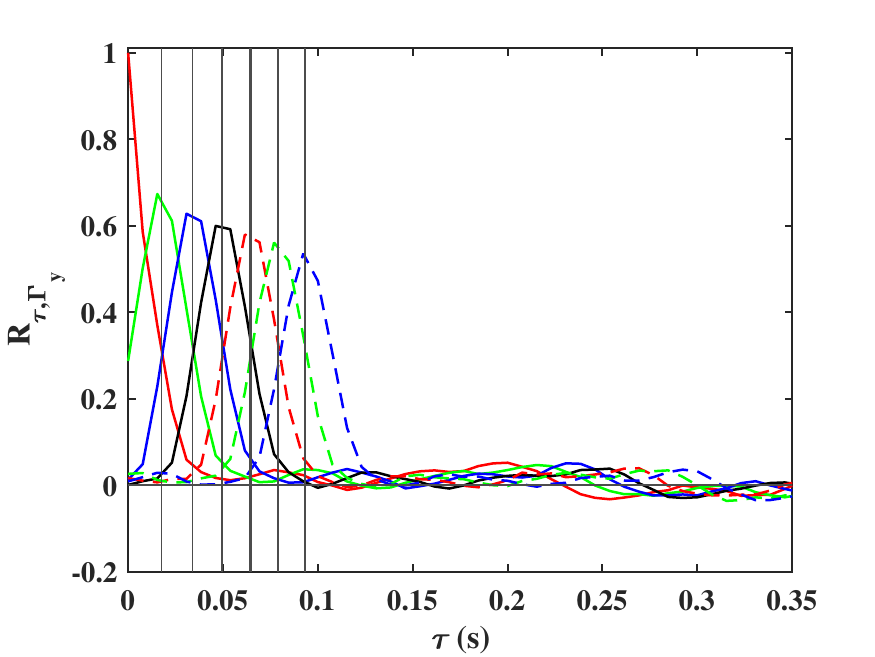}\\
\adjustbox{valign=t}{(c)} & \includegraphics[width=0.48\textwidth,valign=t]{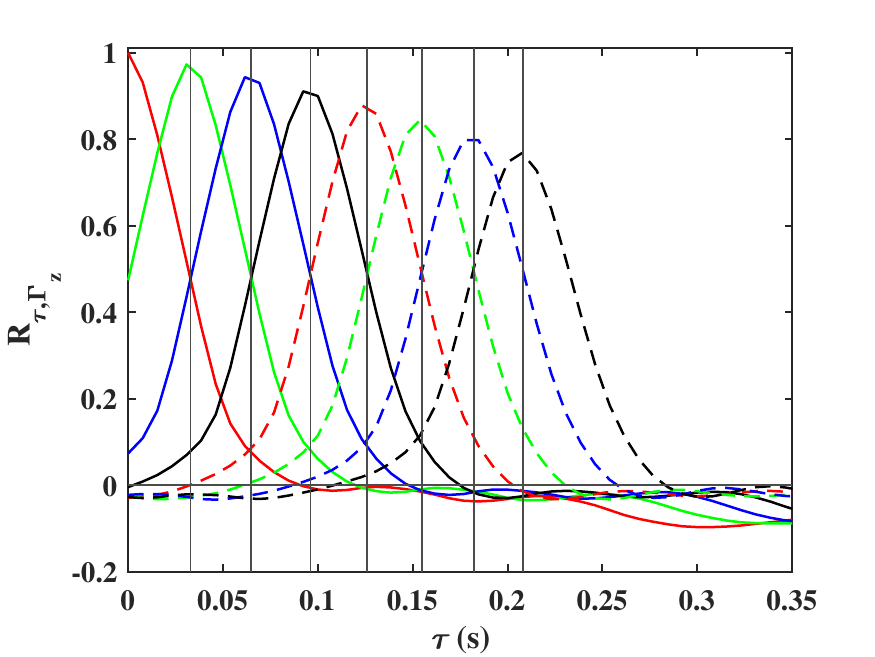}&
\includegraphics[width=0.48\textwidth,valign=t]{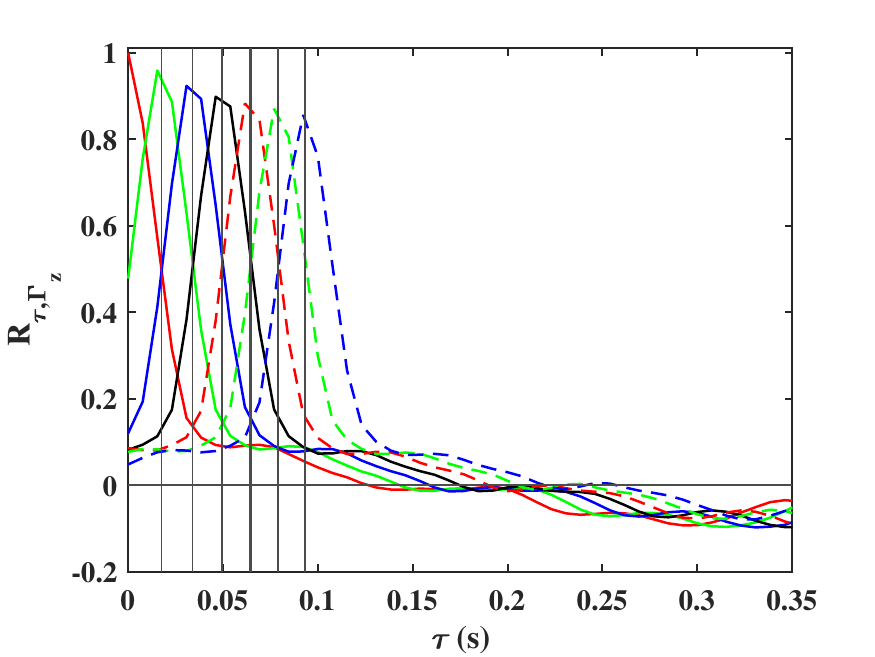}\\
\end{tabular}
\caption{\textcolor{black}{The space-time correlation of circulations: (a) $R_{\tau,\Gamma_x}$, (b) $R_{\tau,\Gamma_y}$, and (c) $R_{\tau,\Gamma_z}$ for the Random grid-rotation mode and $r/\eta=$ 51.1.
For legends refer to the caption of figure \ref{fig:TSCorrS1SQ27}.}}
\label{fig:TSCorrRSQ25}
\end{figure}

We also look at the space-time correlations of the circulation.
The space-time correlation function is defined as
\begin{equation}
R_{\tau,\Gamma_{x}}(x_1,\Delta x)=\frac{\left \langle \gamma_{x}(x_1,t)\gamma_{x}(x_1+\Delta x,t+\tau)\right \rangle}{\left \langle \gamma_{x}(x_1,t)^2\right \rangle^{0.5}\left \langle \gamma_{x}(x_1+\Delta x,t+\tau)^2\right \rangle^{0.5}} ,
\label{eqn:STcorr}
\end{equation}
where $\Delta x$ is the space shift and $\tau$ is the time shift. 
Figure \ref{fig:TSCorrS1SQ27} shows the space-time correlation for $x_1=$ 10 and 130 mm for mode S1 and the largest loop with $r/\eta=$ 61.5.
The space-time correlation is obtained by correlating the fluctuating circulations at two spatially separated points by varying $\tau$ 
 \cite{Favre1965},\cite{Libby1996}. 
For smoothness of the correlation function, we use a time shift that corresponds to ten frames of the high-speed video.
With reference to Figures \ref{fig:TSCorrS1SQ27} and \ref{fig:TSCorrRSQ25} we now look at some of the common characteristics of these space-time correlation functions.
For $\Delta x=$0 this simplifies to the  autocorrelation in time, which by definition becomes 1 when $\tau=$ 0 and decreases to zero at $\tau\approx$ 0.08 s, following which it becomes slightly negative with further increase in $\tau$.
The zero-crossover point of the correlation function gives a characteristic time scale.
On the other hand, keeping $\tau=0$, $R_{\tau=0,\Gamma_{x}}$ we recreate the space-only correlation already presented in Figure \ref{fig:Corr} and encoded on the intersection with the $y-$axis ($\tau_{adv} = 0$) of the curves for various $\Delta x$.

The full space-time correlations, for example for $R_{\tau,\Gamma_{x}}(x_1=10,\Delta x=10.2)$ mm, the correlation grows (from 0.3 when $\tau=$0) with increase in $\tau$ to reach a maximum at $\tau\approx$ 0.03 and then decreases, falling to zero at $\tau\approx$ 0.11 s. The correlation is close to symmetric about this peak.
The location of this peak can be predicted by the Taylor hypothesis of frozen turbulence.
We use an analogy to a passive fluid element transported by the mean flow to understand this.
The spatial position of the fluid element which is initially at $x_1$ at time $\tau=$ 0 can be predicted knowing the mean advection velocity $\left \langle U \right \rangle(x)$ of the transporting flow.
The \textcolor{black}{element} is advected through a distance $dx=\left \langle U \right \rangle d\tau$ in a time period $d\tau$.
In line with this \textcolor{black}{element} transport, the circulation fluctuations are also transported by the mean flow.
The mean velocity profile $\left \langle U \right \rangle(x)$ along the centerline is precisely known from our measurements and hence the advection time can be determined.
The advection time $\tau_{adv}$ for a fluid element to be transported from point $x_1$ to $x_1+\Delta x$ is given by,
\begin{equation}
\tau_{adv}=\int_{x_1}^{x_1+\Delta x}\frac{dx}{\left \langle U \right \rangle(x)}.
\label{eqn:inttau_adv}
\end{equation}
The advection times for the chosen, space-shifted points, with $\Delta x=$10.2 to 71.9 mm are indicated by the vertical lines in the figures.
These lines coincide exactly with the advection times at the peaks in the corresponding correlations $R_{\tau,\Gamma_{i}}$.
The decrease in the peak values is expected from the dynamical evolution of the turbulent rotational structures during the transport by the mean flow.  The measurement errors between sequences of time-volumes also contributes to the reduced correlation peaks.

Figures \ref{fig:TSCorrS1SQ27} and \ref{fig:TSCorrRSQ25} show the space-time correlation results for the Synchronous and Random grid-rotation modes, respectively.
The left columns show results close to the inlet of the contraction where $x_1= 10$ mm and right column towards its exit at $x_1= 130$ mm.
For the latter location, the peaks are spaced by much shorter time intervals as expected due to increase in the mean advection velocities owing to the reduced cross-sectional area.
The most striking non-trivial difference between the three directional circulations is manifest in the width of the correlation curves, particularly near the end of the contraction.  
\textcolor{black}{The width of the $R_{\tau,\Gamma_{x}}$ curves, measured between the first and second crossing,  
is much larger than for the corresponding $R_{\tau,\Gamma_{y}}$ and $R_{\tau,\Gamma_{z}}$.  
To minimize effect of noise we measure the width of the curves around the peak, from the 0.1 correlation levels.} 
Based on the time between first and second crossings of 0.1, we can define a timescale $T_i$, indicating how far in $\Delta \tau$ the rotational structures are correlated for $\Gamma_{x}$, $\Gamma_{y}$ \& $\Gamma_{z}$.
These curves show $T[x,y,z] = [0.11,0.046,0.046]$ s for the Synchronous mode and $T[x,y,z] = [0.18,0.044,0.070]$ s for the random mode.  The correlation time-duration for $\Gamma_x$ is therefore 2-3 times longer than in the other cross-stream directions, strongly supporting the presence of enhanced streamwise coherent vortices.

\begin{figure}
\begin{tabular}{ p{0.06cm} c c }
{} & $x=$ 8 mm & $x=$ 190 mm\\ 
{} & {} & {}\\
\adjustbox{valign=t}{(a)} & \includegraphics[width=0.49\textwidth,valign=t]{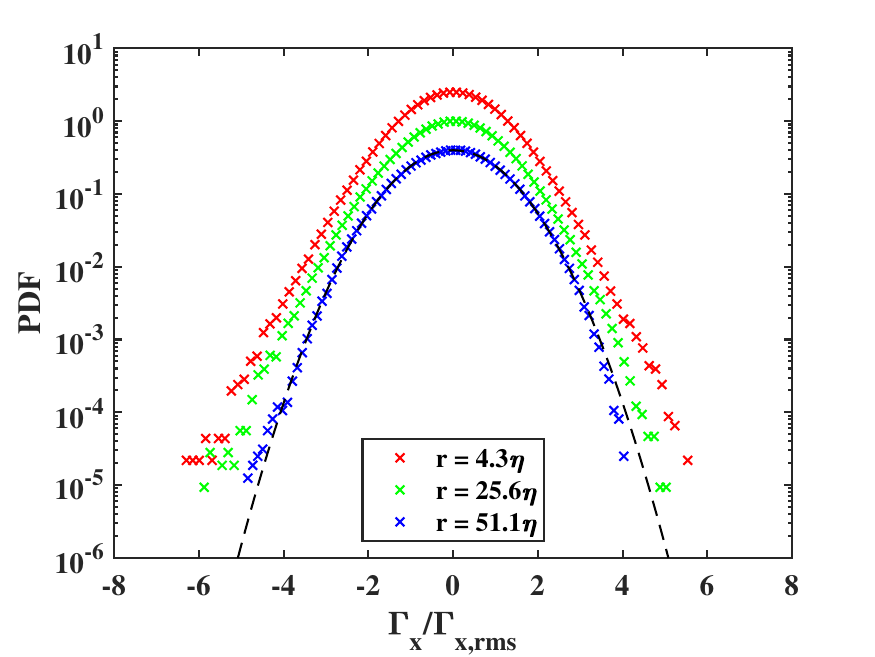} & \includegraphics[width=0.49\textwidth,valign=t]{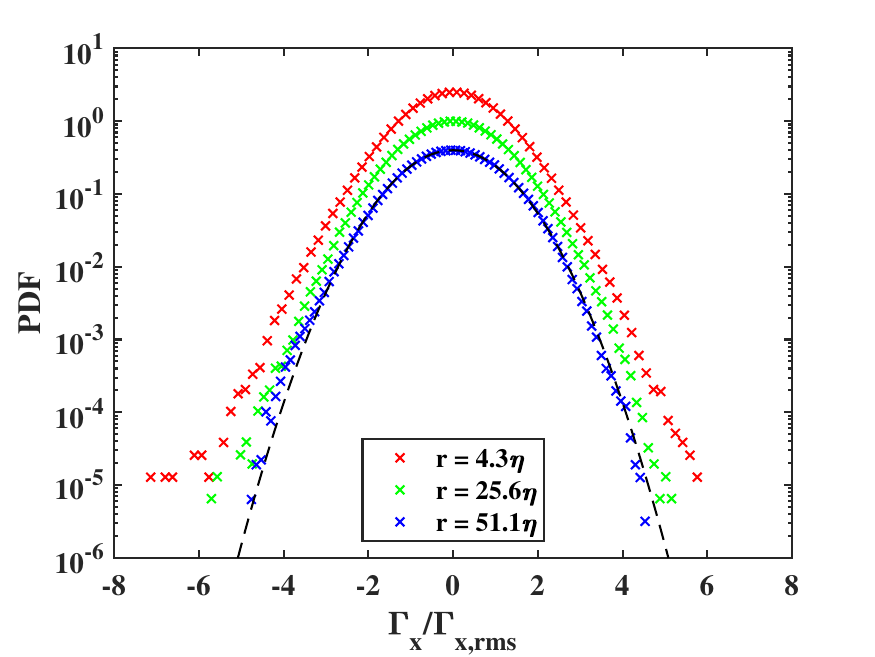}\\
\adjustbox{valign=t}{(b)} & \includegraphics[width=0.49\textwidth,valign=t]{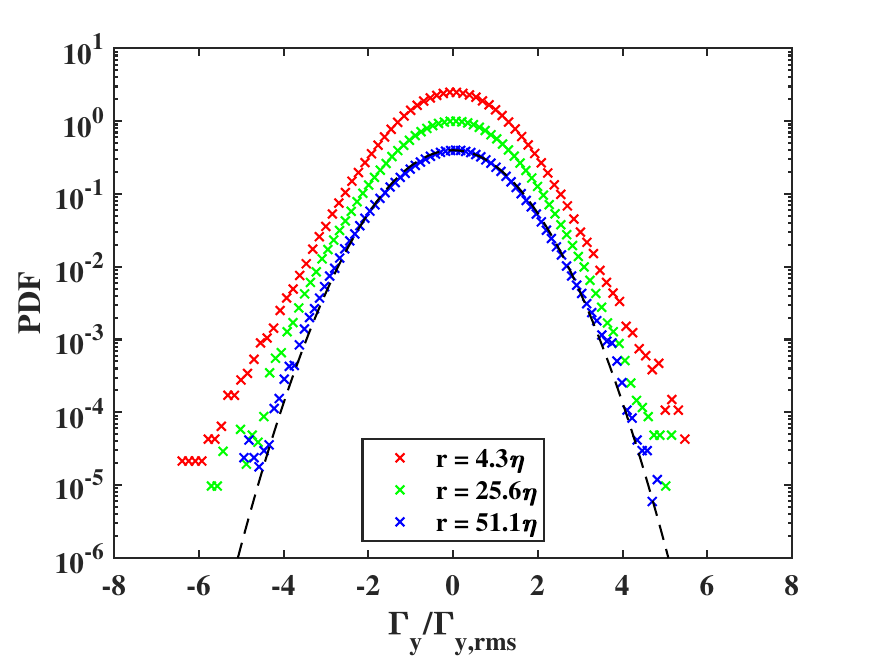} & \includegraphics[width=0.49\textwidth,valign=t]{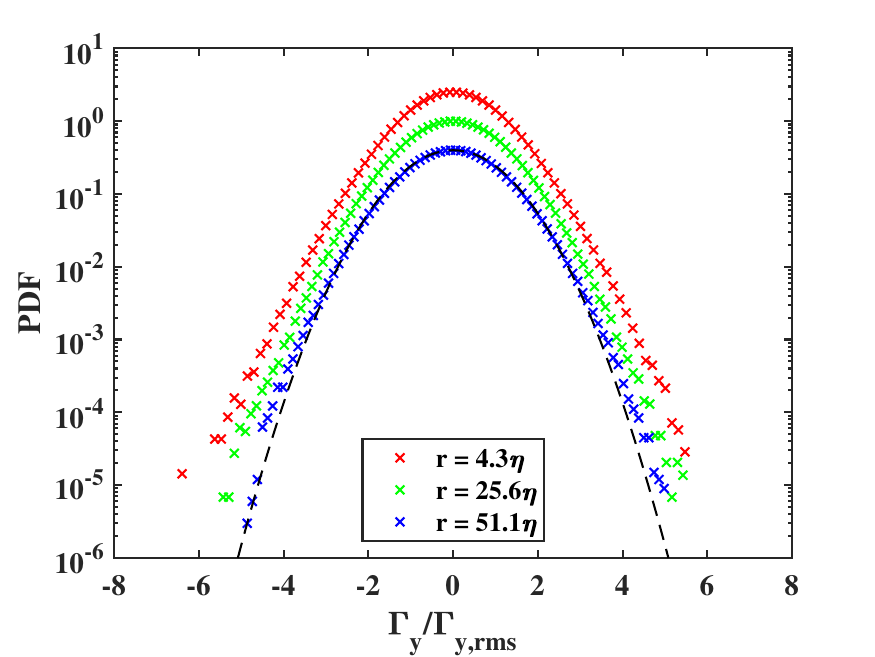} \\
\adjustbox{valign=t}{(c)} & \includegraphics[width=0.49\textwidth,valign=t]{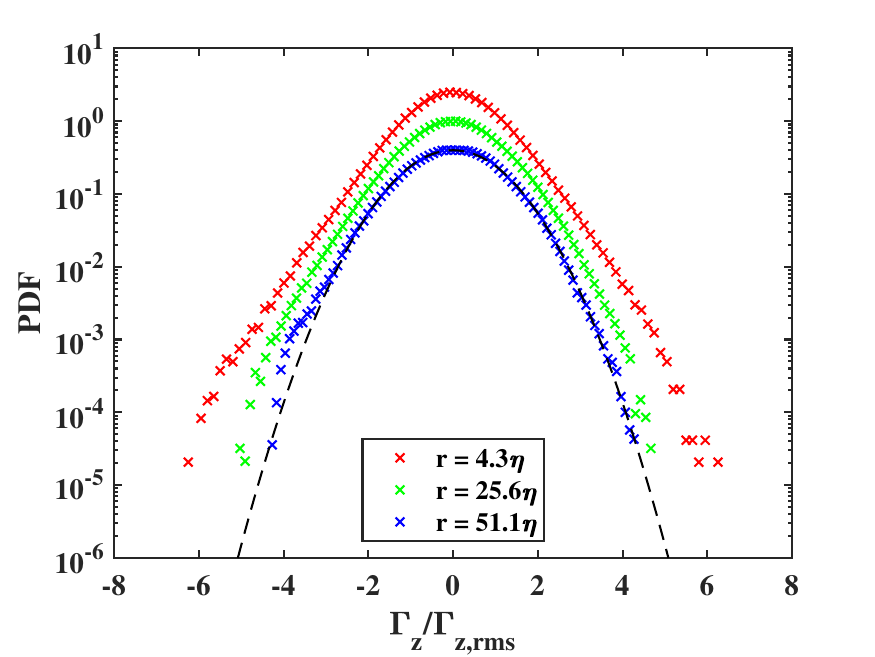} & \includegraphics[width=0.49\textwidth,valign=t]{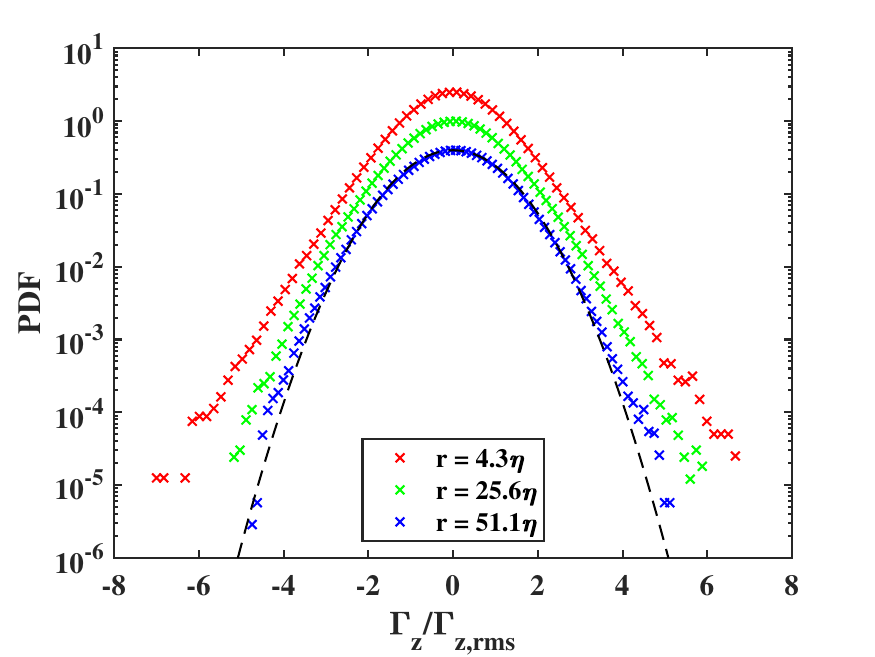} \\
\end{tabular}
\caption{The probability density function (PDF) of circulation $\Gamma_{x}$ (a), $\Gamma_{y}$ (b) and $\Gamma_{z}$ (c) measured with synchronous grid-oscillation mode at inlet $x=$ 8 mm (left) and outlet $x=$ 190 mm (right). The circulation is computed for square loops of size $r\sim$ 4$\eta$ (near the dissipative scale), 25$\eta$ (the inertial scale) and 50$\eta$ (the large scale). The PDF is computed for circulation after subtracting its mean value using 100 bins. The bin values are normalized by the corresponding r.m.s value. 
For clarity, two of the curves are shifted vertically by a relative factor of 2.5.
The Gaussian distribution is shown by the continuous dashed curves.}
\label{fig:PDFCirc}
\end{figure}

\subsection{Probability density function of circulation}
The PDFs of circulation computed at two different streamwise locations, $x=$ 8 mm, downstream of $SOC$ and $x=$ 190 mm downstream of $EOC$ are shown in Figure \ref{fig:PDFCirc}. 
The circulation is computed using square loops around the centerline of the measurement volume. 
The size of the loop is varied from 3 $-$ 27 nodes which corresponds to 1.3 mm to 17 mm. 
The maximum size of the loop is limited by the depth of the measurement volume which is approximately 18-19 mm in our case. 
\textcolor{black}{Circulation computed over nine square loops around the chosen streamwise location is used in each time-step for the PDF calculations.}
The circulation data obtained from 4$-$6 runs conducted on the same day and for the same grid-rotation protocol are combined together to increase the tail resolution of the PDFs. 
With each run having about 50,000 time steps, the PDFs are generated with $\sim$ 2 million square loops.

Figure \ref{fig:PDFCirc} shows PDFs obtained with the synchronous mode, for loops of three different sizes, i.e. $r\sim$ 4$\eta$, near the dissipative scale; 25$\eta$ which is the typical inertial scale (based on the values of $\lambda$ in Table \ref{tab:parameters}); and 50$\eta$ which corresponds to the large scale. 
The PDFs for the different loops sizes have been shifted to minimize clutter, in such a manner that the peak values of two of them are shifted up vertically by the same distance, i.e. by a factor of 2.5. 
\textcolor{black}{The PDFs with the random mode are not shown herein, but are qualitatively similar to the PDFs obtained with the synchronous mode.}

The circulation PDFs, \textcolor{black}{shown} in Figure \ref{fig:PDFCirc}, all show Gaussian behavior for the largest loop size, but deviate from the Gaussian, having wider tails, as the loop is reduced in size to approach the dissipation scales.  The widest tails are measured for $\Gamma_z$, as characterized below by the flatness factor. 
The PDFs of circulation on all three planes exhibit this behavior, though the degree of deviation from Gaussianity are different.
This transition from non-Gaussian to Gaussian with $r$ is more prominent for $\Gamma_{z}$, see Figure \ref{fig:PDFCirc} (c).
The deviation from the Gaussian implies intermittency of the vortex structures \cite{Sreenivasan1996},\cite{Zhou2008}.
This behavior of the PDF has been reported in PIV measurements in the cylinder wake \cite{Sreenivasan1995}, in buoyancy-driven turbulence \cite{Zhou2008} and DNS of isotropic turbulence \cite{Sreenivasan1996},\cite{Iyer2019}. 
Cao et al. \cite{Sreenivasan1996} fitted the stretched tails to exponential forms with exponents varying with the scales. 
Iyer et al. \cite{Iyer2019} showed that the PDFs collapse into a single curve exhibiting a self-similarity when the dimensions of the rectangular loops lie in the inertial range.
\textcolor{black}{We look at the time traces of the instantaneous circulations normalized by their respective r.m.s values, plotted in Figure D1.} 
\textcolor{black}{
Let us look at Figure D1(a) 
which shows the time variation for the synchronous mode at $x=$ 8 mm, with left and right panel corresponding to $r\sim$ 4$\eta$ and 50$\eta$ respectively, for one realization of 50,000 volumes.
Note that there is frequent occurrence of extreme values of $\Gamma_z$ (with $|\Gamma_z/\Gamma_{z,rms}|> $3).
Though such extreme values are seen with $\Gamma_x$ and $\Gamma_y$, they are relatively less frequent with $\Gamma_x$ and even fewer for $\Gamma_y$.
This indicates that $\Gamma_z$ is more intermittent compared to the other two components which manifests as widened PDF tails, with the biggest departure from Gaussian.
With all circulation components, we see the rate of variations reduces with increase in $r$.
The frequent occurrence of extreme values with $\Gamma_z$ with $r\sim$ 4$\eta$ do not exist with $r\sim$ 50$\eta$, which explains its prominent transition from the non-Gaussian to Gaussian behavior in the PDF.
A similar observation with prominent transition in $\Gamma_z$ can be made at $x=$ 190 mm shown in Figure D1(b). }

\begin{figure}
\begin{tabular}{ p{0.06cm} c c }
{} & $x=$ 130 mm & $x=$ 190 mm\\
{} & {} & {}\\
\adjustbox{valign=t}{(a)} & \includegraphics[width=0.49\textwidth,valign=t]{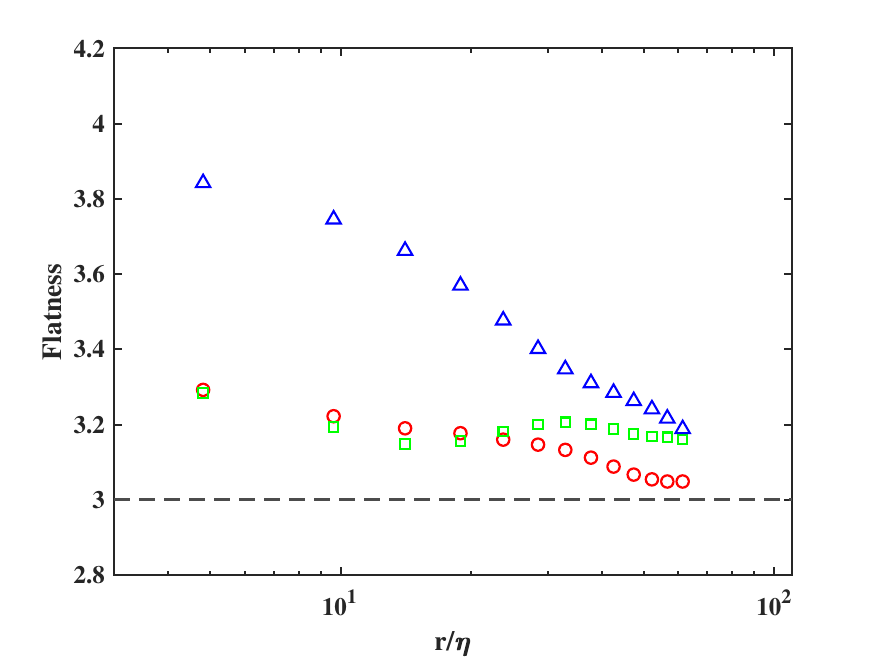} & \includegraphics[width=0.49\textwidth,valign=t]{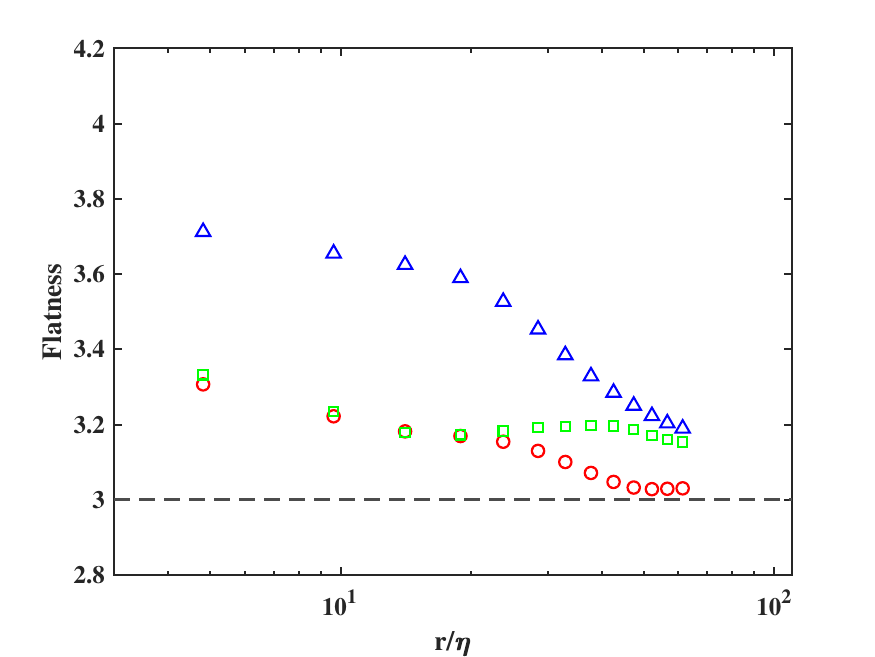}\\
\adjustbox{valign=t}{(b)} & \includegraphics[width=0.49\textwidth,valign=t]{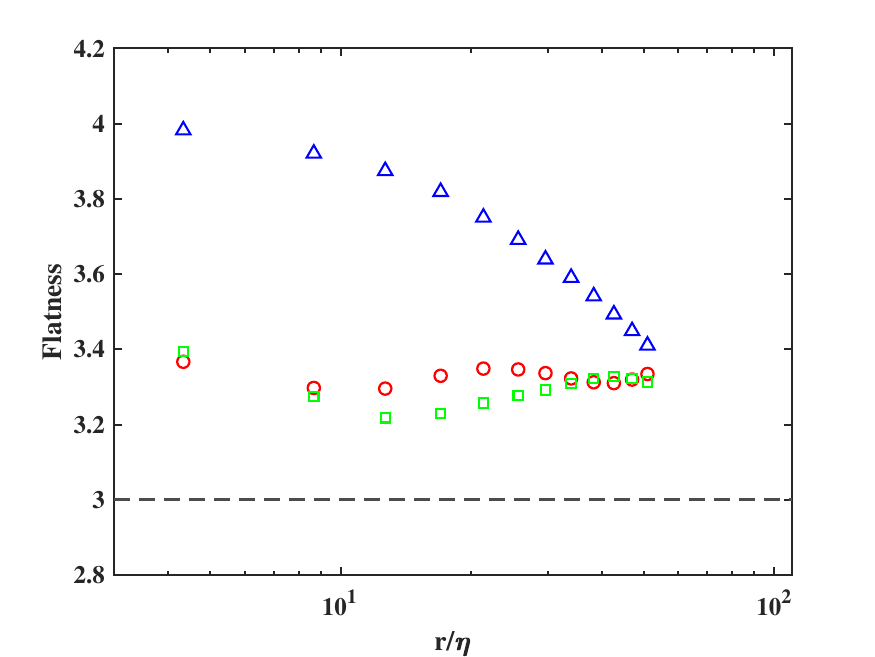} & \includegraphics[width=0.49\textwidth,valign=t]{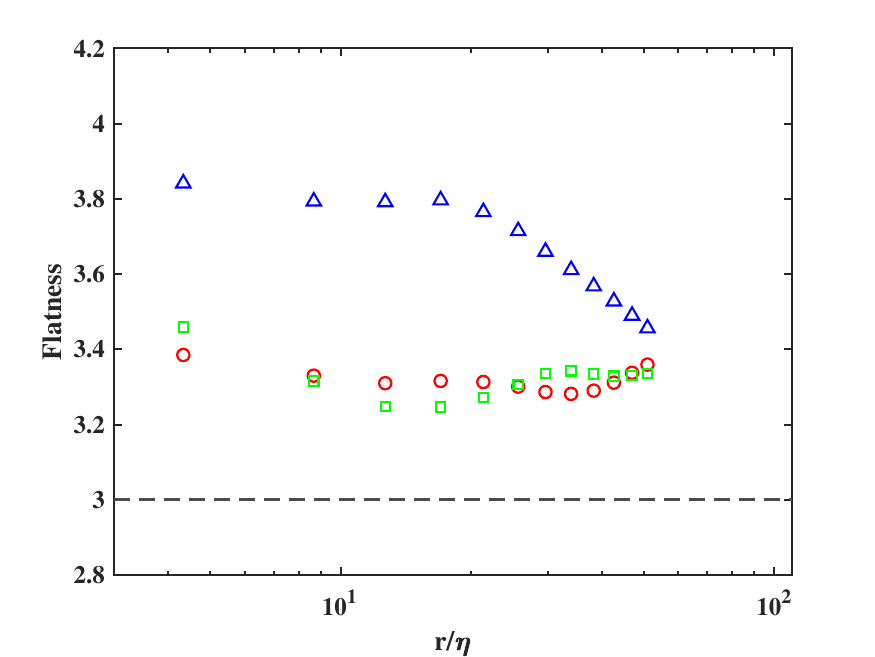} \\
\end{tabular}
\caption{The flatness for probability density function (PDF) of circulation  $\Gamma_{x}$ ($\circ$), $\Gamma_{y}$ ($\square$) and $\Gamma_{z}$ ($\triangle$) measured at $x=$ 130 mm (left) and 190 mm (right) with the synchronous grid-oscillation mode (a) and the random mode (b). The flatness is computed as the fourth moment of the distribution normalized the r.m.s value, $\left | \left \langle \Gamma^{4}  \right \rangle \right |/\left | \left \langle \Gamma^{2}  \right \rangle \right |^2$. The Gaussian value is indicated by the horizontal dashed line.}
\label{fig:PDFFlatness}
\end{figure}

To quantify the deviation from Gaussian, we plot the flatness of the obtained distribution for the entire range of scales for both the grid-rotation modes in Figure \ref{fig:PDFFlatness}. 
The values are plotted at two locations after straining: $x=$ 130 mm which is slightly downstream of the maximum strain location, and at $x=$ 190 mm slightly after the exit of the contraction.
The Gaussian flatness equals 3 and is indicated in the figure by dashed lines.
\textcolor{black}{With $\Gamma_{z}$, we see the expected behavior of flatness tending to the Gaussian value as the loop size increases. 
Due to its higher intermittency, a notable decrease in its flatness is seen with increase in $r$ and this continues till the largest loop.
For $r\sim$ 50$\eta$ the flatness is $\approx$ 3.2 \& $\approx$ 3.4 with the synchronous \& random modes respectively, as seen in Figure \ref{fig:PDFFlatness}.}
However, with circulations in the streamwise and transverse planes ($\Gamma_{x}$ and $\Gamma_{y}$), there is marginal or no decrease in the flatness values with change in the loop sizes.
This indicates weak intermittency in $\Gamma_x$ and $\Gamma_y$ and is inline with their time signals in Figure D1. 
The tendency of flatness of circulation towards constancy in the IR has been observed by Iyer {\it et al.} \cite{Iyer2019} with increase in $Re_{\lambda}$ compared to the flatness of the velocity increments.
Based on this they conclude that $\Gamma$ is weakly intermittent and the velocity increments are highly intermittent.
This could be due to the stronger effect of straining on circulation in these two planes.
We have seen that the vorticity component in the streamwise and transverse directions undergo maximum variation due to straining in the same contraction \cite{Mugundhan2020}.
Higher values of flatness were observed close to the $SOC$, which indicates higher intermittency, which is due to the influence of the grid protocol and slightly higher $Re_\lambda$. 
A greater tendency for the PDFs towards Gaussian at lower $Re_\lambda$ has been reported by Cao et al. (\cite{Sreenivasan1996}).

\begin{figure}
\begin{tabular}{ p{0.06cm} c c }
{} & Synchronous Mode & Random Mode\\
{} & {} & {}\\ 
\adjustbox{valign=t}{(a)} & \includegraphics[width=0.49\textwidth,valign=t]{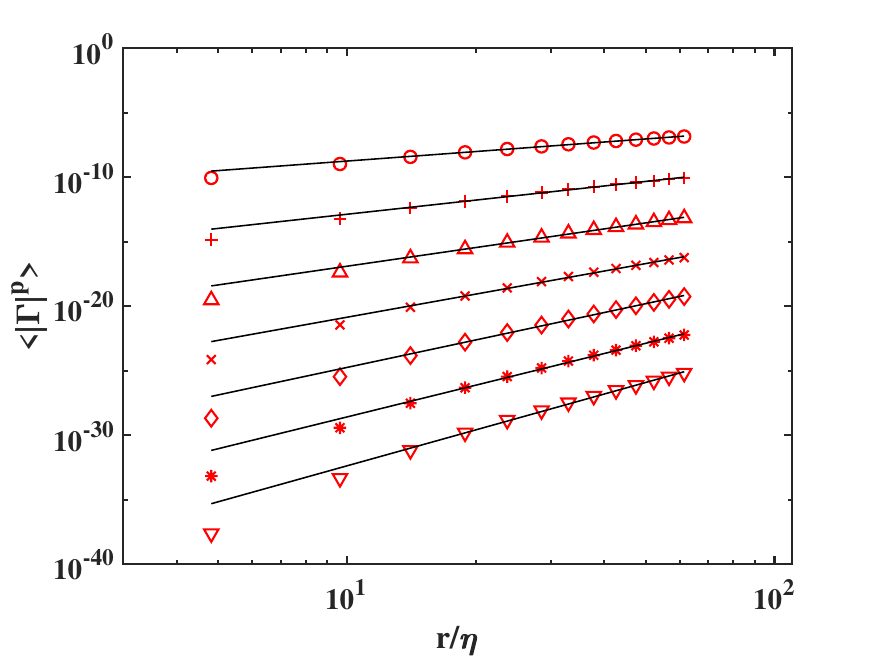} & \includegraphics[width=0.49\textwidth,valign=t]{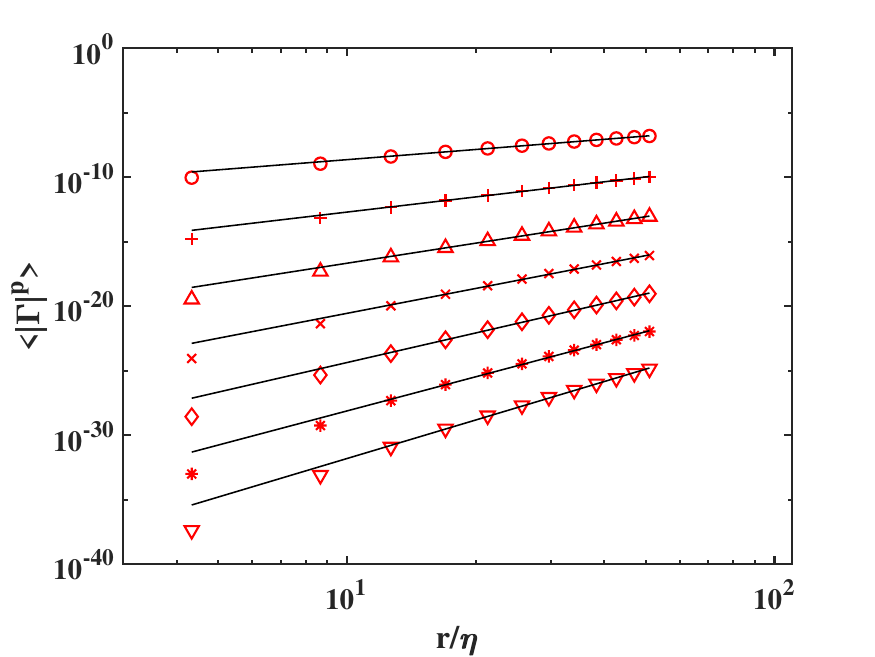} \\
\adjustbox{valign=t}{(b)} & \includegraphics[width=0.49\textwidth,valign=t]{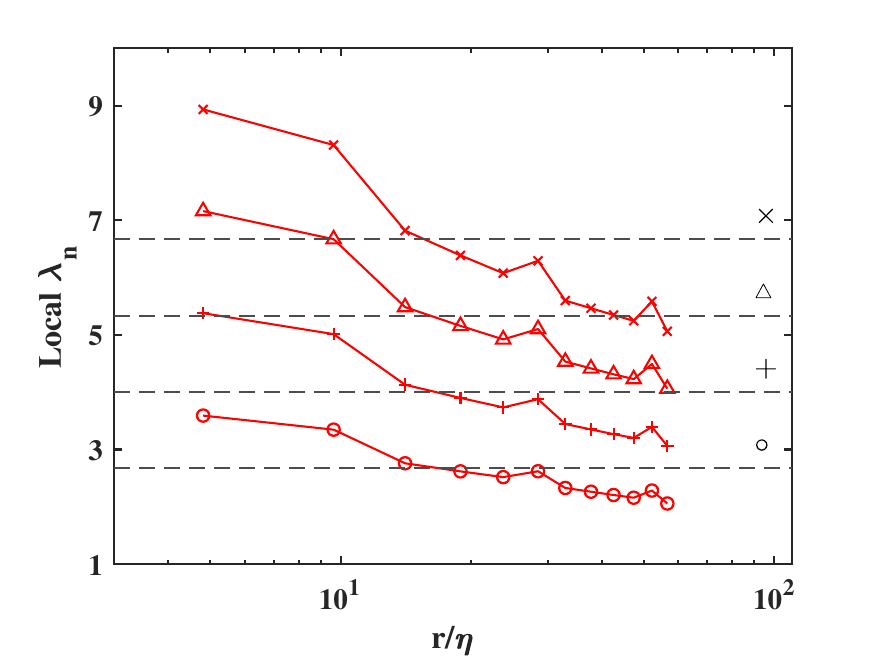} & \includegraphics[width=0.49\textwidth,valign=t]{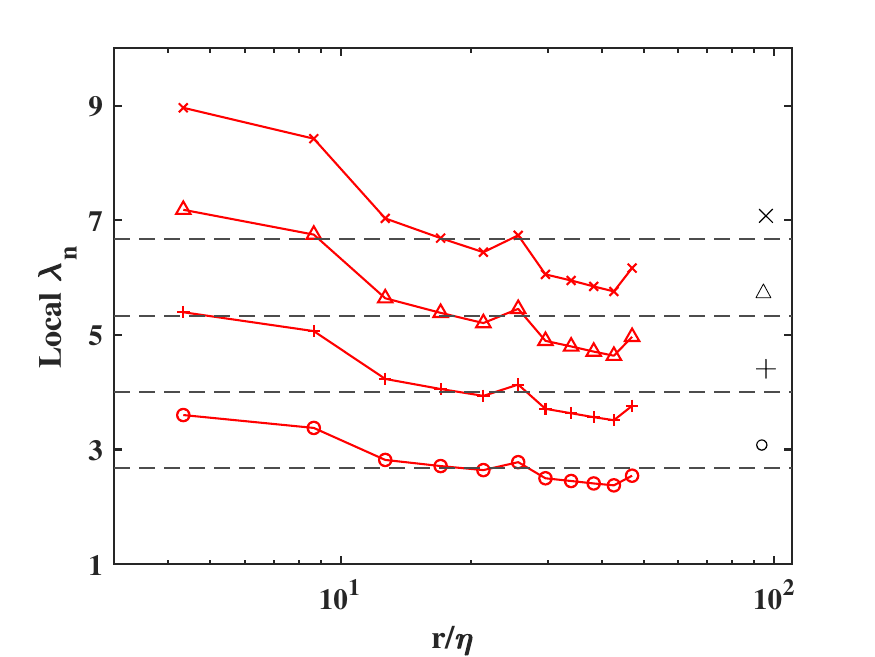} \\
\adjustbox{valign=t}{(c)} & \includegraphics[width=0.49\textwidth,valign=t]{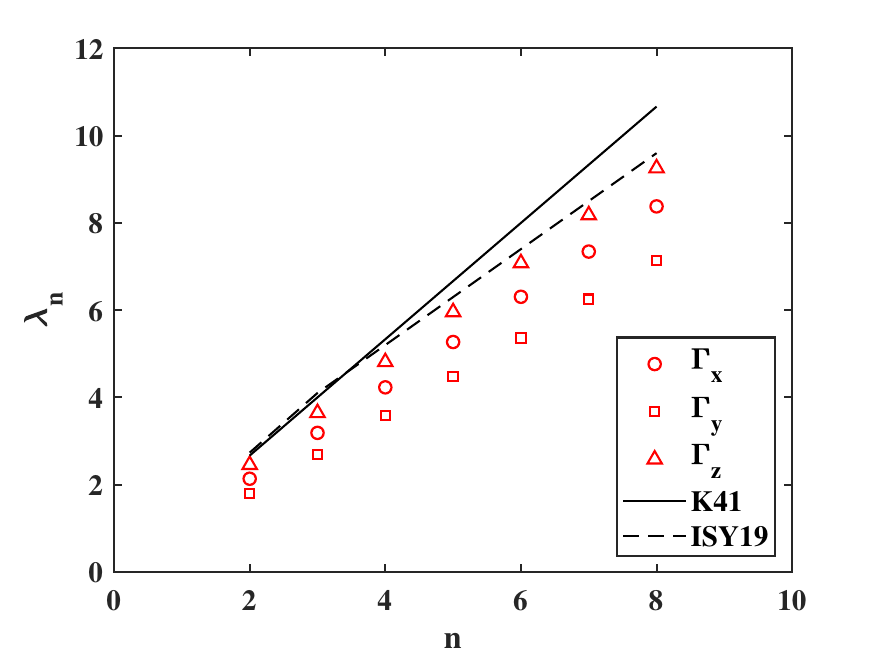} & \includegraphics[width=0.49\textwidth,valign=t]{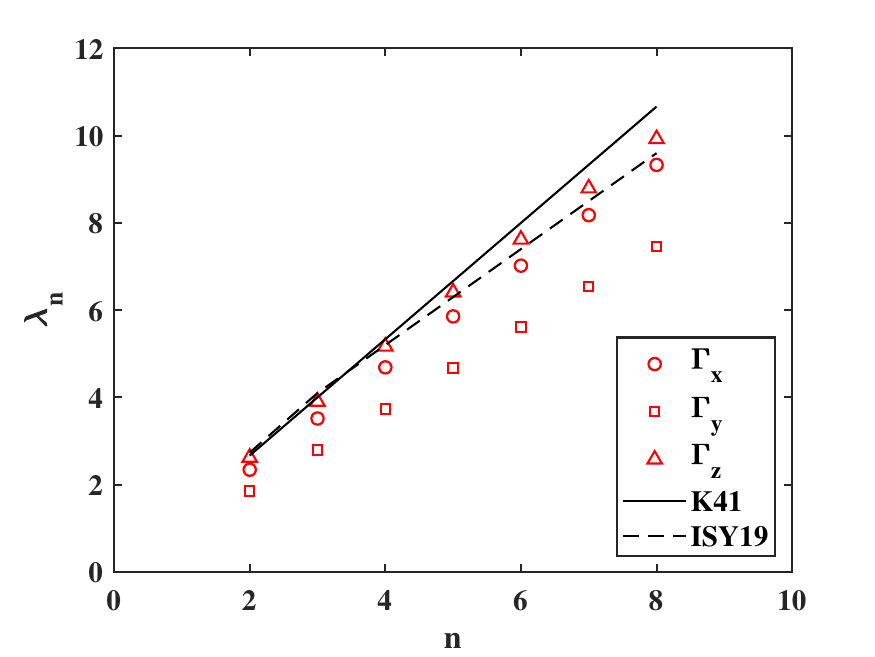} \\
\end{tabular}
\caption{The moments of circulation-PDF, $ \left \langle \left |\Gamma \right |^{n} \right \rangle$ and its scaling computed at $x=$ 190 mm with mode S1 (left) and the random mode (right). 
(a) The moments of $\Gamma_{z}$ of order $n=$ 2 ($\circ$), 3 ($+$), 4 ($\bigtriangleup$), 5 ($\times$), 6 ($\diamond$), 7 ($\ast$) and 8 ($\bigtriangledown$). The black line represents the best power-law fit of the form $\left \langle \left |\Gamma \right |^{n} \right \rangle\sim \left ( r/\eta \right )^{\lambda_n}$ for the data in the intermediate scales. 
(b) The local slopes of the data shown in (a) which is computed as $\Delta [log \left \langle \left |\Gamma \right |^{n} \right \rangle]/\Delta[log \left ( r/\eta \right )]$. The local slopes are shown for the moments of order $n=$ 2$-$5. The horizonal dashed lines with the symbol on it represent the corresponding K41 value. 
(c) The values of power-law exponent for moments of order $n=$ 2$-$8. 
The solid line indicates the K41 scaling, $\lambda_n=4n/3$.
The dotted lines indicate the fits proposed by ISY19 (Iyer {\it et al.} (2019) \cite{Iyer2019}) i.e., for $n<$ 3, ${\lambda_n}=1.367n$ and $n>$ 3, ${\lambda_n}=1.1n+(3-D)$, where $D$ is 2.2.}
\label{fig:PDFMom}
\end{figure}

\subsection{Moments of the circulation}

We look at the scaling properties of the circulation moments, which are computed as 
(\ref{eqn:gamma_mom}). 

\begin{equation}
\left \langle \left |\Gamma \right |^{n} \right \rangle=\int \left |\Gamma \right |^{n} f(v)dv.
\label{eqn:gamma_mom}
\end{equation}

Here, $f$ is the circulation PDF, and the absolute value of the circulation is used to include the odd-order moments in the scaling.
Iyer {\it et al.} \cite{Iyer2019} suggest the scaling improves if the absolute value is used
and we follow their example.
Keep in mind that our geometry retains symmetries,  along the centerline, so the mean values of all components should be zero.
These moments are also referred to as the circulation structure functions \cite{Zhou2008}, definition being in line with the velocity structure functions.
According to the Kolmogorov theory (K41) for homogeneous isotropic turbulence, the circulation structure functions have a power-law scaling, $\sim (r/\eta)^{\lambda_n}$, where the exponent $\lambda_n=$ 4$n/$3 \cite{K41}.
The velocity structure function has a lower value for its exponent which is $\zeta_n=$ $n/$3.
However, based on the simulations at high Re$_\lambda \sim$ 1300, Iyer {\it et al.} \cite{Iyer2019} proposed two fits for these exponents based on the value of $n$.
The proposed fits are included in the caption of Figure \ref{fig:PDFMom}.
The moments of circulation-PDF is plotted in Figure \ref {fig:PDFMom}(a) for order $n=$ 2 to 8 and for both the grid-rotation modes. 
We restrict to only order 8, considering the low accuracy of the technique in prediction of higher order statistics. 
The power-law fit that approximates the data in the intermediate scales ($r/\eta >$ 10) are also shown. 
The corresponding local slopes of the data shown in Figure \ref{fig:PDFMom}(a) is plotted in Figure \ref{fig:PDFMom}(b) assuming a linear variation between adjacent points. 
We see that the local slopes seen with $n=$ 2, remains fairly constant over a range of scales with $r/\eta$ above 10.
Hence we chose this range of scales to obtain the power-law fits. 
The exponent of the fits for both the modes and circulation in the three planes at \textcolor{black}{$x=$ 190 mm} are plotted in Figure \ref {fig:PDFMom}(c). 
If the PDF has a Gaussian distribution, the moments should follow the K41 scaling. 
It is seen that in all cases, the exponents have sub-Kolmogorov scaling which has been seen with experiments and simulations reported in the literature \cite{Sreenivasan1995,Sreenivasan1996,Iyer2019}.
\textcolor{black}{The deviation from K41 increases with increase in $n$.}
It can be seen from figures \ref{fig:PDFMom}(c) that of the three circulations the $\Gamma_{z}$ most closely follows the K41 (the solid lines in the figure), followed by the streamwise $\Gamma_{x}$ and the transverse $\Gamma_{y}$, which fall below the prediction. 
\textcolor{black}{The exponents of the fits near the contraction inlet are shown in Figure D2. 
Here, $\Gamma_{z}$ follows K41, however there is a change in the deviation from K41 with the other components.
Larger departure from K41 is seen for $\Gamma_{x}$. 
Zhou {\it et al.} \cite{Zhou2008} note that such deviation from K41 could be due to intermittency or the anisotropy of the coherent structures.}
\textcolor{black}{The moments of the spanwise $\Gamma_z$ do not undergo much change between inlet and outlet of the contraction.
On the other hand, the compression of the vortical structures manifests as the increase in the deviation of the moments of $\Gamma_y$.}
This is inline with the flatness shown by the streamwise and transverse circulations, which is attributed to the intermittency.
As pointed out previously, in the large scale, the vortex filaments experience extension and compression along these directions respectively.
\textcolor{black}{The streamwise $\Gamma_x$ and the transverse $\Gamma_y$ undergo the straining effects.}
Of the two grid-rotation modes, the random mode shows lesser deviations from K41 as compared to the mode S1.
However for both modes, the exponents are closer to the scaling given by ISY19 (Iyer et al. (2019) \cite{Iyer2019}), which are based on the DNS data of homogeneous isotropic turbulence. 

\begin{table}
\centering
\begin{tabular}{ p{3.8cm} p{1.0cm} p{1.0cm} p{1.0cm} p{1.0cm} p{1.0cm} p{1.0cm} }
\hline \\
{} & \multicolumn{3}{c}{$\Delta x=$ 10 mm} & \multicolumn{3}{c}{$\Delta x=$ 21 mm}\\
\cmidrule(lr){2-4}\cmidrule(lr){5-7}
\vspace{-0.005in}\\
{Condition} & \centering{$R_{\Gamma_x}$} & \centering{$R_{\Gamma_y}$} & \centering{$R_{\Gamma_z}$} & \centering{$R_{\Gamma_x}$} & \centering{$R_{\Gamma_y}$} & {\hspace{0.25cm}$R_{\Gamma_z}$}\\
\vspace{-0.25cm}\\
\hline \\
No deformation & 0.53 & 0.25 & 0.35 & 0.27 & $-$0.01 & $-$0.06 \\
With deformation & 0.46 & 0.32 & 0.39 & 0.22 & $-$0.03 & $-$0.05 \\
\vspace{0.005in}\\
\hline \\
{} & \multicolumn{6}{c}{\includegraphics[width=9cm,valign=c]{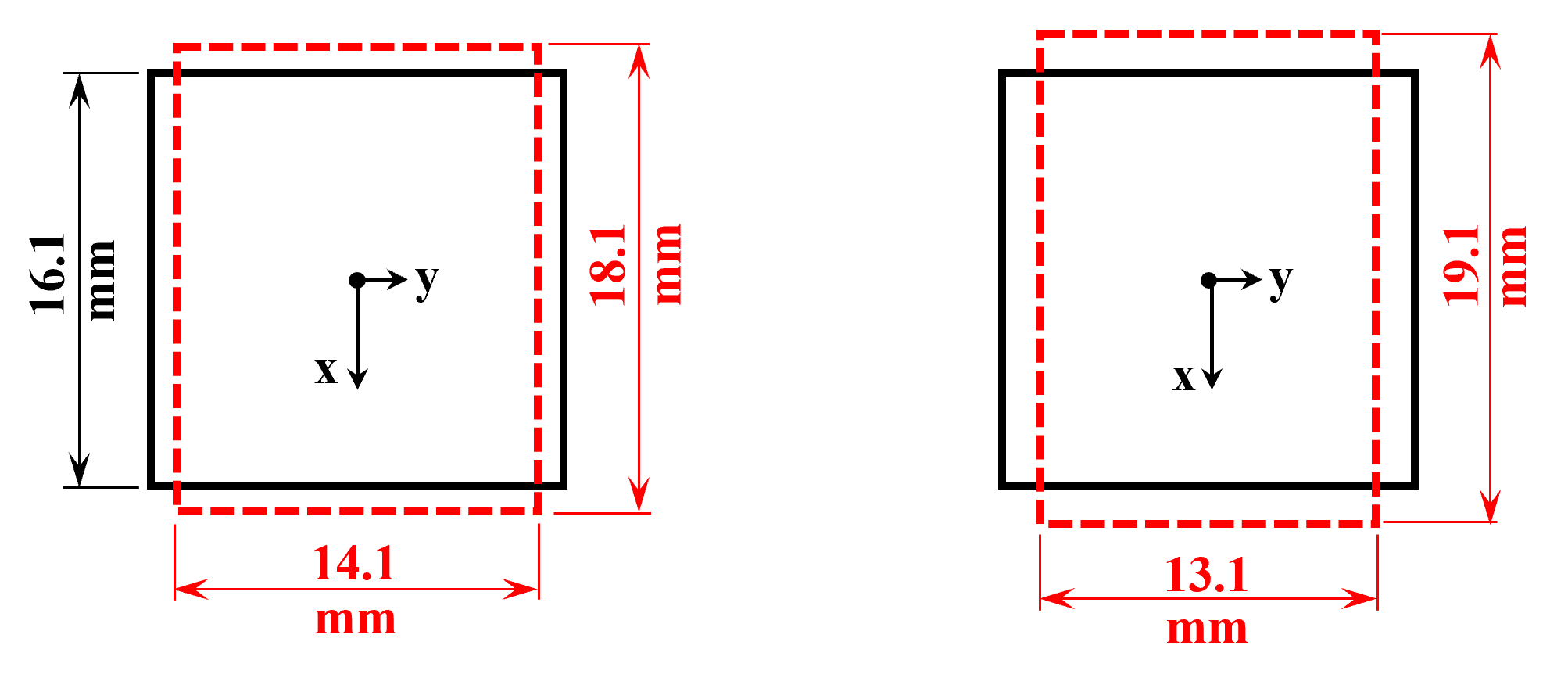}} \\
\hline \\
\end{tabular}
\vspace{0.1in}\\
\caption{Comparison of the correlation values $R_{\Gamma_x}$, $R_{\Gamma_y}$ and $R_{\Gamma_z}$ obtained with loops with and without considering deformations from the mean strain, calculated for the synchronous grid-rotation mode.  The correlation is computed with loops which are initially a 16.1-mm square at $x_1=$ 130 mm. The variation of the correlation with $\Delta x$ for loops without deformation is shown in Figure \ref{fig:Corr}.
The loop after deformation during the respective space shifts is indicated by red dashed lines.}
\label{tab:corr_comp}
\end{table}

\subsection{Circulation with loop deformation}

The spatial correlations of the circulations were computed using a fixed square loop shape and area inside the contraction. 
However, the material elements and loops get deformed by the mean strain as they are advected through the contraction.
To estimate the effect of this on the correlation functions, we compare circulation statistics with an without deforming the loop-shape, while keeping their area constant.
Table \ref{tab:corr_comp} sketches the deformation of an initially 16.1-mm square-loop at $x_1=$ 130 mm, which gets stretched by 2 mm in the $x$ direction as it is advected by 10 mm downstream; and by 3 mm as it is advected by 21 mm, 
with a corresponding squeezing in the $y-$direction and no change in its $z-$direction.
Table \ref{tab:corr_comp} lists the resulting changes in the correlation values at $x=10$ mm of
($-$0.07, 0.07, 0.04) for ($R_{\Gamma_x}$, $R_{\Gamma_y}$, $R_{\Gamma_z}$) respectively.
Further downstream, for $\Delta x=$ 21 mm, the corresponding differences were ($-$0.05, $-$0.02, 0.01), where 
$R_{\Gamma_x}$ remains strong, while $R_{\Gamma_y}$ and $R_{\Gamma_z}$ have values close to zero.
We conclude that the rather modest deformation of the loops does not alter our conclusions.

\section{Discussion and conclusions}

\begin{figure}
\begin{tabular}{ c c }
{(a)} & {(b)}\\ 
\includegraphics[width=0.5\textwidth,valign=t]{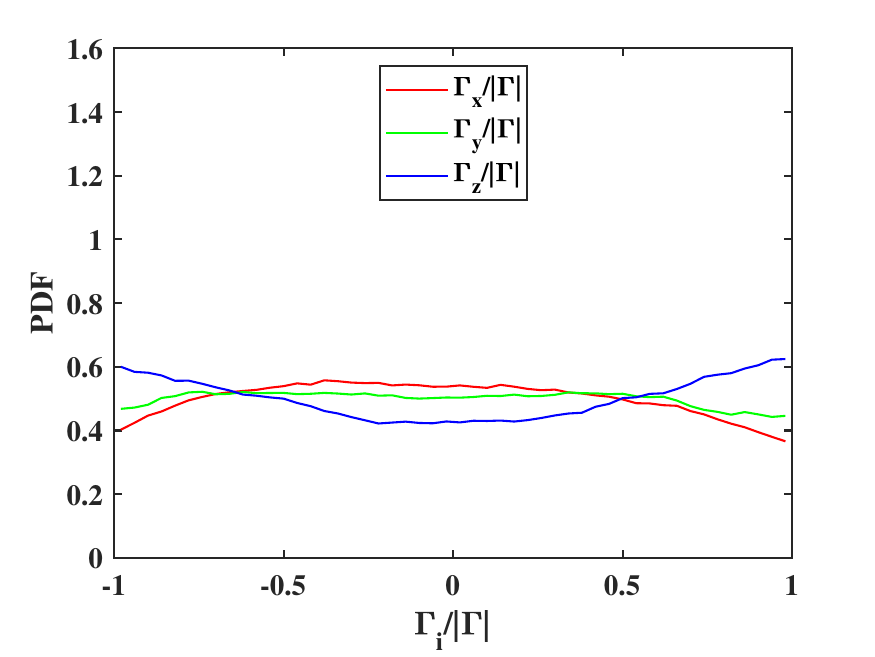} &  \includegraphics[width=0.5\textwidth,valign=t]{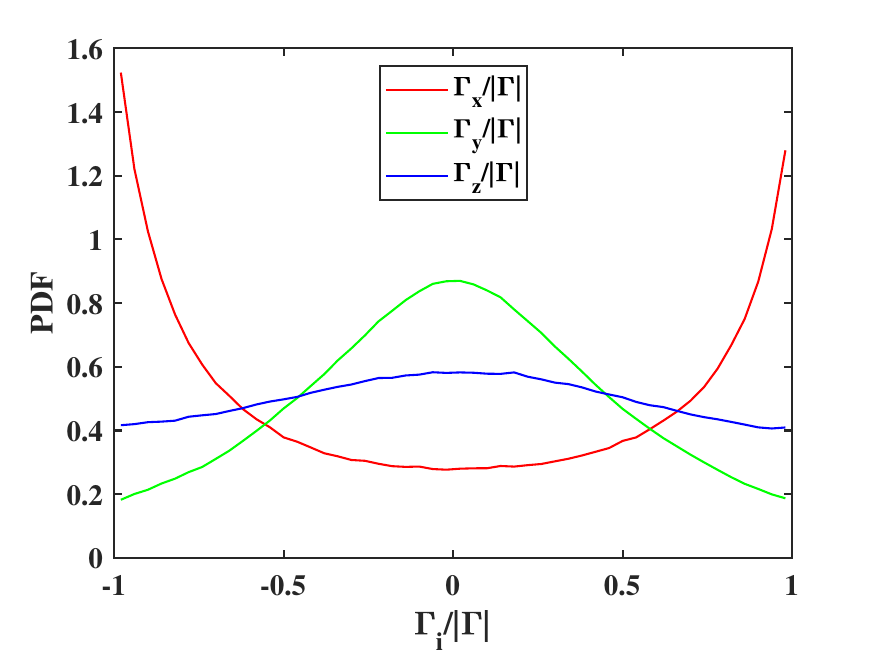}\\
\end{tabular}
\caption{The probability density function (PDF) of the relative circulation strengths at (a) $x=$ 8 mm which is near the start of contraction, and (b) $x=$ 190 mm which is after the exit. The PDFs are measured with the random mode for loop size $r\sim$ 50$\eta$ using 50 bins. At the inlet all the three circulation components are about evenly distributed, while at the exit the circulation is strongest in the streamwise stretching direction $\Gamma_x$ and weakest in the transverse compression direction $\Gamma_y$.}
\label{fig:PDFCircSt_R_YG}
\end{figure}

Herein we have experimentally determined the statistical evolution of the three components of circulation in turbulent flow that is convected through a two-dimensional smooth contraction.
Turbulent fluctuations were injected into the flow using an active grid operated in either synchronous or random rotation modes, to reach $Re_{\lambda}\approx$ 220.
We have used the latest \textcolor{black}{3-D LPT} with the STB algorithm to measure the time-resolved three-dimensional velocity fields in volume-slices inside the contraction.
This enabled us to calculate the simultaneous circulations in loops contained in three mutually perpendicular planes, \textcolor{black}{casting these circulations in terms of a vector.
This allow us to pinpoint the effects of the mean strain on the circulation components and their effective orientation. Figure \ref{fig:PDFCircSt_R} reveals this orientation by comparing the PDF of the individual directional components $\Gamma_x$, $\Gamma_y$ and $\Gamma_z$ normalized by the instantaneous absolute value $\left| \Gamma \right|$ of the vector.
While the $\Gamma_x$ circulation around a fixed-size loop, will clearly increase from the streamwise stretching and enhancement of vorticity within the loop, the circulation will also increase by the enhanced streamwise alignment of the vortices, see Mugundhan {\it et al.} (2020).
We interpret the orientation of the circulation vector as indicating the corresponding orientation of large-scale coherent vortices.}  This is highlighted in Figure \ref{fig:PDFCircSt_R_YG}},
which compares the three PDFs at the inlet in (a) and the outlet of the contraction in (b).   At the inlet all the three circulation components are evenly distributed, with a nearly flat distribution, while at the exit the component PDFs vary greatly.  
\textcolor{black}{They show that the straining of the contraction causes orientation with $\Gamma_x$ to be the strongest (by the largest probability at $\Gamma_x/\left| \Gamma \right| = \pm 1$), while the weakest orientation is in the contraction direction, i.e.  $\Gamma_y/\left| \Gamma \right|$ peaking around 0.  }

This effective orientation can also be seen with the behavior of the streamwise integral length scales $L_{\Gamma}$ deduced from the streamwise spatial correlation of the circulation components.
\textcolor{black}{The straining causes a strong increase in $L_{\Gamma_{x}}$ while the other components are mostly unchanged, as shown in Figure \ref{fig:IntScale}.}
We also see that the random grid-oscillation mode has larger integral scales than the synchronous mode.
This is due to the lower grid rotation speed used in the random mode, 3.5 vs 4.0 rps.
Low grid-rotation speeds enables longer flap-coherent structure interaction time and hence results in larger integral scales.
Similar observations were made by Larssen \& Devenport \cite{Larssen2011}.
\textcolor{black}{The length scales based on the two-point velocity correlations in active-grid generated turbulence in wind tunnel have been reported \cite{Larssen2011,Thormann2014} but are} \textcolor{black}{ based on point measurements using Taylor's hypothesis of frozen turbulence.
Larssen \& Devenport \cite{Larssen2011} used a macro length scale $=k^{3/2}/\epsilon$ as a substitute for integral length scale.
They observed larger macroscales at lower rotation speeds of the active grid.}

The circulation PDFs all show Gaussian behavior for the largest loop sizes (Figure \ref{fig:PDFCirc}), but deviate from the Gaussian, with wider tails, as the loop is reduced in size towards the dissipative scales.  The widest tails are measured for $\Gamma_z$.
This deviation is characterized by the flatness factor (Figure \ref{fig:PDFFlatness}), which
shows the clearest transition away from Gaussian for $\Gamma_z$ evaluated on small loops.
On the other hand, the flatness of $\Gamma_x$ and $\Gamma_y$ remain fairly near the Gaussian value between $3-3.4$.
\textcolor{black}{This may indicate that $\Gamma_x$ and $\Gamma_y$ are weakly intermittent compared to the spanwise $\Gamma_z$.}
\textcolor{black}{The power-law scaling of the moments of $\Gamma$ shows sub-Kolmogorov scaling for all three components.
Large deviations are seen in the compressed direction of $\Gamma_y$ at the exit of the contraction.}
\textcolor{black}{The PDF satisfies approximately the area rule with only 10\% deviations in the rms values, which we discuss in Appendix F.
A larger sample would be needed to verify this rule, which is beyond the current experiments.}

\textcolor{black}{In closing, while the commonly used statistical characterizations of circulation, such as pdf and scaling exponents, have proven powerful to characterize the dynamics of homogenous turbulence, we find that for turbulence exposed to large-scale strain it is more effective to measure the simultaneous circulations in three perpendicular planes.  This allows us to construct a circulation vector and pinpoint how the strain enhances and orients coherent vortical structures, by instantaneously following the relative strength of the components, as is implied by the PDFs in Figure 13.  Keep in mind that this has only become possible experimentally with the recently available volumetric measurements of velocity, employed herein.}

\section*{Acknowledgements}

This study was supported by King Abdullah University of Science and Technology (KAUST) under grant BAS/1352-01-01.  

\newpage
\appendix
\FloatBarrier


\section{Relative strength of circulation components}

\begin{figure}[h!]
\begin{tabular}{ p{0.06cm} c c }
{} & $r\sim$ 28$\eta$ & $r\sim$ 60$\eta$ \\
{} & {} & {}\\ 
\adjustbox{valign=t}{(a)} & \includegraphics[width=0.49\textwidth,valign=t]{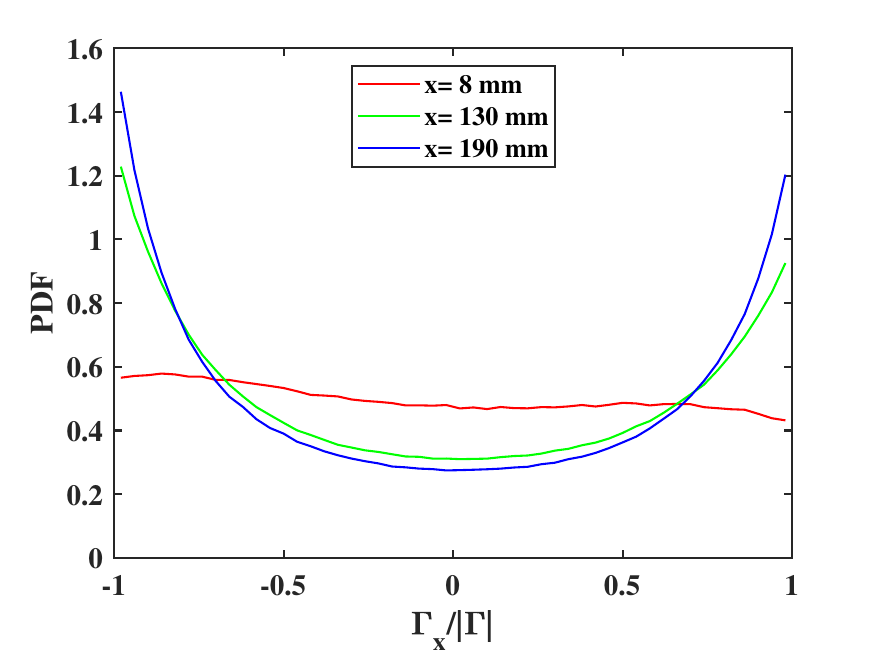} &  \includegraphics[width=0.49\textwidth,valign=t]{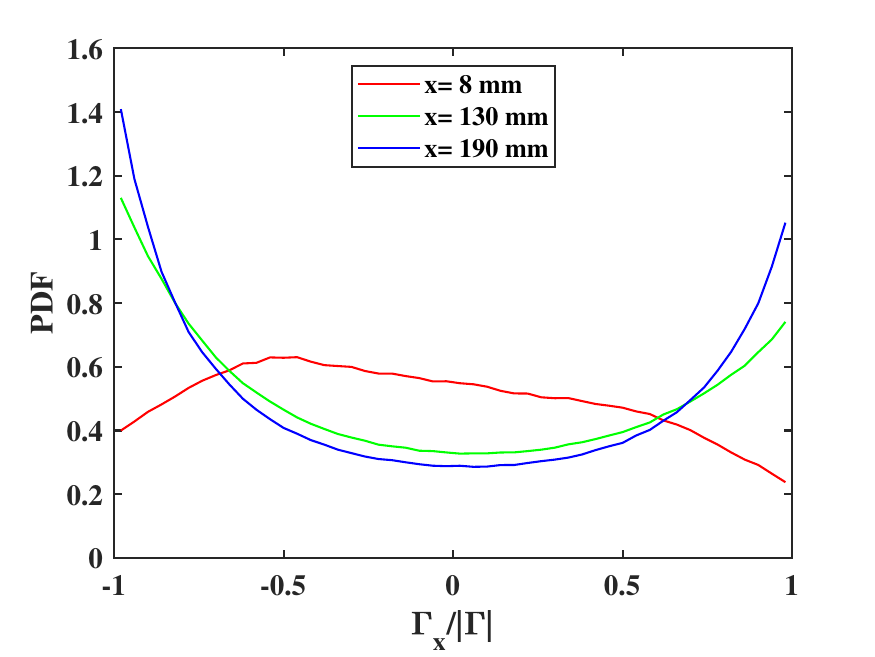}\\
\adjustbox{valign=t}{(b)} & \includegraphics[width=0.49\textwidth,valign=t]{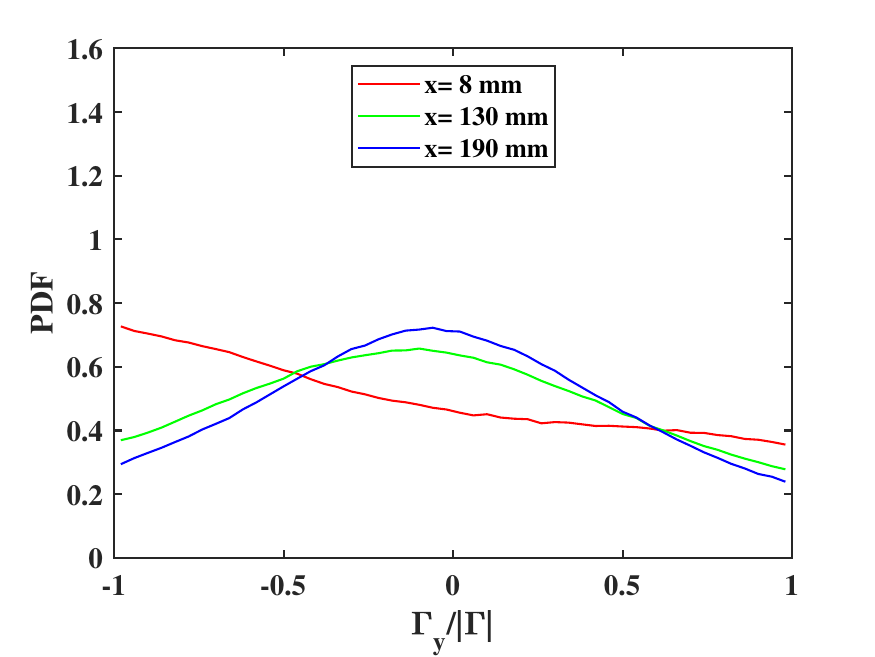} & \includegraphics[width=0.49\textwidth,valign=t]{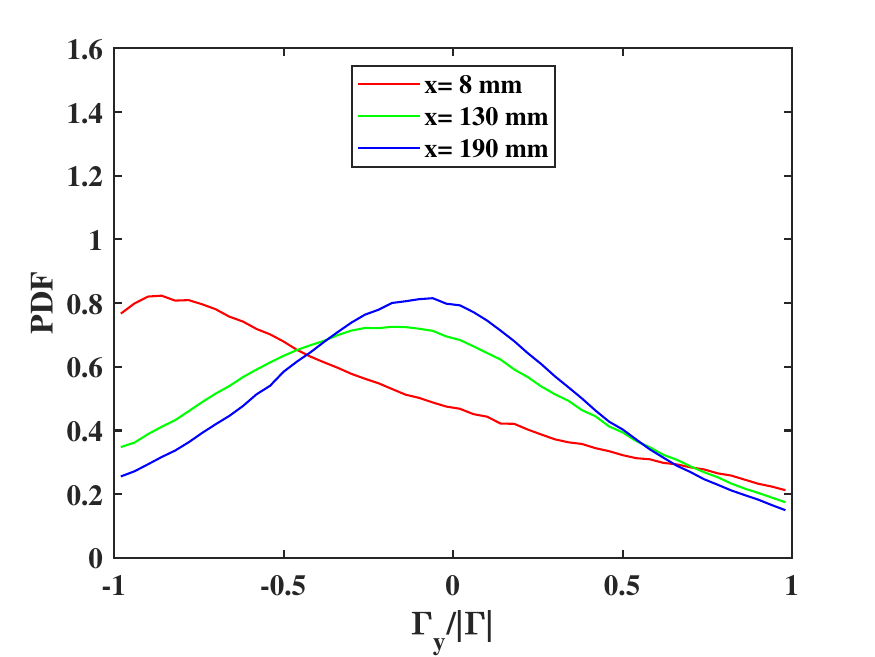}\\
\adjustbox{valign=t}{(c)} & \includegraphics[width=0.49\textwidth,valign=t]{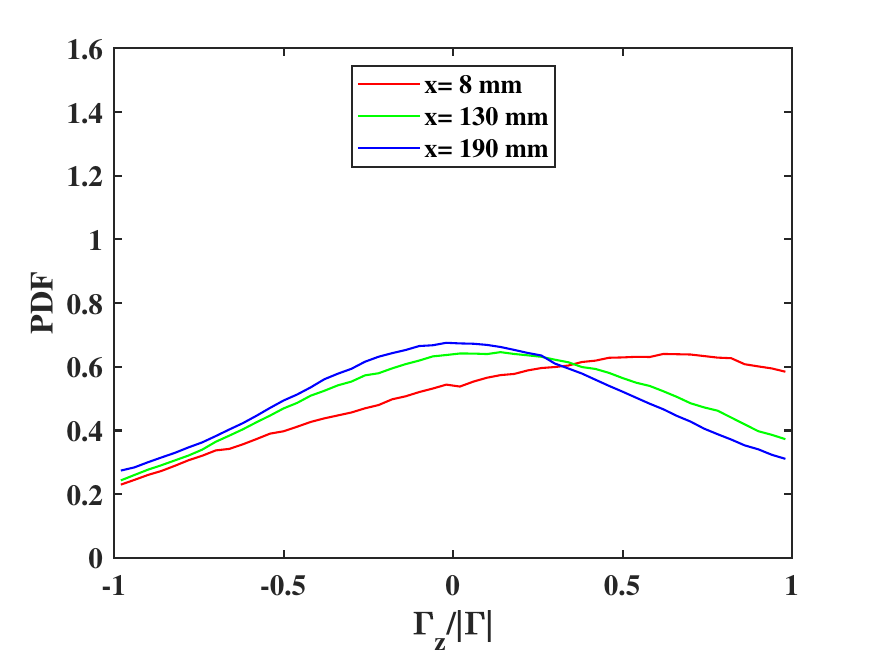} & \includegraphics[width=0.49\textwidth,valign=t]{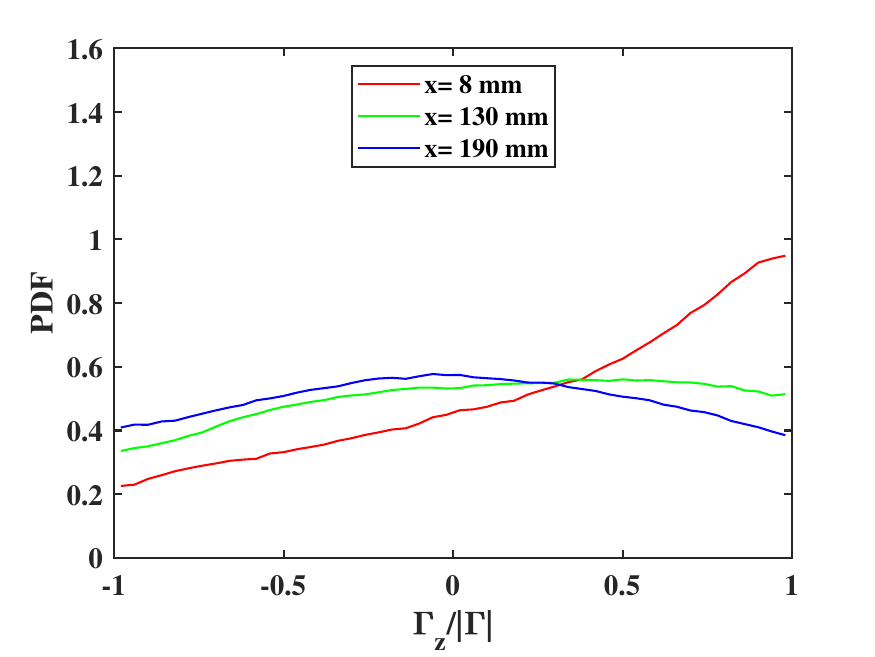}\\
\end{tabular}
\caption{The asymmetry of the circulations for the synchronous grid-oscillation mode. The streamwise variation of probability density function (PDF) of relative circulation strengths $\Gamma_{x}/\left | \Gamma \right |$ (a), $\Gamma_{y}/\left | \Gamma \right |$ (b) and $\Gamma_{z}/\left | \Gamma \right |$ (c) measured with  grid-rotation mode S1 for loop sizes $r\sim$ 28$\eta$ (left) and $r\sim$ 60$\eta$ (right). Refer to caption of Figure 5 
for more details. Note the much larger asymmetry here for mode S1 than the random mode in Figure 5.} 
\label{fig:PDFCircSt_S1}
\end{figure}

\newpage
\section{Homogeneity of the mean and r.m.s vorticity field}
\FloatBarrier

\begin{figure}[h!]
\begin{center}
\begin{tabular}{ p{0.5cm} c c }
{} & Synchronous Mode & Random Mode\\
\vspace{-2mm}
\adjustbox{valign=c}{$\left \langle\omega_{x}\right \rangle$} & \includegraphics[width=0.375\textwidth,valign=c]{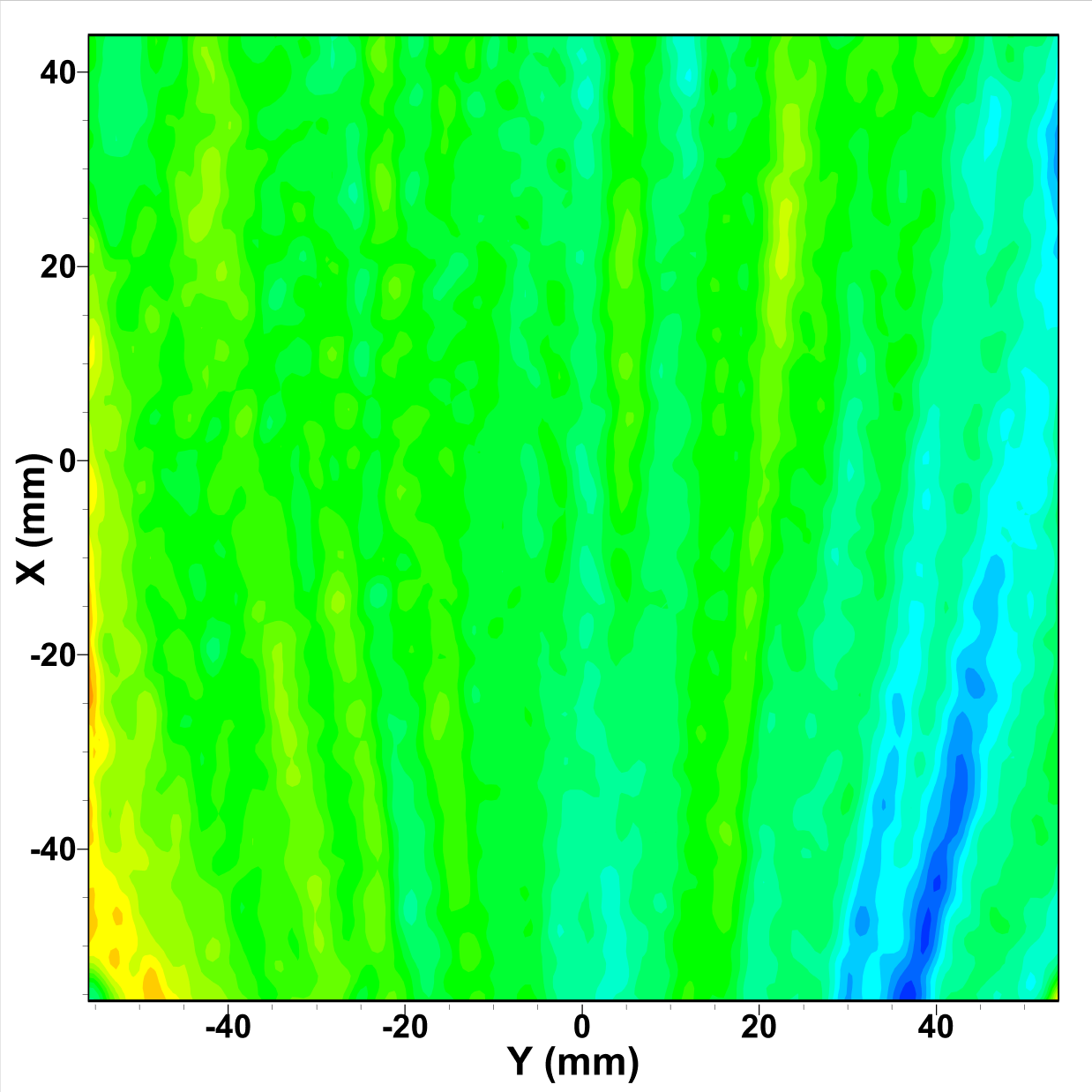}&
\includegraphics[width=0.375\textwidth,valign=c]{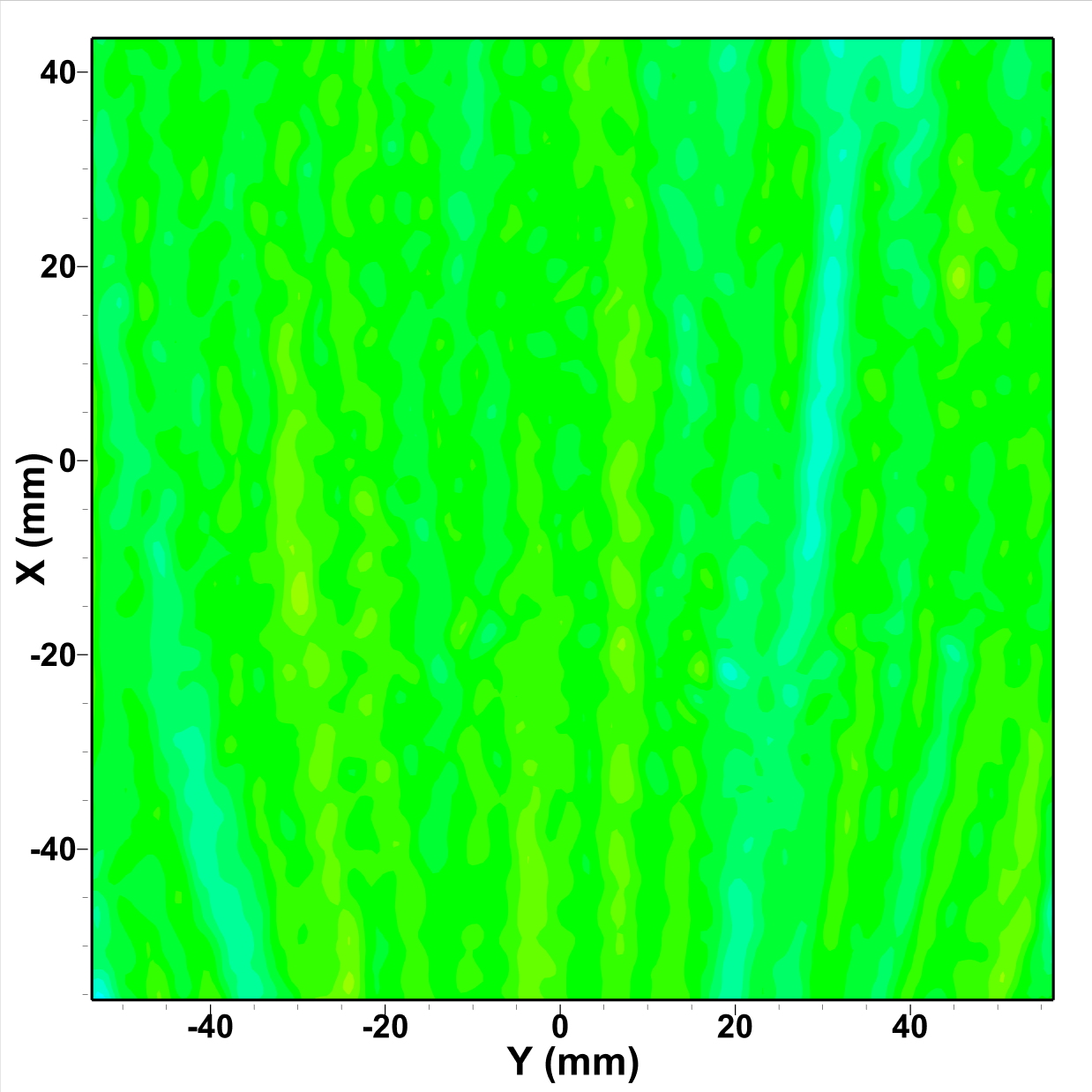}\\
\vspace{-2mm}
\adjustbox{valign=c}{$\left \langle\omega_{y}\right \rangle$} & \includegraphics[width=0.375\textwidth,valign=c]{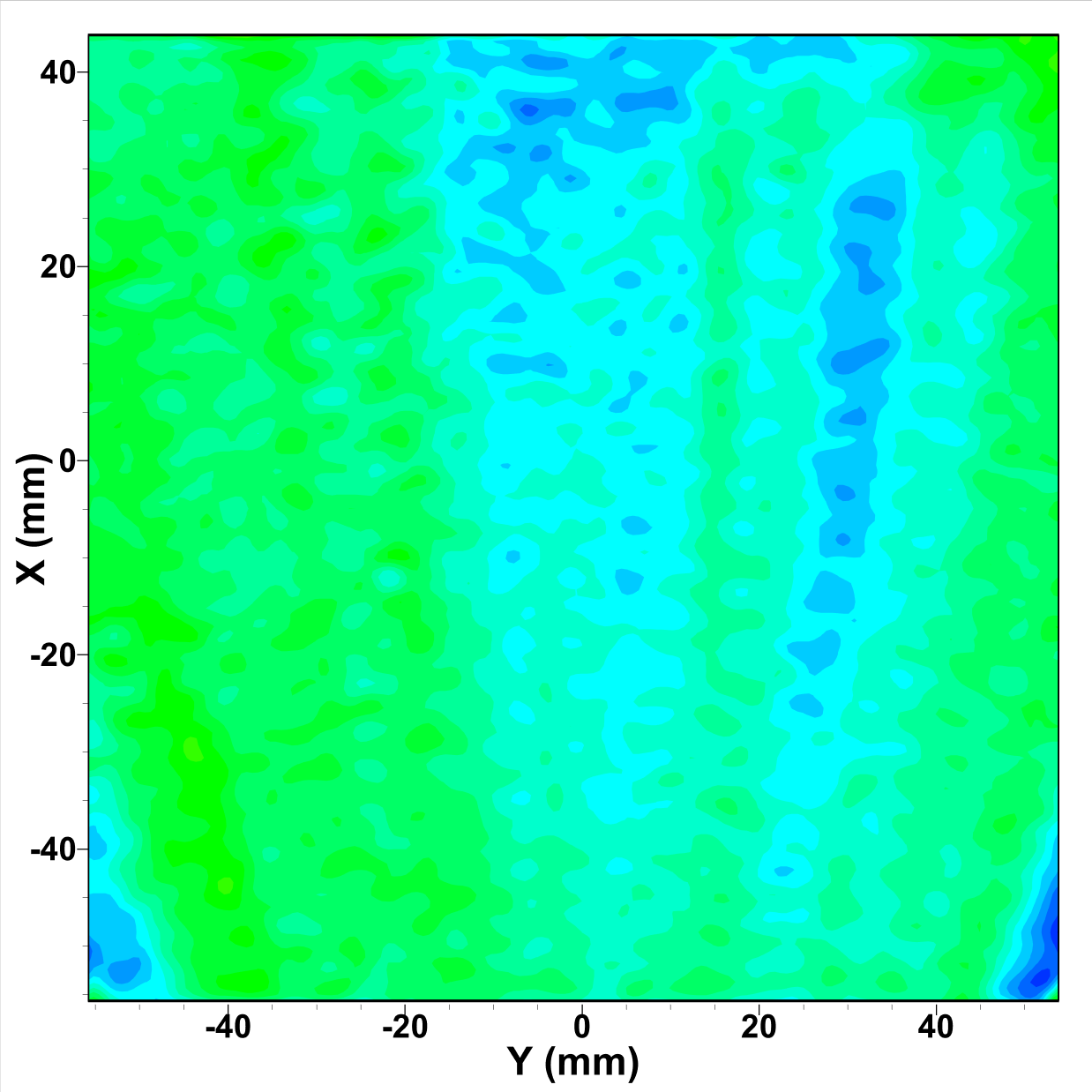}&
\includegraphics[width=0.375\textwidth,valign=c]{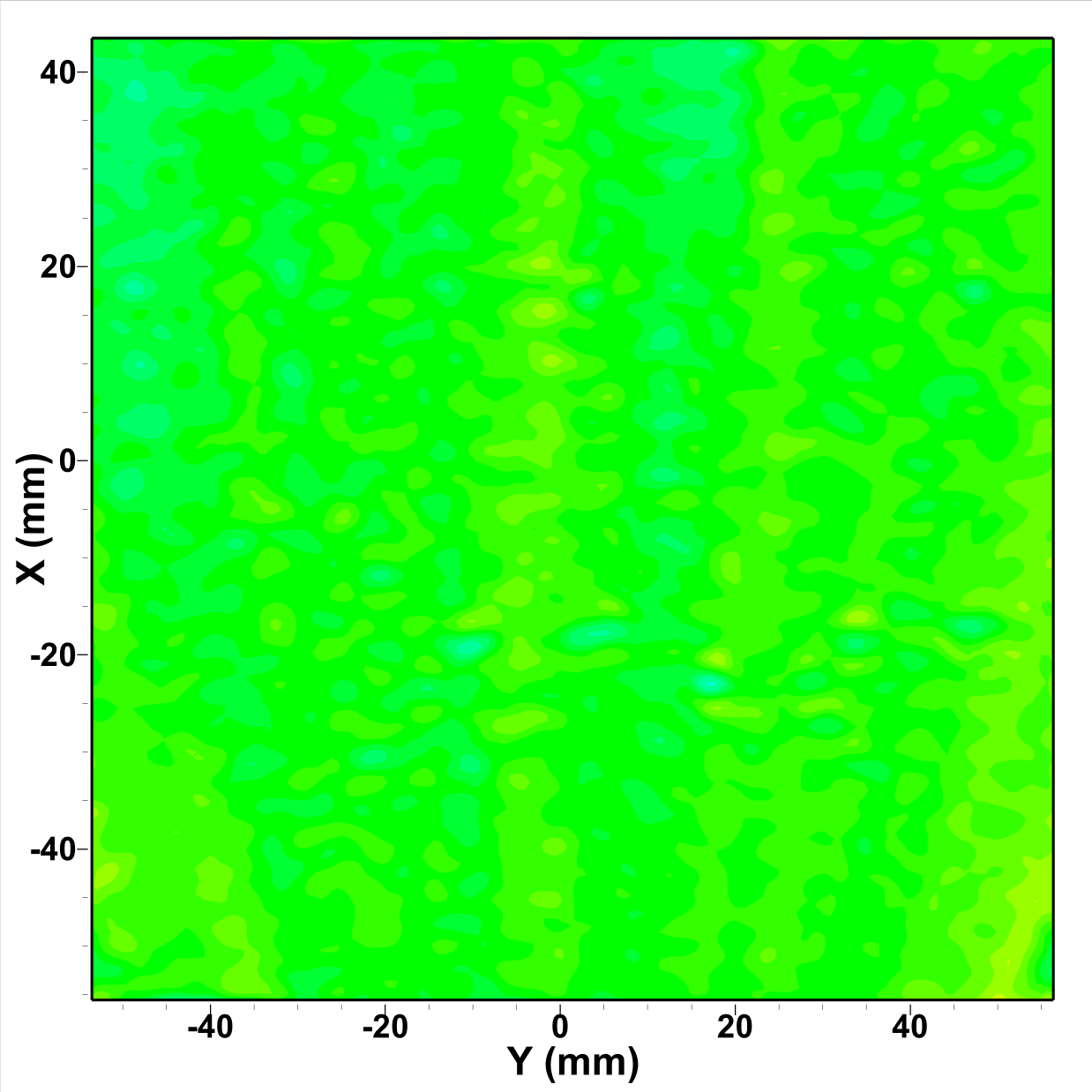}\\
\vspace{-2mm}
\adjustbox{valign=c}{$\left \langle\omega_{z}\right \rangle$} & \includegraphics[width=0.375\textwidth,valign=c]{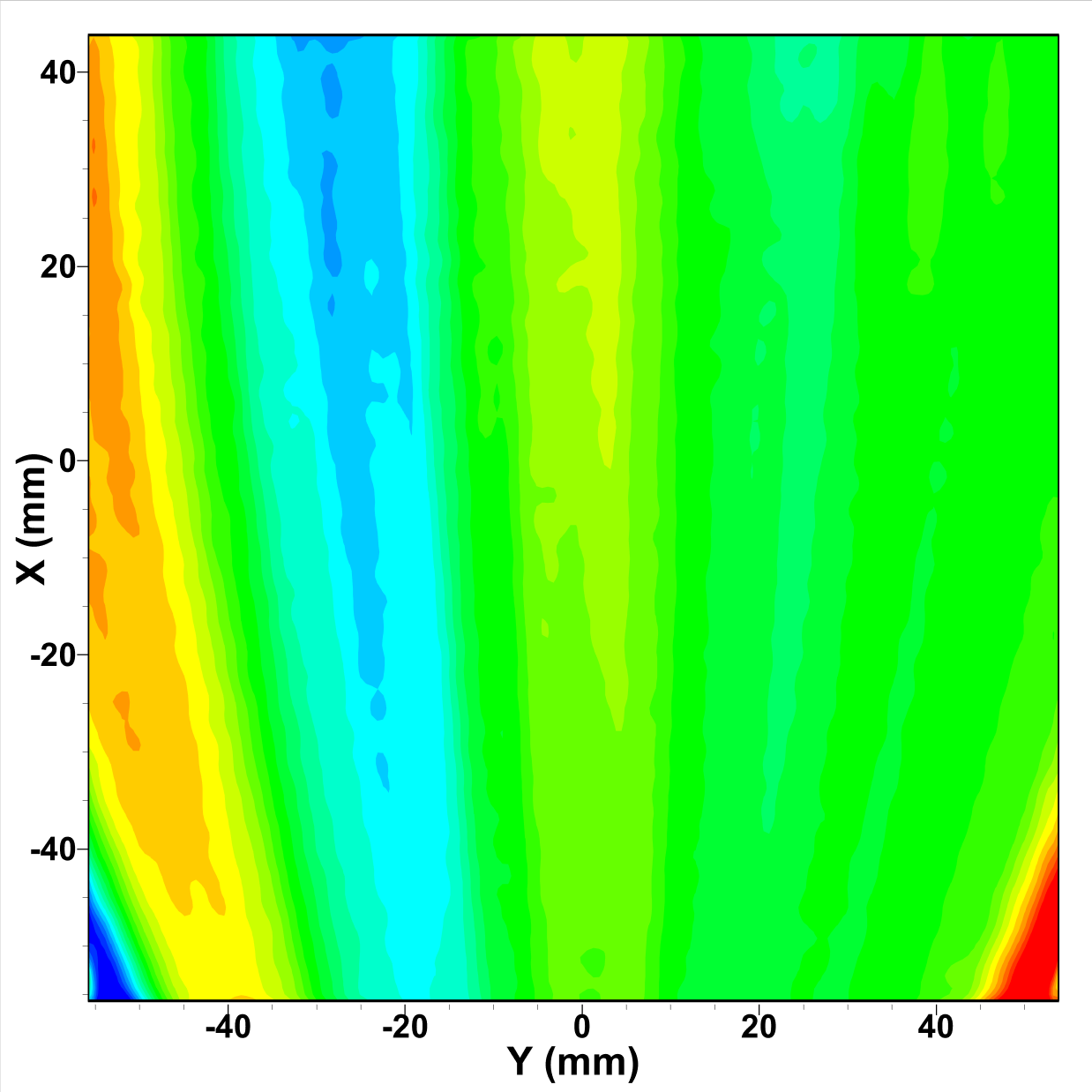}&
\includegraphics[width=0.375\textwidth,valign=c]{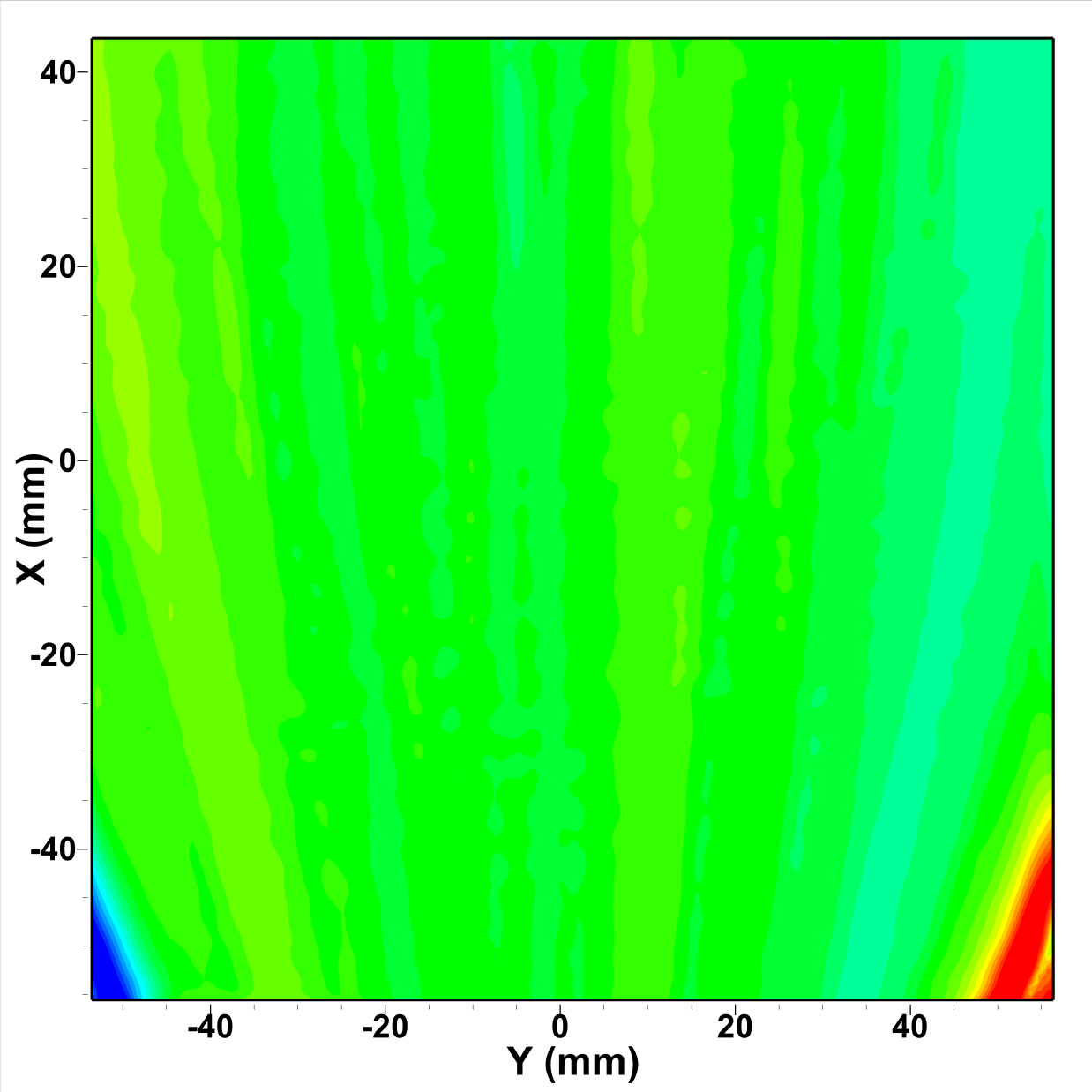}\\
\multicolumn{3}{c}{\includegraphics[width=0.65\textwidth,valign=c]{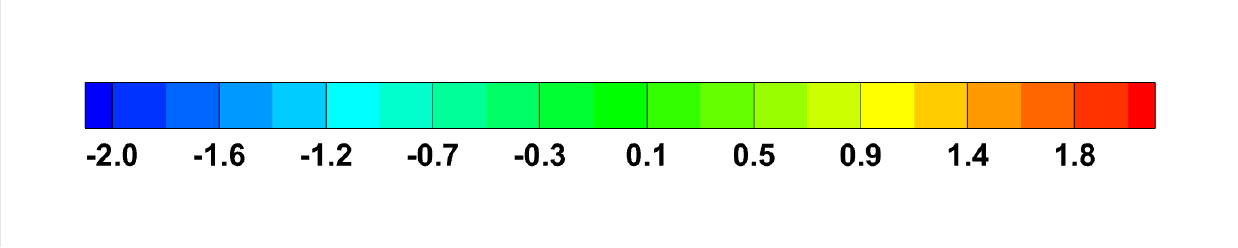}}\\
\end{tabular}
\vspace{-9mm}
\caption{Contour plots of the time-averaged vorticity field ($s^{-1}$) in the mid-$z$ plane of the measurement volume $P1$. The top, middle and bottom rows correspond to the streamwise component $\left \langle\omega_{x}\right \rangle$, the transverse component $\left \langle\omega_{y}\right \rangle$ and the spanwise component $\left \langle\omega_{z}\right \rangle$ respectively.
The  $X-Y$ co-ordinates shown in the figure are local to the measurement region as used in the calibration. 
The origin (0,0) here is the local coordinates and it corresponds to (42,0) in the general coordinate system $x-y$ shown in Figure 1. 
The range of the vertical scale $X=$ [44,$-$56] thereby corresponds to $x=$ [$-$2,99] in the general coordinate system.
The time averaging is done over 50,000 images captured at 1300 frames-per-second.
The overall magnitude values on the centerline are close to zero as expected from $x-$symmetry.
However, more inhomogeneity is observed with the synchronous mode.}
\label{fig:AvgVort}
\end{center}
\end{figure}

\begin{figure}[h!]
\begin{center}
\begin{tabular}{ p{1.0cm} c c }
{} & Synchronous Mode & Random Mode\\
\vspace{-2mm}
\adjustbox{valign=c}{$\omega_{x,rms}$} & \includegraphics[width=0.375\textwidth,valign=c]{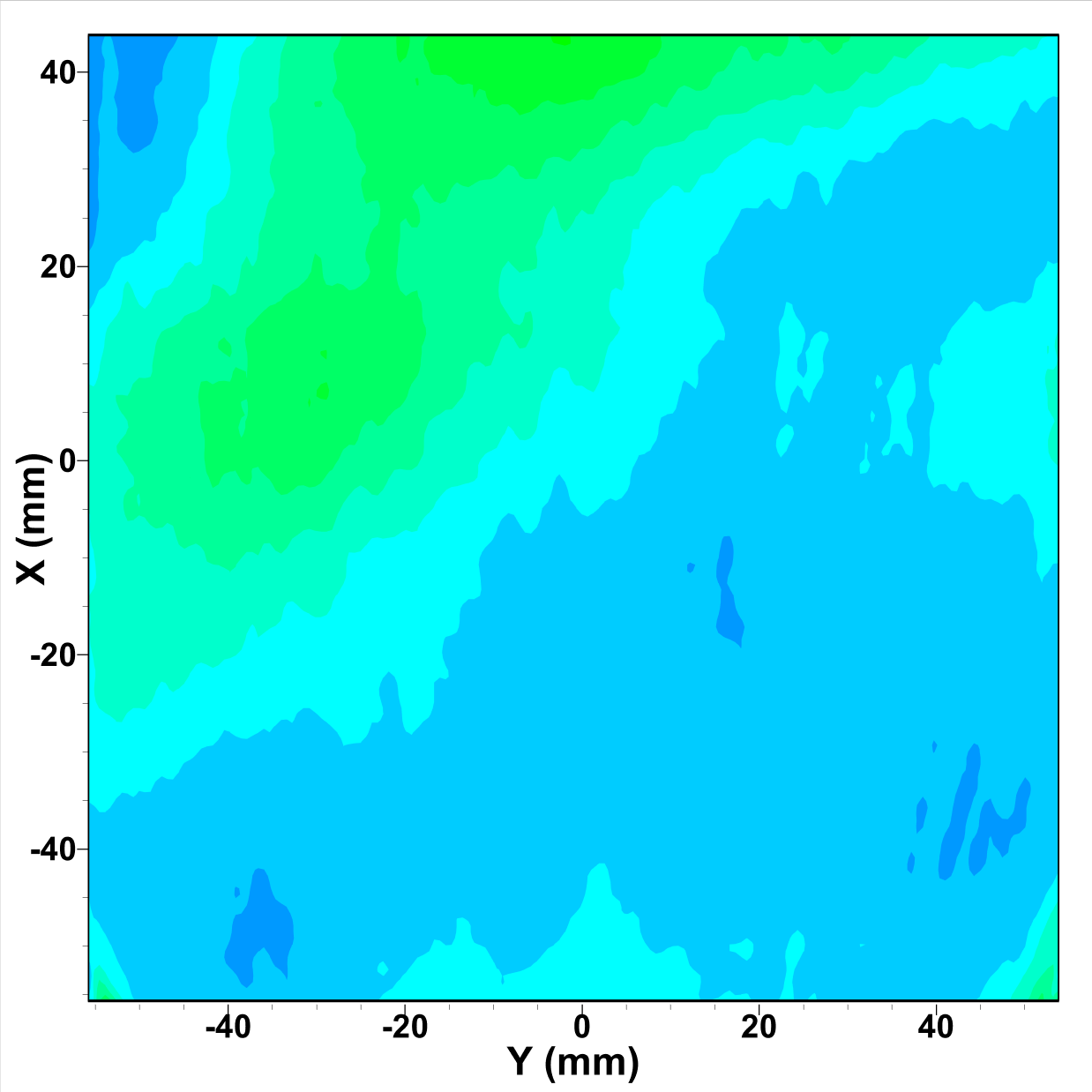}&
\includegraphics[width=0.375\textwidth,valign=c]{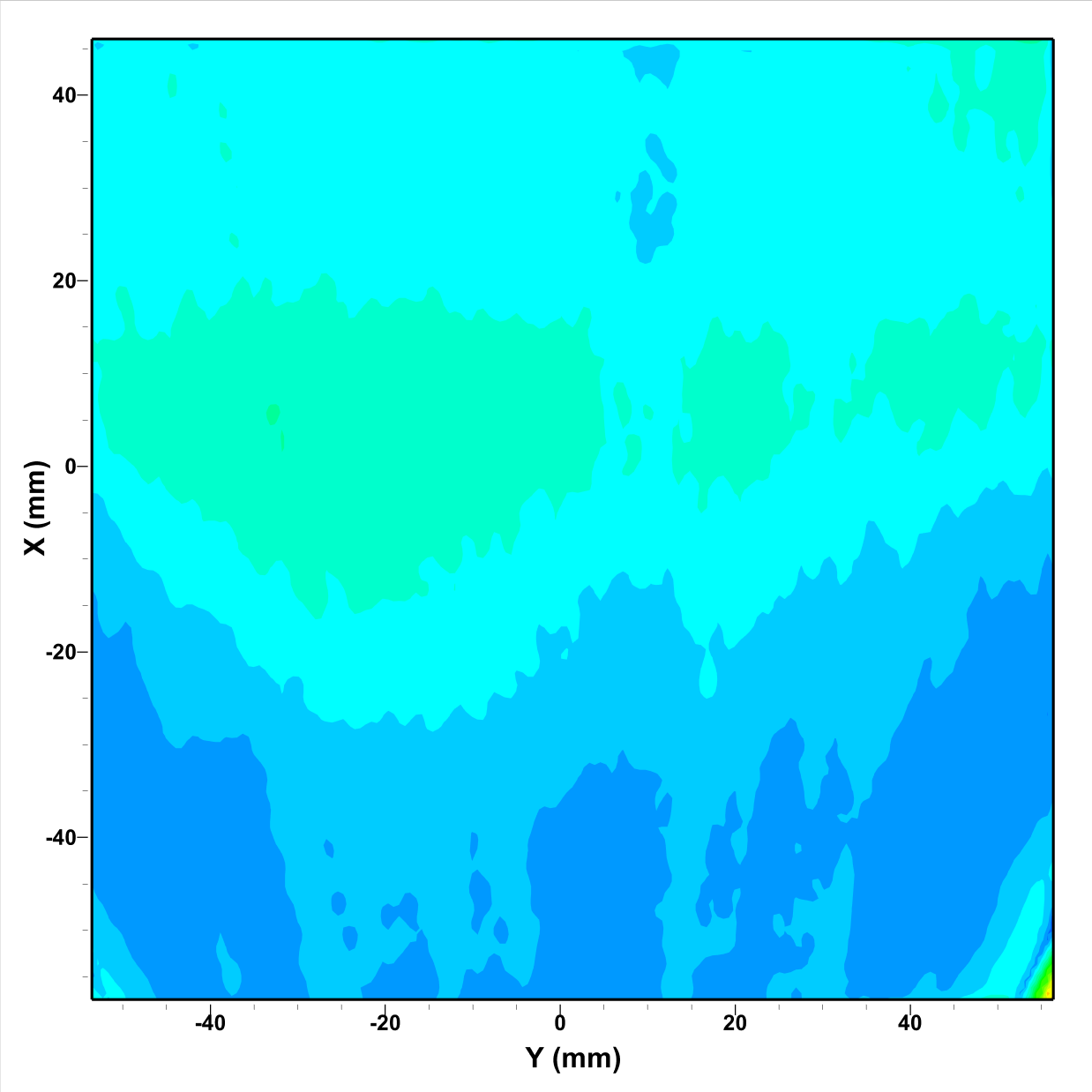}\\
\vspace{-2mm}
\adjustbox{valign=c}{$\omega_{y,rms}$} & \includegraphics[width=0.375\textwidth,valign=c]{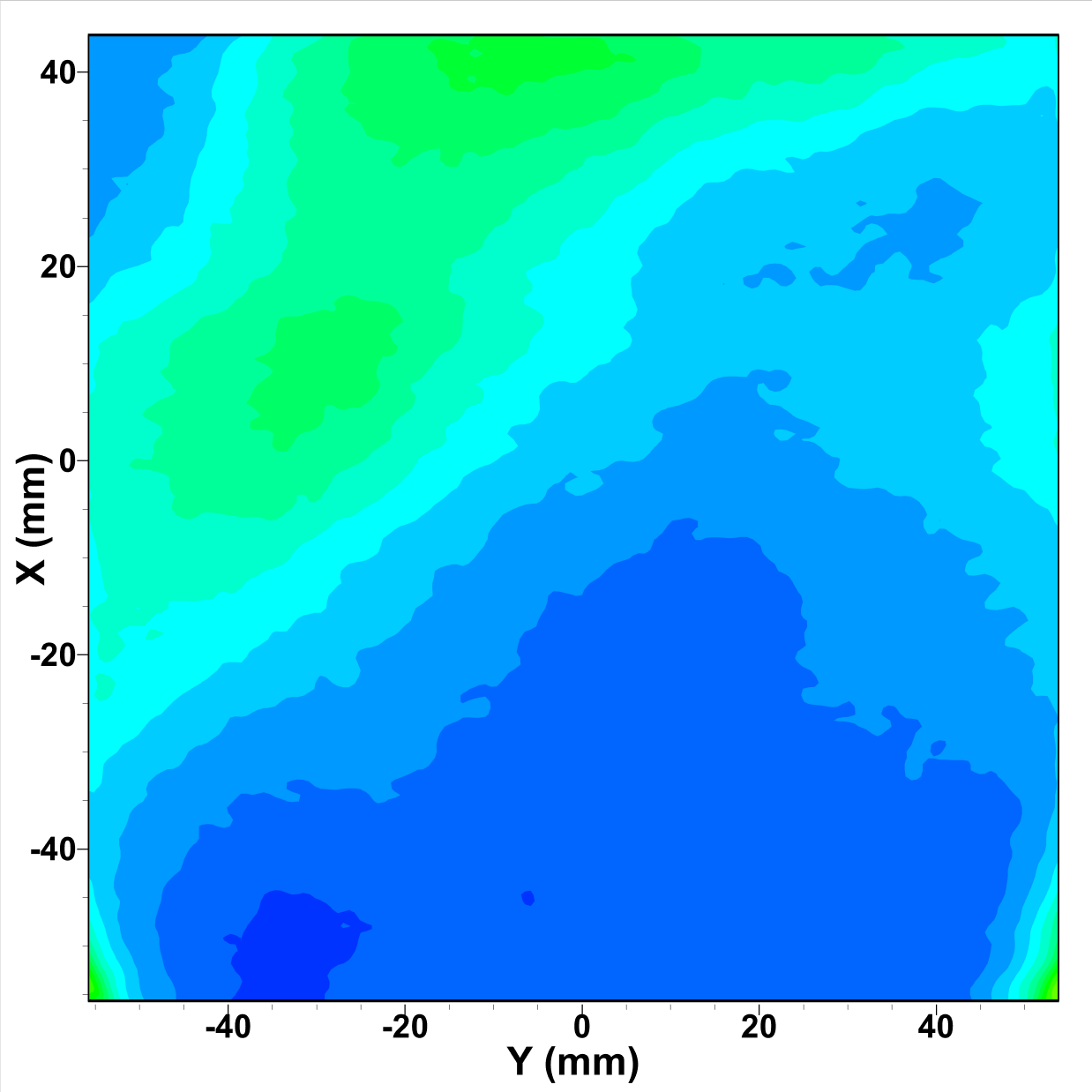}&
\includegraphics[width=0.375\textwidth,valign=c]{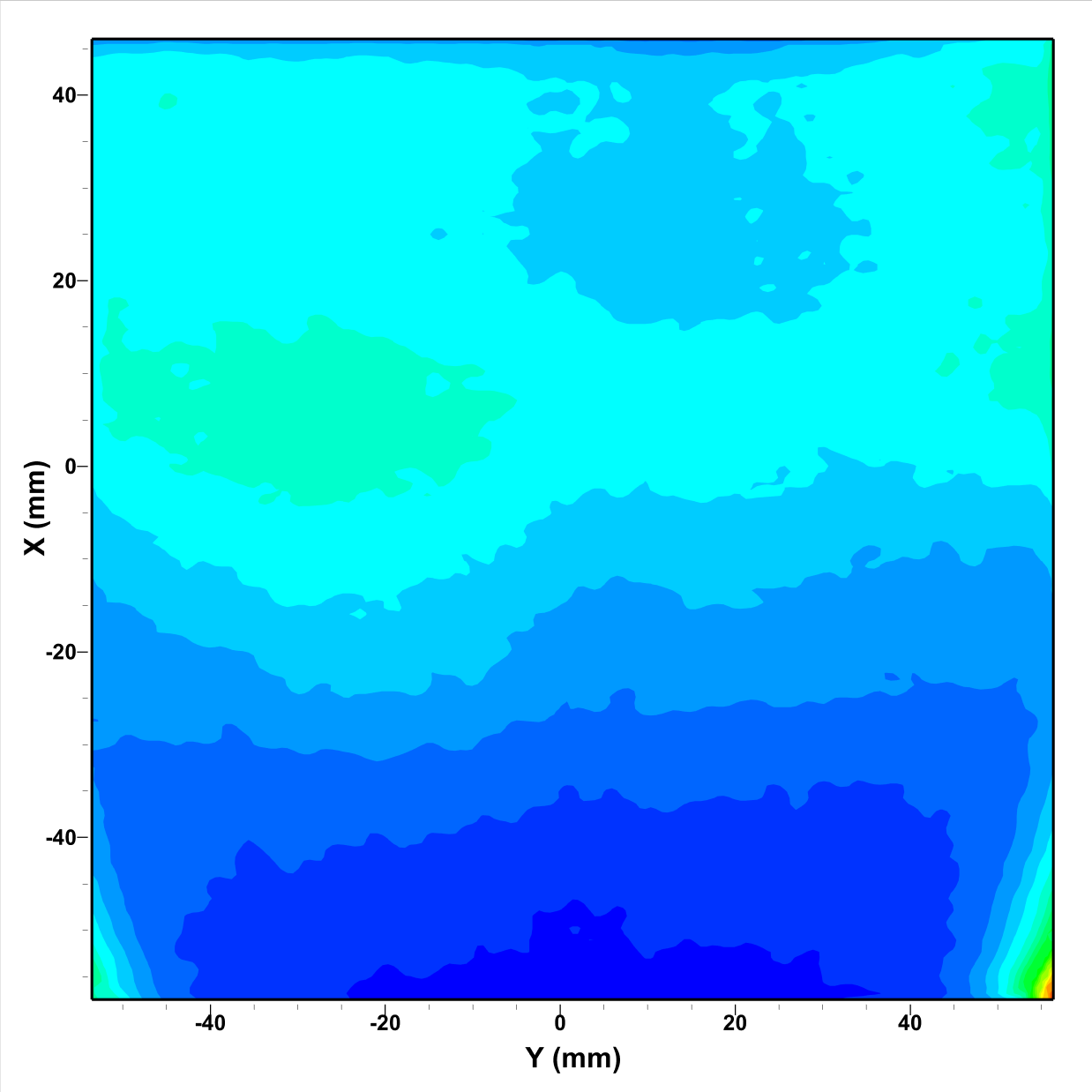}\\
\vspace{-2mm}
\adjustbox{valign=c}{$\omega_{z,rms}$} & \includegraphics[width=0.375\textwidth,valign=c]{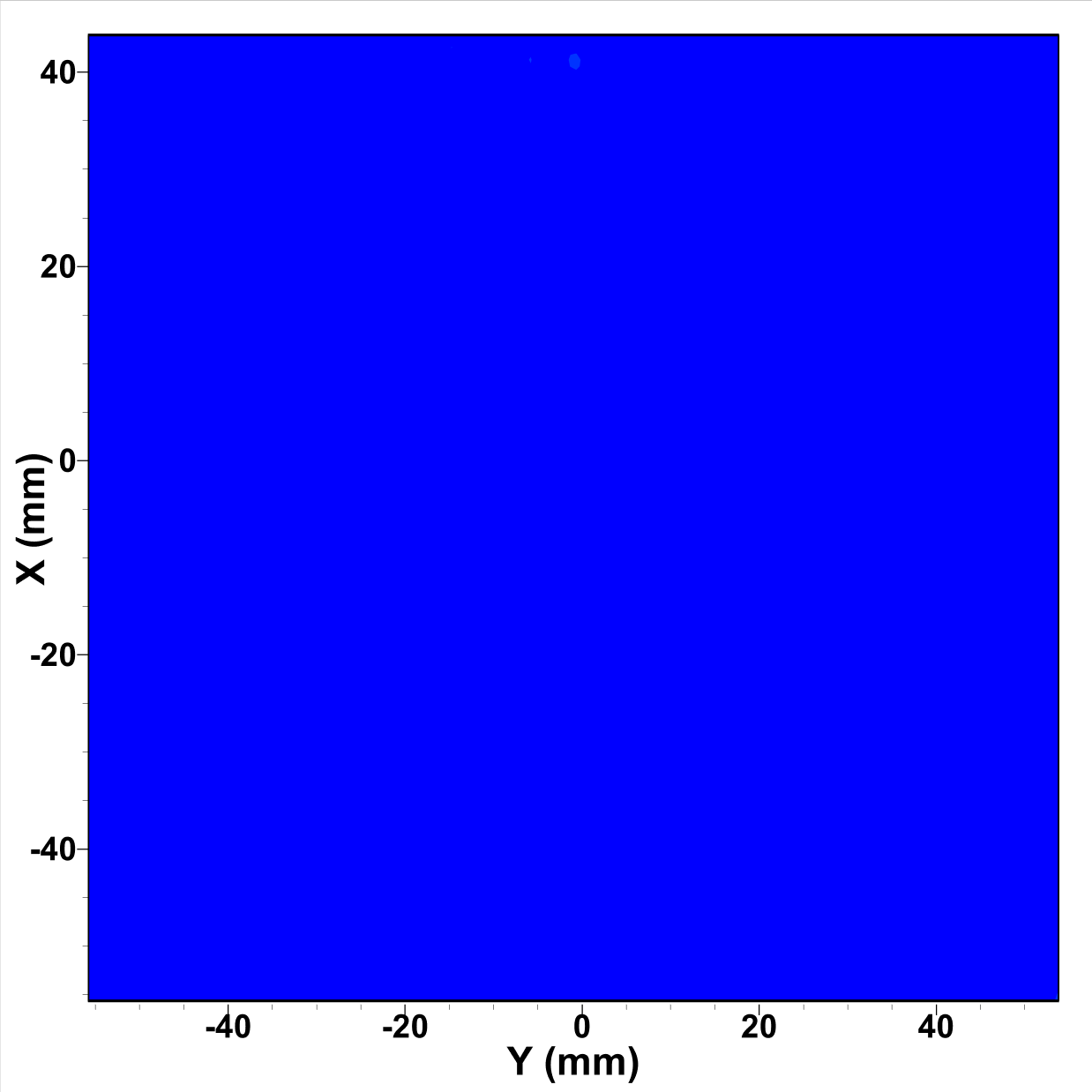}&
\includegraphics[width=0.375\textwidth,valign=c]{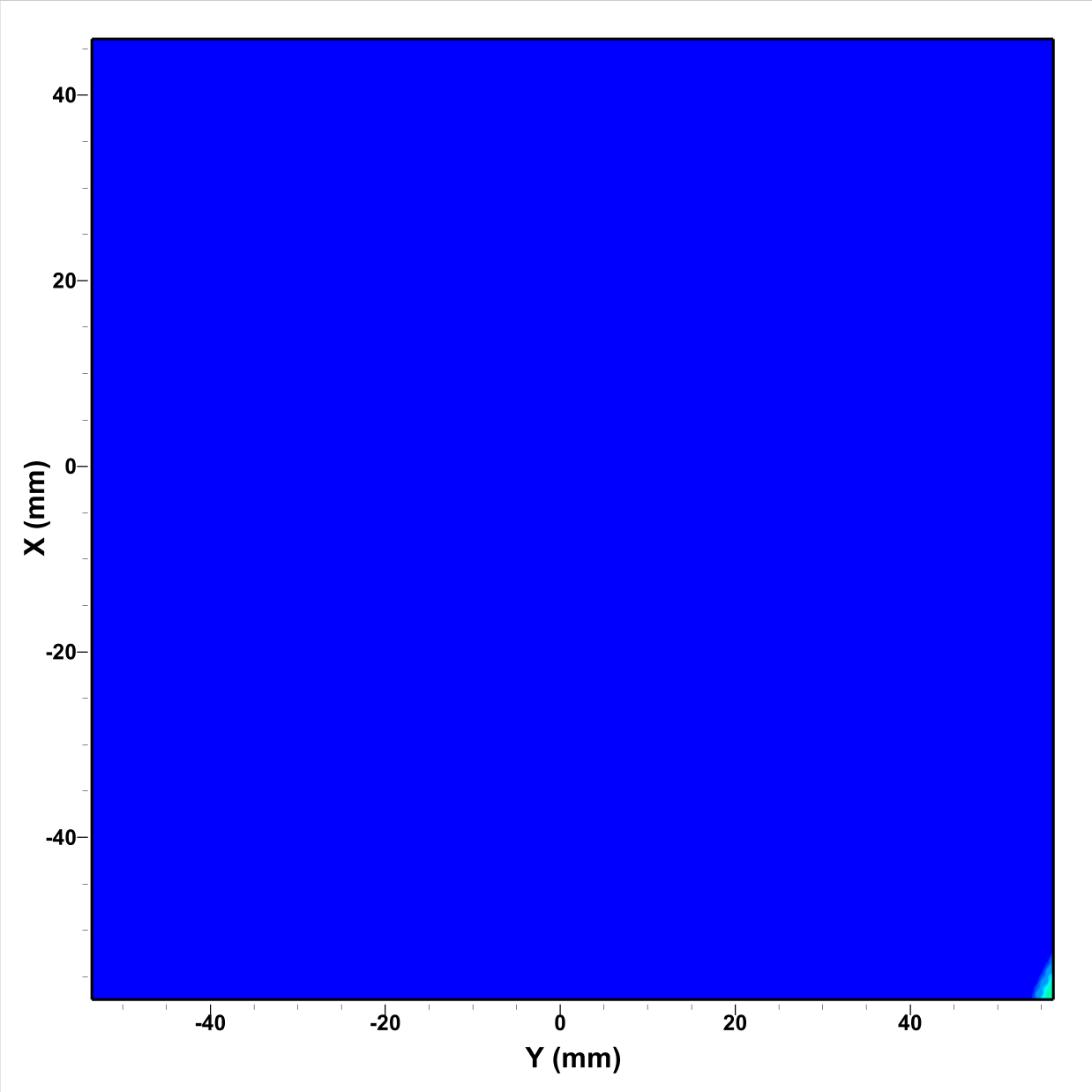}\\
\multicolumn{3}{c}{\includegraphics[width=0.65\textwidth,valign=c]{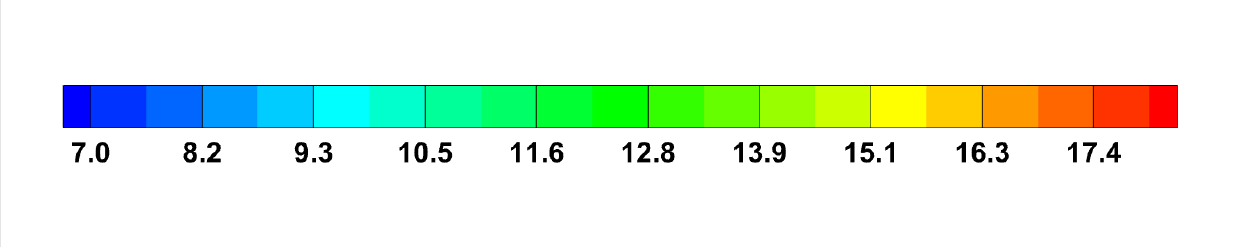}}\\
\end{tabular}
\vspace{-8mm}
\caption{Contour plots of the r.m.s vorticity field ($s^{-1}$) in the mid-$z$ plane of the measurement region $P1$. The top, middle and bottom rows correspond to the streamwise component $\omega_{x,rms}$, the transverse component $\omega_{y,rms}$ and the spanwise component $\omega_{z,rms}$ respectively.
Refer to Figure \ref{fig:AvgVort} for details on the coordinates.
The synchronous mode shows much larger asymmetries in $\omega_{x,rms}$ and $\omega_{y,rms}$, than the random mode.} 
\label{fig:RMSVort}
\end{center}
\end{figure}

\newpage
\section{\textcolor{black}{Mean velocity field}}
\label{sec:meanvelfield}
\FloatBarrier

\begin{figure}[h!]
\begin{center}
\begin{tabular}{ c c }
Synchronous Mode & Random Mode\\
\includegraphics[width=0.485\textwidth,valign=c]{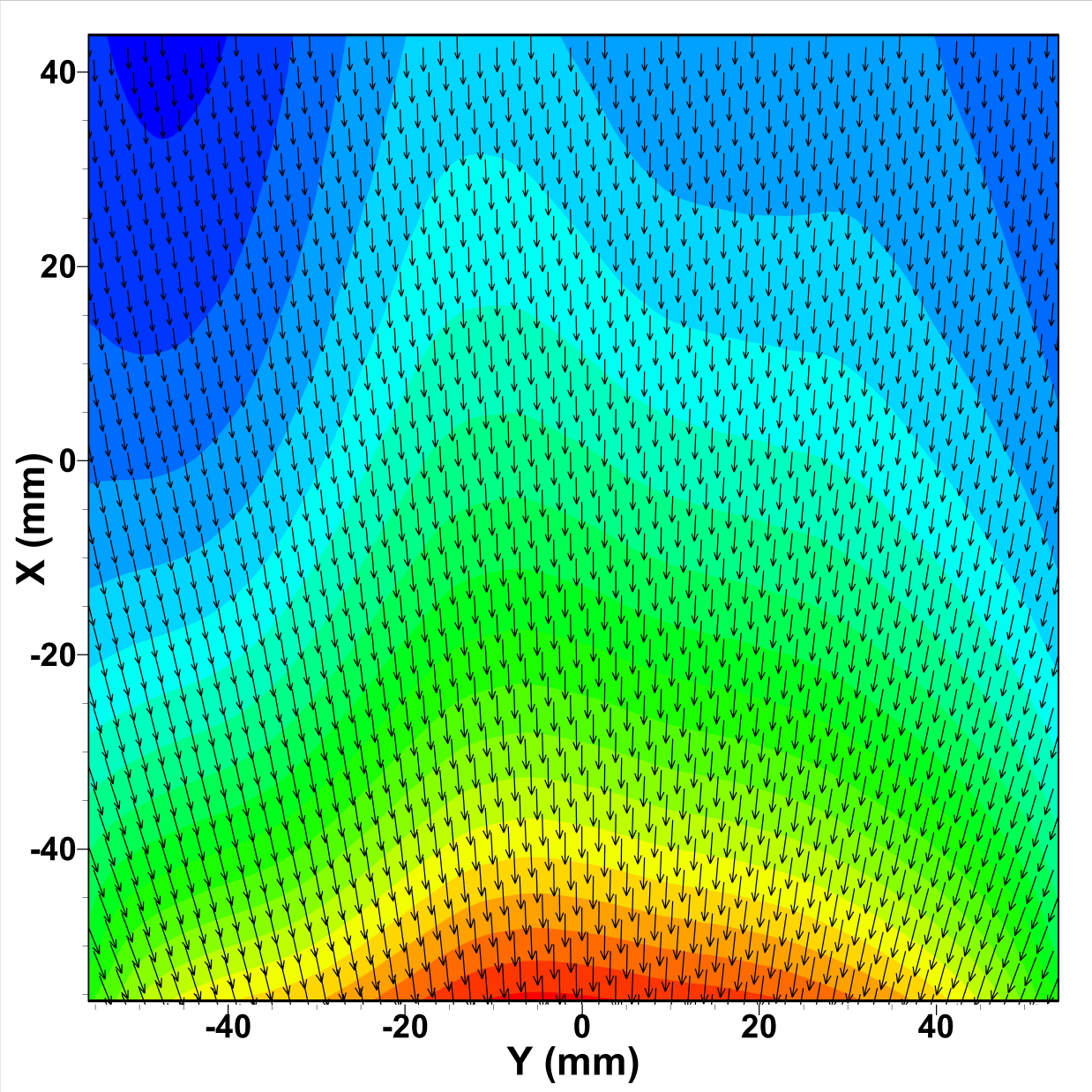} &
\includegraphics[width=0.485\textwidth,valign=c]{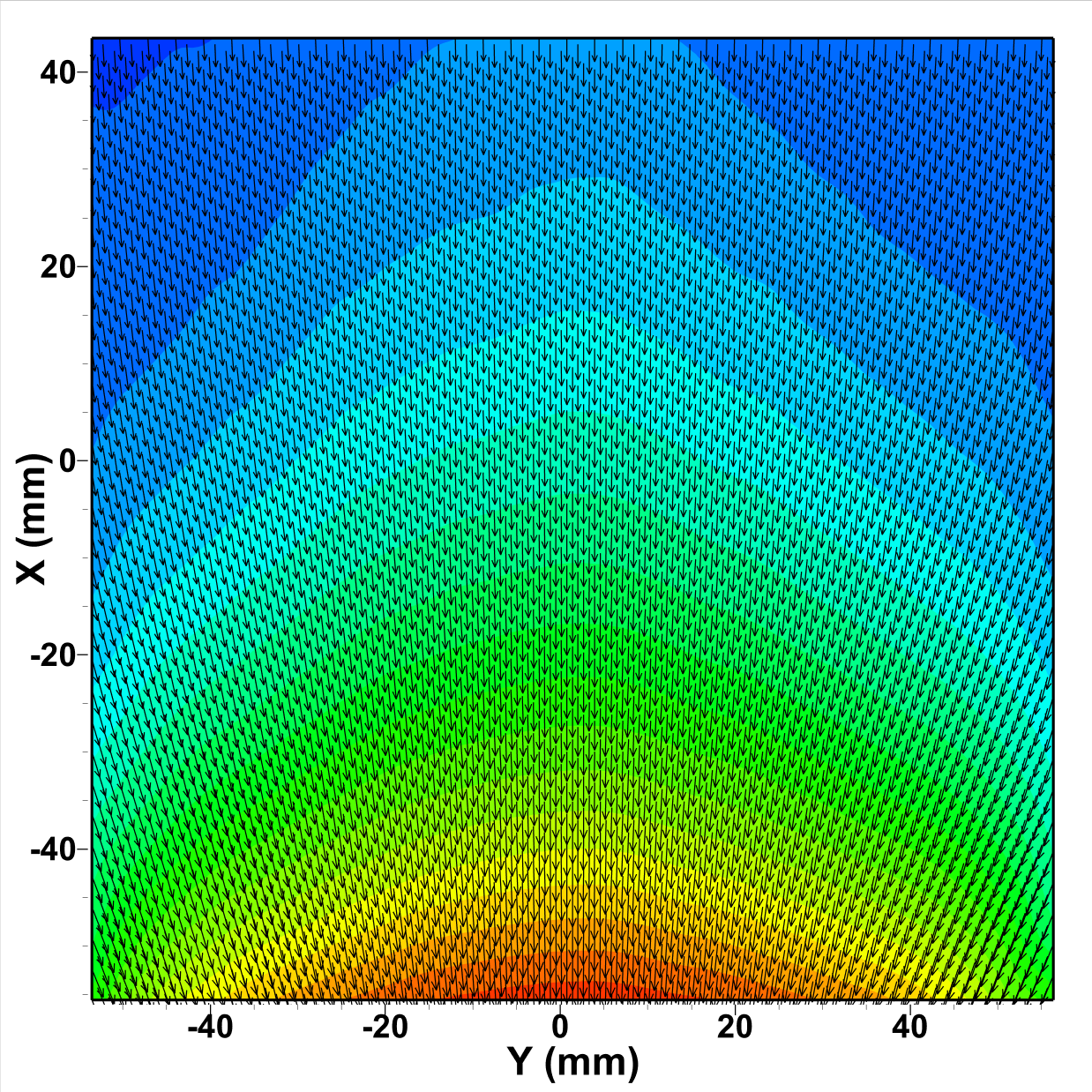}\\
\multicolumn{2}{c}{\includegraphics[width=0.45\textwidth,valign=c]{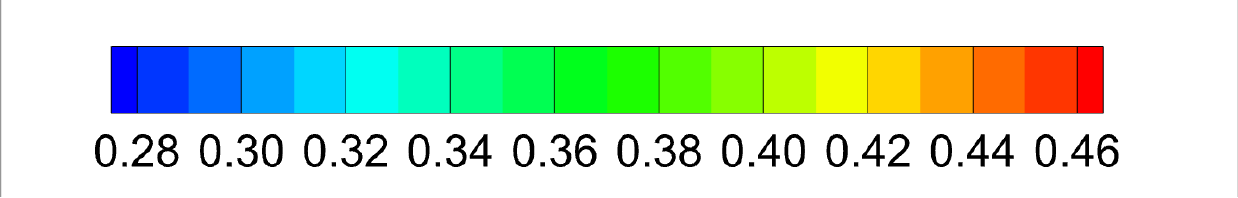}}\\
\includegraphics[width=0.2\textwidth,valign=c]{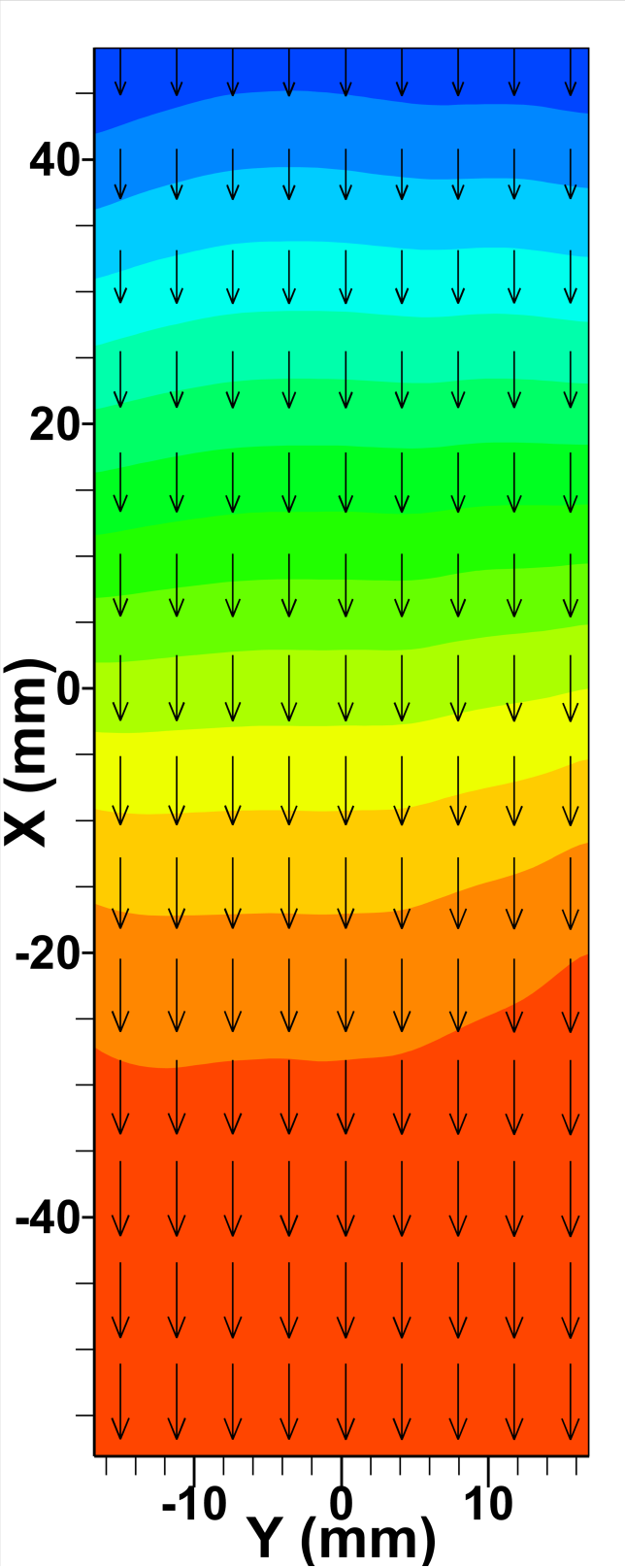} &
\includegraphics[width=0.2\textwidth,valign=c]{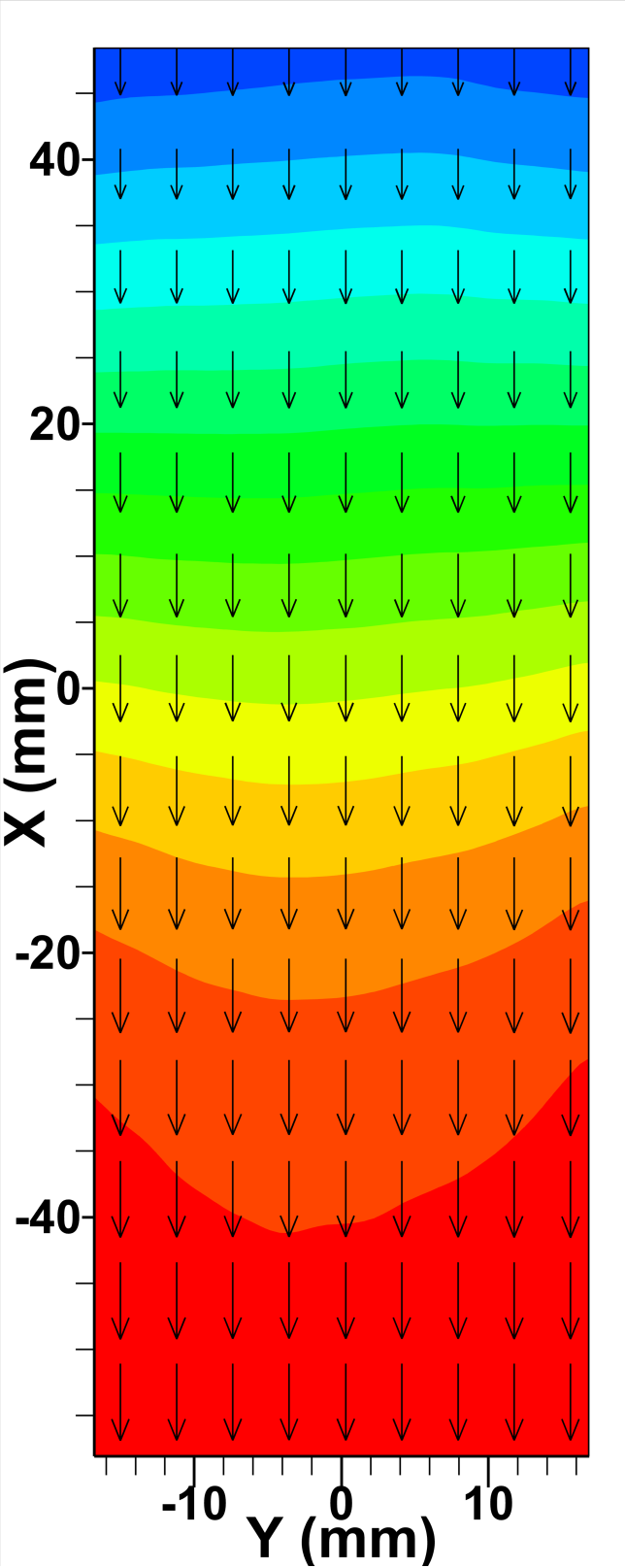}\\
\multicolumn{2}{c}{\includegraphics[width=0.35\textwidth,valign=c]{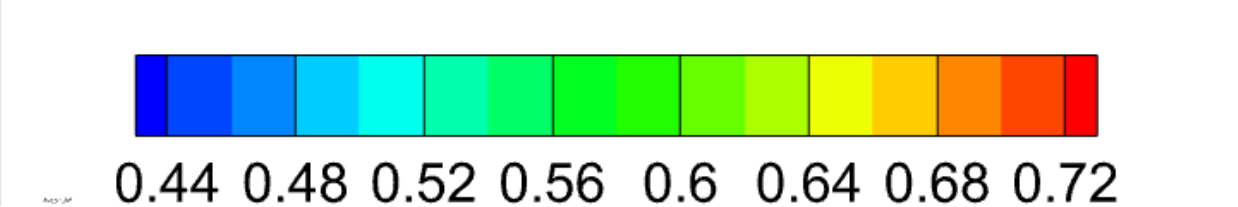}}\\
\end{tabular}
\caption{\textcolor{black}{Contours of the time-averaged streamwise velocity field $\left \langle U \right \rangle$ ($ms^{-1}$)  and vector plots of the time-averaged velocity field $\left \langle (U,V) \right \rangle$ ($ms^{-1}$) in the mid-$z$ plane of the measurement volumes, P1 (top row) and P2 (bottom row). 
The $X-Y$ co-ordinates shown in the figure are local to the measurement region, as used in the calibration. 
In P1, the origin (0,0) here is the local coordinates and it corresponds to (42,0) in the general coordinate system $x-y$; and in P2 the origin corresponds to (147,0) in the general coordinate system shown in Figure 1. 
Thus, the range of the vertical scale in P1, $X=$ [44,$-$56] corresponds to $x=$ [$-$2,99]; and in P2 $X=$ [48,$-$58] corresponds to $x=$ [100,206]. 
Only a few of the vectors are shown for clarity.
For P1, vectors are skipped for every (12,12) grid points and in P2 it is skipped for every (6,12) grid points.}}
\label{fig:AvgUVect}
\end{center}
\end{figure}

\newpage
\section{\textcolor{black}{Intermittency of Circulation \& Scaling Exponents}}
\label{sec:Intermittency}
\FloatBarrier


\begin{figure}[h!]

\begin{tabular}{c c}
\qquad\qquad\qquad\qquad$r\sim$ 4$\eta$ & \qquad\qquad\qquad\qquad\qquad$r\sim$ 50$\eta$\\
\multicolumn{2}{c}{(a)}\\
\multicolumn{2}{c}{\noindent\makebox[0.94\textwidth]{\includegraphics[width=1.2\textwidth,trim={0 2cm 0 2cm},clip]{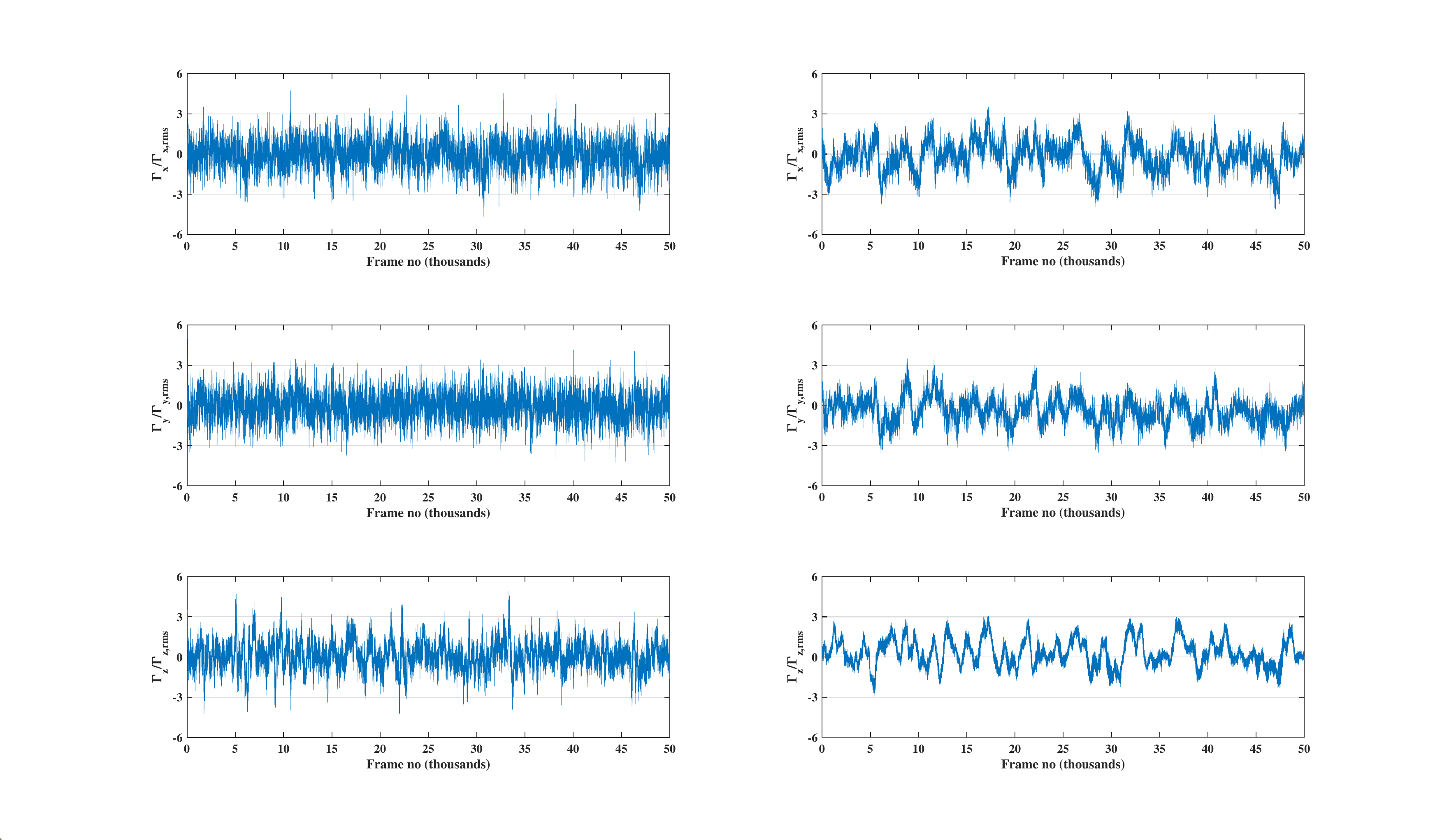}}}\\
\end{tabular}
\begin{tabular}{c c}
\multicolumn{2}{c}{(b)}\\
\multicolumn{2}{c}{\noindent\makebox[0.94\textwidth]{\includegraphics[width=1.2\textwidth,trim={0 2cm 0 2cm},clip]{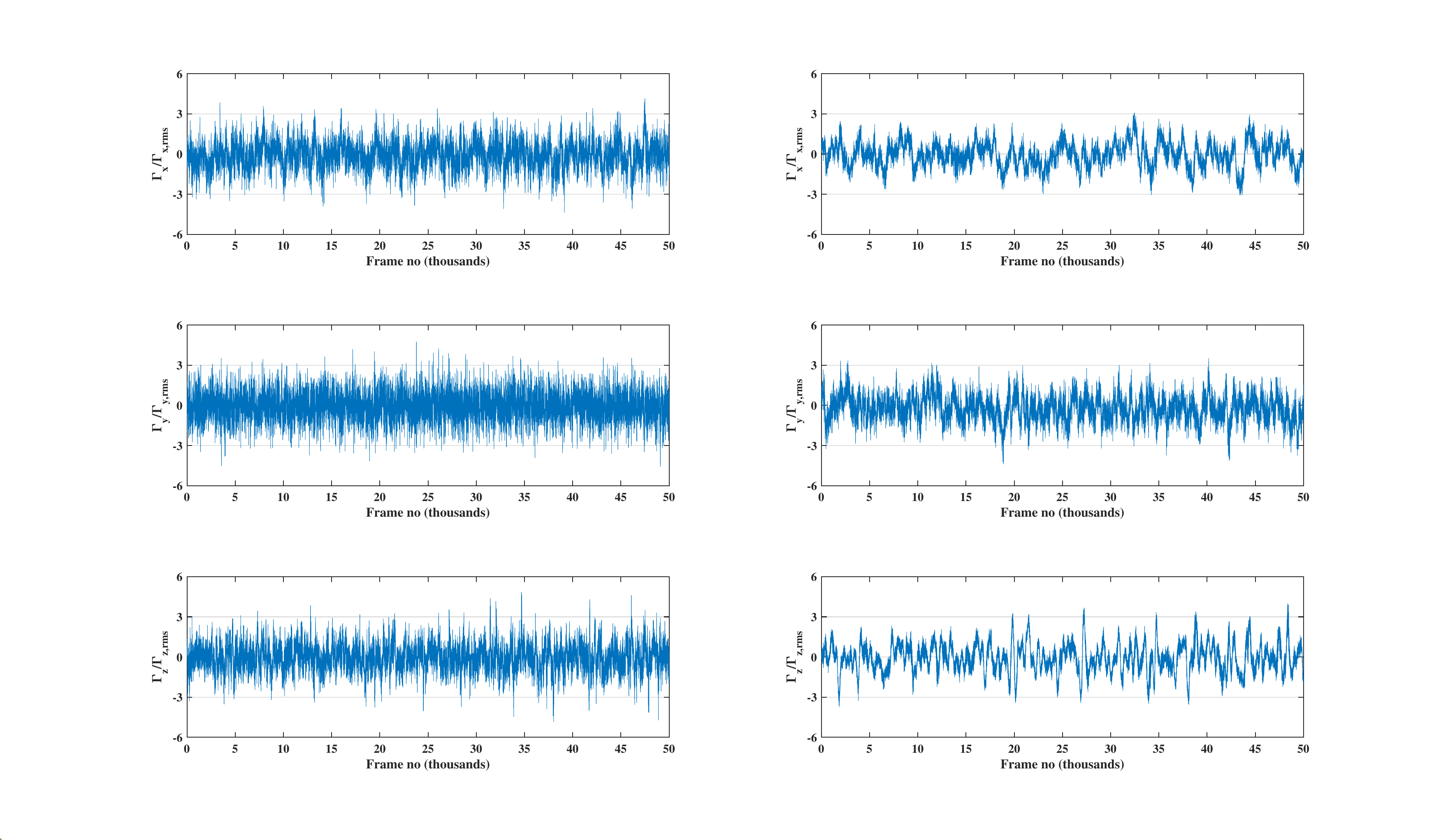}}}\\
\end{tabular}
\caption{\textcolor{black}{The time traces of instantaneous circulation normalized by the respective r.m.s, $\Gamma_i/\Gamma_{i,rms}$ computed for square loops around the centerline of the contraction of size $r\sim$ 4$\eta$ (left) and $r\sim$ 50$\eta$ (right) for the synchronous mode at (a) $x=$ 8 mm, and (b) $x=$ 190 mm. The $x-$axis represents the frame number and each frame is spaced by $\approx$ 0.8 ms.}}
\label{fig:timeseries}
\end{figure}

\begin{figure}[h!]
\begin{tabular}{ c c }
{(a)} & {(b)}\\ 
\includegraphics[width=0.5\textwidth,valign=t]{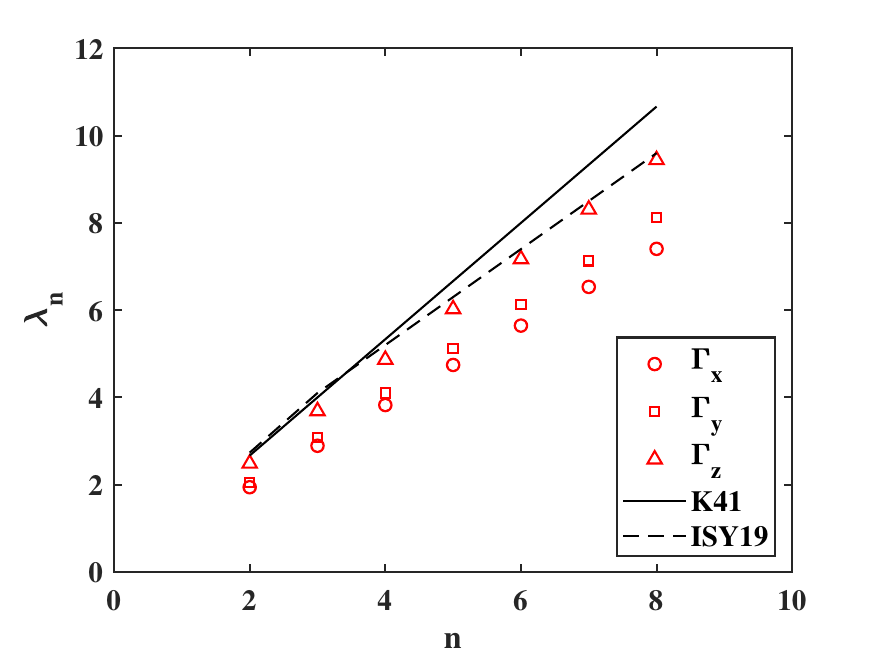} &  \includegraphics[width=0.5\textwidth,valign=t]{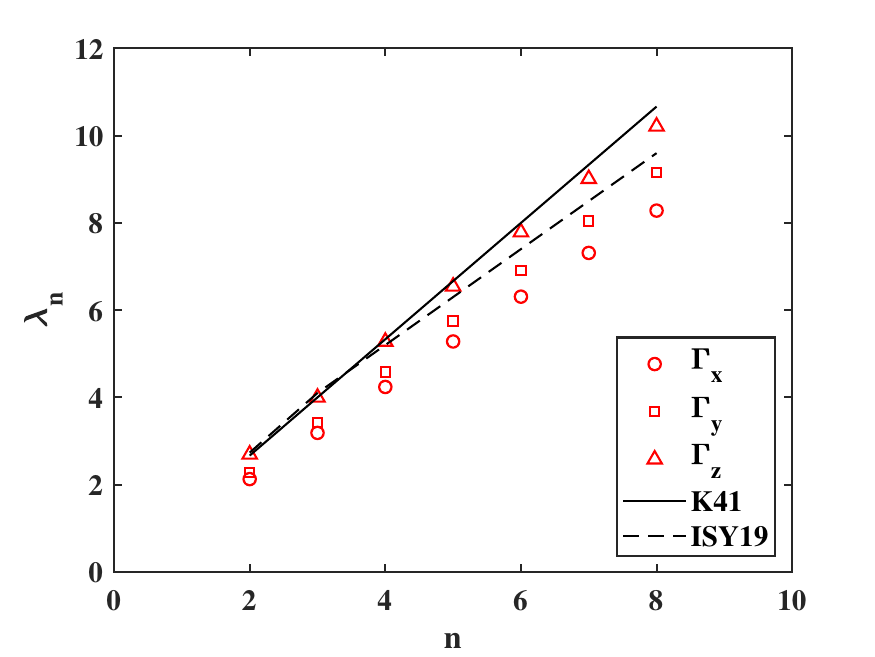}\\
\end{tabular}
\caption{\textcolor{black}{The power-law exponent for moments of circulation-PDF, $ \left \langle \left |\Gamma \right |^{n} \right \rangle$, at $x=$ 8 mm with synchronous (a) and the random mode (b). 
Refer to caption of Figure 12 
for details.}}
\label{fig:PDFmomx8}
\end{figure}\vspace{0.25in}

\section{\textcolor{black}{Uncertainty and Error quantification}}
\label{sec:Error}
\begin{figure}[h!]
\begin{tabular}{ c c }
{(a)} & {(b)}\\ 
\includegraphics[width=0.5\textwidth,valign=t]{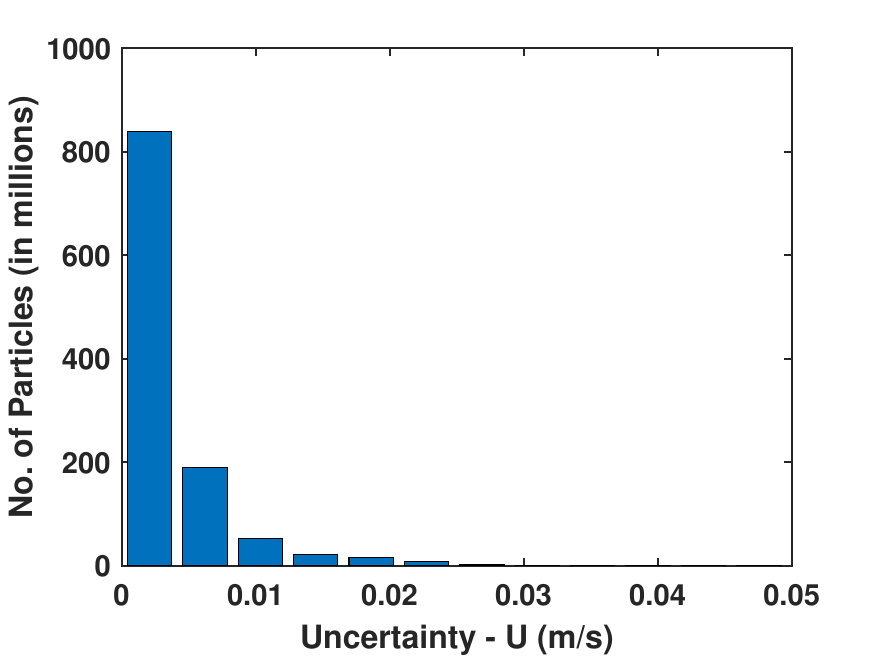} &  \includegraphics[width=0.5\textwidth,valign=t]{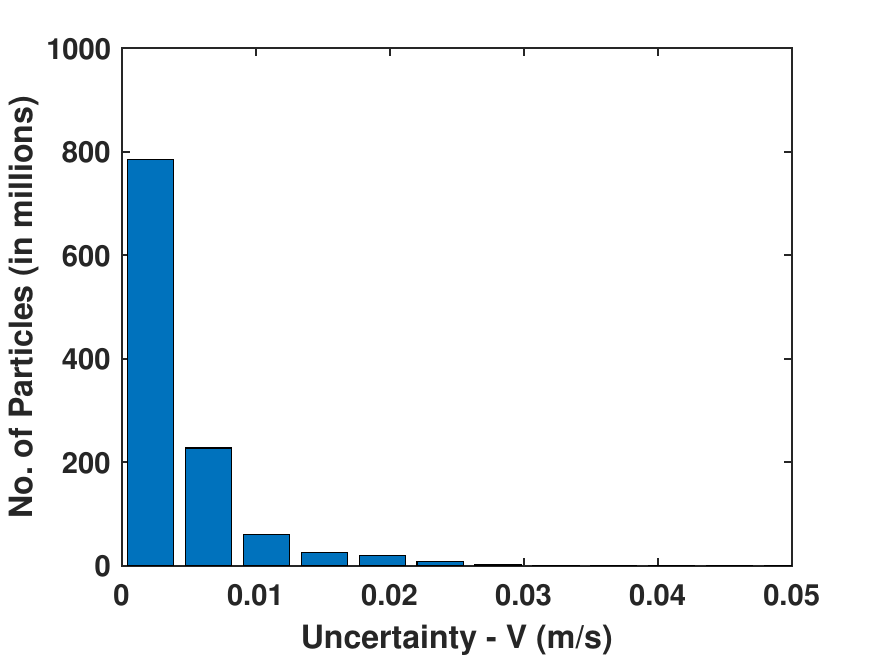}\\
\multicolumn{2}{c}{(c)}\\
\multicolumn{2}{c}{\includegraphics[width=0.5\textwidth,valign=c]{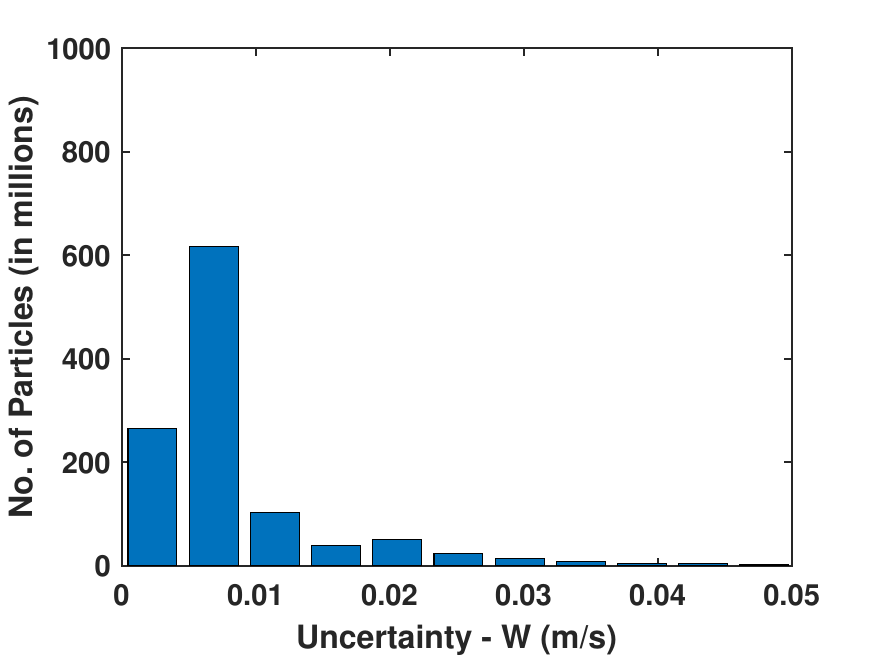}}\\
\end{tabular}
\caption{\textcolor{black}{The histogram of uncertainty in the particle velocities $u$ (a), $v$ (b) and $w$ (c) predicted using the particle trajectories in the measurement region P1.
    The uncertainty is obtained using $\sim$10$^9$ particle samples over 20,000 volumes.}}
\label{fig:uncert}
\end{figure}

\textcolor{black}{We present the uncertainty calculated using the DaVis 10.2 program from LaVision.
The uncertainty is estimated for the particle tracks and the velocities.
The estimation is based on the method by Janke and Michaelis \cite{Janke2021}, wherein they use linear regression analysis to calculate confidence bands of the polynomial fits of the particle trajectories to estimate the uncertainties in the particle positions and velocity. 
The histogram of uncertainty values from using $\sim 10^9$ particles from about 20,000 volumes is shown in Figure \ref{fig:uncert}. 
The 1-sigma values of uncertainties in the measurement regions, as a percentage of the mean inlet velocity are:\\
\\
(1) P1: velocity $u$, $v$ \& $w$ are 1.4\%, 1.5\% \& 2.5\% respectively.\\
(2) P2: velocity $u$, $v$ \& $w$ are 1.6\%, 1.8\% \& 4.0\% respectively.\\}
\\
\textcolor{black}{As the flow is accelerating, the percentage is based on the mean velocity at the inlet to the contraction, which gives a conservative local effect.}

\textcolor{black}{How does this compare to other studies in the literature?  Sciacchitano et al. \cite{Sciacchitano2021} compared the accuracy of the various Lagrangian Particle Tracking algorithms by various research groups, which included the LaVision algorithm that we use in the current study.  
The comparisons were done using synthetic experimental data of a turbulent wall-bounded flow in the wake of a cylinder.
The algorithms accuracy in predicting velocities were in the range 0.3 $-$ 2.5\% for particle concentration varying between 0.005 and 0.016 particles-per-pixel (ppp), which is in line with what we see herein.}

\textcolor{black}{The actual uncertainties in the particle position are 0.08, 0.1 and 0.2 pixel in the $x$, $y$ and $z$ directions which is also inline with the Lagrangian particle tracking measurements using shake-the-box algorithm in the literature. 
Rowin and Ghaemi \cite{Rowin2019} measured near-wall turbulent flow over a superhydrophobic surface using 3-D LPT.
They used DaVis 8.4 to implement the shake-the-box algorithm and report in-plane \& out-of-plane position accuracy of 0.1 \& 0.2 pixel respectively.
Schröder and Schanz \cite{Schroder2023} in their recent review of the latest 3-D Lagrangian particle tracking methods, report that the position accuracy of the particle in LPT is typically around 0.1 pixel.}

\textcolor{black}{We quantify the uncertainty in the circulation based on the uncertainty in the velocities by performing Monte Carlo simulations.
Circulation is computed for a loop with an error imposed on the measured velocities using random numbers as $\Gamma_{i,e}$.
The random number imposed on each velocity component follows a normal distribution with zero mean and the corresponding standard deviation reported above. 
The error in the circulation is computed as $\Gamma_{i,e}-\Gamma_{i}$, where $\Gamma_{i}$ is the actual circulation without the imposed error.
Considering the 75\% overlap of adjacent grid points, relative to the interpolation from particle tracks, the random error in every fourth velocity vector is assumed to be independent and thus conservatively the random number is kept the same over 4 consecutive grid points.
Therefore, for a contour of length $N=$16 $dx$, we estimate the error $\epsilon_{\Gamma} = \epsilon_{u} / \sqrt{4}$. 
These calculations are performed over 1000 samples to obtain the RMS of the errors.
The RMS thus obtained for an $r=$ 10 mm square loop, for $\Gamma_{x}$, $\Gamma_{y}$ and $\Gamma_{z}$ are about 2.1\%, 2.2\% and 1.1\% respectively.
In line with the velocity error quantification, these percentages are based on mean velocity at the inlet to the contraction and the loop size $r$.  The reduced error from $u$ to $\Gamma$ is consistent with 
Sciacchitano and Wieneke \cite{Sciacchitano2016} who discuss the uncertainty of vorticity for PIV measurements, suggesting that the uncertainty in vorticity can be reduced by computing it using circulation around a small loop, rather than taking velocity derivatives.}

\begin{figure}[h!]
\begin{tabular}{ c c }
{(a)} & {(b)}\\ 
\includegraphics[width=0.5\textwidth,valign=t]{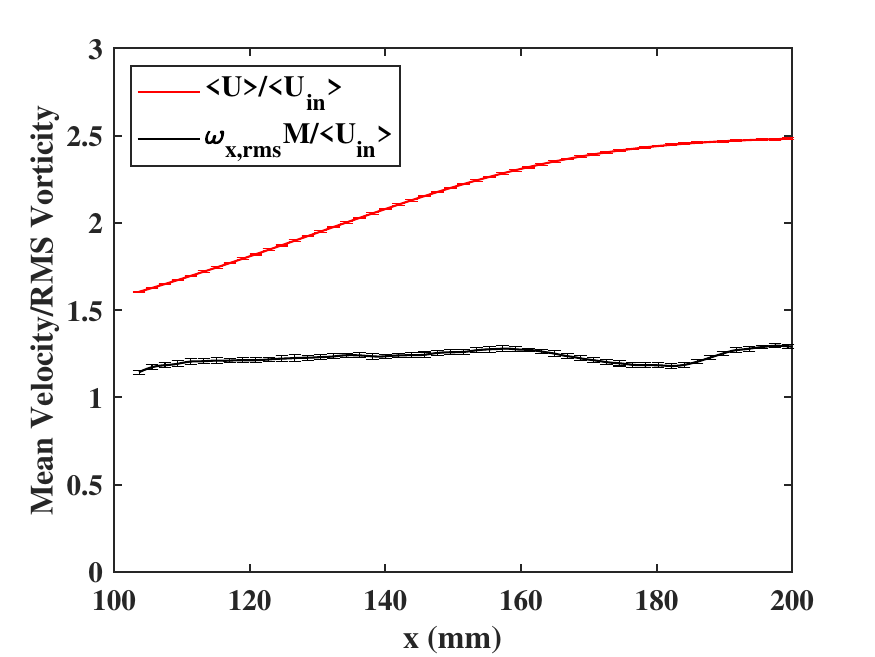} &  \includegraphics[width=0.5\textwidth,valign=t]{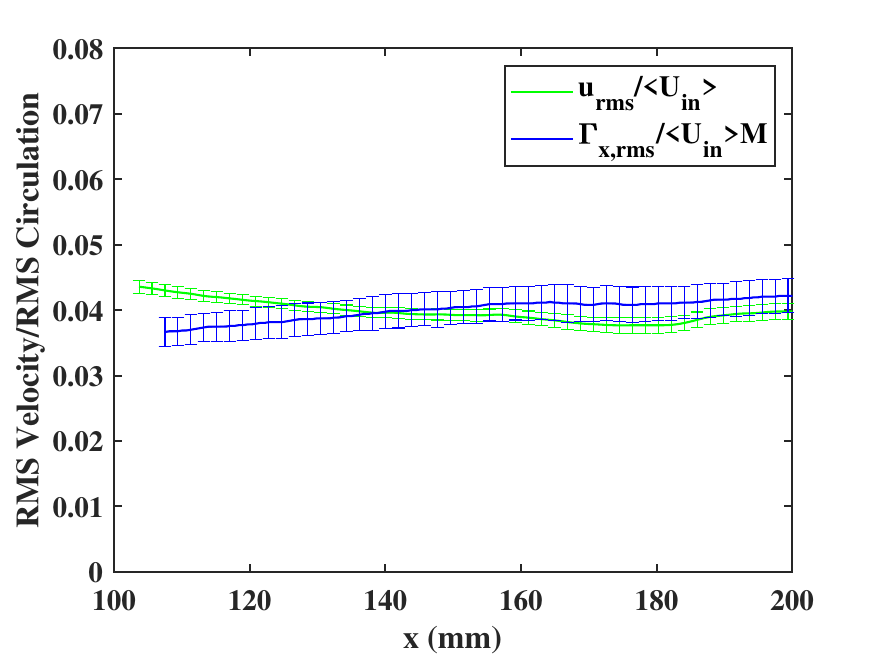}\\
\end{tabular}
\caption{\textcolor{black}{(a) Mean velocity and RMS vorticity with error bars. (b) RMS velocity and RMS circulation in region P2, for the synchronous mode. 
Error bars correspond to the RMS variation between the six experimental runs. 
For clarity, bands have been shown for every third point in $x$.}}
\label{fig:errors}
\end{figure}

\textcolor{black}{The typical measurement convergence error bars associated with the mean and RMS quantities is shown in Figure \ref{fig:errors}.
The convergence error computations are made using the six independent realizations for the synchronous mode, inside the contraction in region P2. 
Error bands shown in the figure correspond to the RMS of the measurements in six different experimental runs.}

\textcolor{black}{The average values for RMS as a percentage of the corresponding mean values are:\\
(1) Mean velocity $\left \langle U \right \rangle$ is 0.2\%;\\
(2) RMS velocities $u_{rms}$, $v_{rms}$, $w_{rms}$ are 2.0\%, 1.6\%, and 1.1\%; and\\
(3) RMS vorticity components $\omega_{x,rms}$, $\omega_{y,rms}$, $\omega_{z,rms}$ are 1.4\%, 1.3\%, and 1.2\% respectively.}

\textcolor{black}{For the circulations computed for an \textcolor{black}{$r=10$ mm} square loop, these values are:\\
(4) RMS circulation $\Gamma_{x,rms}$, $\Gamma_{y,rms}$, $\Gamma_{z,rms}$ are 4.5\%, 2.0\%, and 3.2\% respectively.\vspace{0.25in}}

\section{\textcolor{black}{Testing the Area Rule}}
\label{sec:Area_rule}

\begin{figure}[ht!]
\begin{tabular}{ c c }
{(a)} & {(b)}\\ 
\includegraphics[width=0.5\textwidth,valign=t]{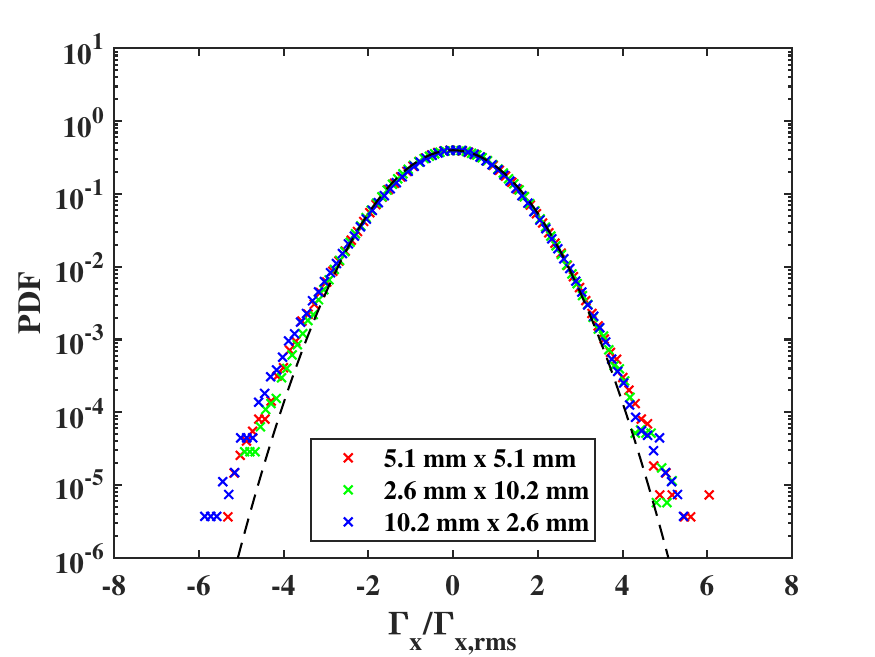} &  \includegraphics[width=0.5\textwidth,valign=t]{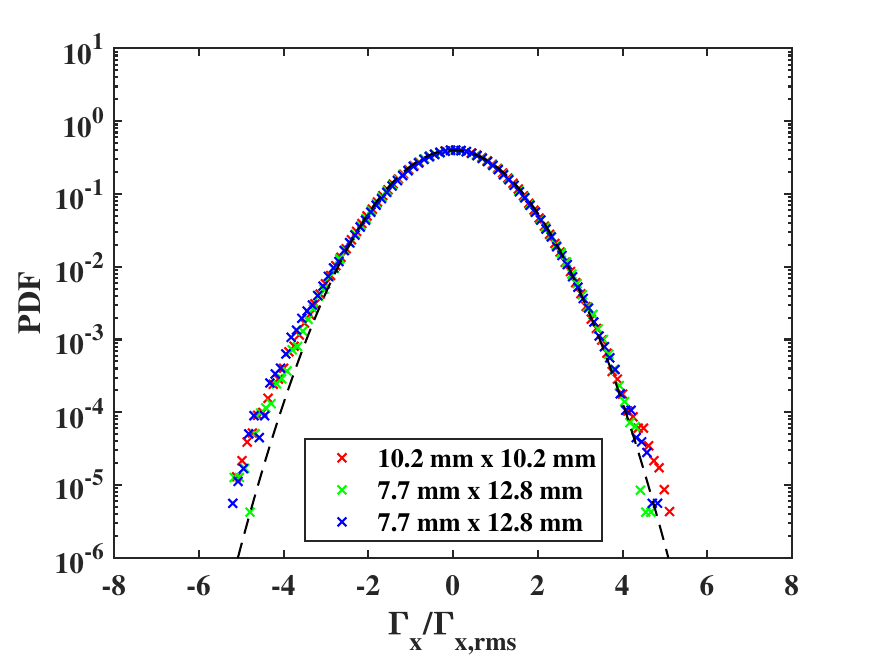}\\
\end{tabular}
\caption{\textcolor{black}{Testing the {\it area rule} for $\Gamma_{x}$.  Comparison of circulation PDFs for 
square and rectangular loops having the same enclosed area. The area of the loops are approximately equal to (a) 25 mm$^2$ and (b) 100 mm$^2$. 
The bin values are normalized by the corresponding r.m.s value.
The Gaussian distribution is shown by the continuous dashed curves.
Data of $\Gamma_{x}$ with synchronous grid-oscillation mode, at inlet $x=$ 8 mm.}}\vspace{5.0in}
\label{fig:arearule}
\end{figure}

\textcolor{black}{Earlier numerical and theoretical studies have focused on the so-called {\it area rule} of the circulation \cite{Migdal1994,Iyer2019}.
Our volumetric data can potentially allow us to test this experimentally.
For this purpose we have computed the PDF of the streamwise circulation $\Gamma_{x}$ with both square  and rectangular loops, which have the same area.  For example loops with outer dimensions of $5.1 \times 5.1$ mm$^2$ vs $2.6 \times 10.2$ mm$^2$.
The loops are taken around the center of the measurement region, just inside the start of the contraction, at $x=$ 8 mm. 
The circulation PDFs obtained for loops of area $A\approx$ 25 and 100 mm$^2$, but three different aspect ratios, are shown in Figure \ref{fig:arearule}.
The normalized PDF for the different loop shapes matches to a great extent, while having some deviations at the tails, which supports the area rule.  
However, the deviations of the tails are due to the limitation in the sampling volume, which in our case are limited by the memory of the cameras.
The larger loop area shows better match of the PDFs, as seen in Figure \ref{fig:arearule}(b).
Similar match of PDFs is seen after the exit of the contraction at $x=$ 190 mm. 
On the other hand, the r.m.s values of $\Gamma_{x}$ varied between 9\%$-$20\% for the 25 mm$^2$ loops, and 2\%$-$7\% for 100 mm$^2$ loops.
A larger sample is thus needed to verify the area rule to cover extreme probabilities at the tails, which is beyond the current experiments.}

\end{document}